%% file: CARS_HE1353_1917_Husemann.tex
\documentclass{aa}
 \usepackage[T1]{fontenc}
 \usepackage[varg]{txfonts}
 \usepackage{multirow}
 \usepackage{color}
 \usepackage{times}
 \usepackage{amssymb}
 \usepackage{verbatim}

 \newcommand{\Ox}{\textsc{[oiii]}}

 \newcommand{\QDeb}{\textsc{qdeblend${}^{\mathrm{3D}}$}}

\newcommand{\farc}{\mbox{\ensuremath{.\!\!^{\prime\prime}}}}

\bibpunct{(}{)}{;}{a}{}{,}

\begin{document}

\title{The Close AGN Reference Survey (CARS).}
\subtitle{A massive multi-phase outflow impacting the edge-on galaxy HE~1353$-$1917}

\author{B.~Husemann\inst{1} 
\and J.~Scharw\"achter\inst{2} 
\and T.~A.~Davis\inst{3} 
\and M.~Pérez-Torres \inst{4,5} 
\and I.~Smirnova-Pinchukova\inst{1} 
\and G.~R.~Tremblay\inst{6} 
\and M.~Krumpe\inst{7}
\and F.~Combes\inst{8}
\and S.~A.~Baum\inst{9,10}
\and G.~Busch\inst{11}
\and T.~Connor\inst{12}
\and S.~M.~Croom\inst{13}
\and M.~Gaspari\thanks{\textit{Lyman Spitzer Jr.} Fellow}\inst{14}
\and R.~P.~Kraft\inst{6}
\and C.~P.~O'Dea\inst{9,10}
\and M.~Powell\inst{15} 
\and M.~Singha\inst{9}
\and T.~Urrutia \inst{7}
}
\institute{
Max-Planck-Institut f\"ur Astronomie, K\"onigstuhl 17, D-69117 Heidelberg, Germany\\
\email{husemann@mpia.de}
\and
Gemini Observatory, Northern Operations Center, 670 N. A’ohoku Pl., Hilo, Hawaii, 96720, USA
\and 
School of Physics \& Astronomy, Cardiff University, Queens Buildings, The Parade, Cardiff, CF24 3AA, UK
\and
Instituto de Astrofísica de Andaluc\'{i}a, Glorieta de las Astronom\'{i}a s/n, 18008 Granada, Spain
\and 
Departamento de F\'{\i}sica Te\'orica, Facultad de Ciencias, Universidad de Zaragoza, E-50009 Zaragoza, Spain
\and 
Center for Astrophysics $|$ Harvard \& Smithsonian, 60 Garden St., Cambridge, MA 02138, USA
\and
Leibniz-Institut f\"ur Astrophysik Potsdam, An der Sternwarte 16, 14482 Potsdam, Germany
\and
LERMA, Observatoire de Paris, PSL Research Univ., Coll\`ege de France, CNRS, Sorbonne Univ., UPMC, Paris, France
\and
Department of Physics \& Astronomy, University of Manitoba, Winnipeg, MB R3T 2N2, Canada
\and
School of Physics \& Astronomy, Rochester Institute of Technology, 84 Lomb Memorial Drive, Rochester, NY 14623, USA
\and
I. Physikalisches Institut der Universität zu Köln, Zülpicher Str. 77, 50937 Köln, Germany
\and
The Observatories of the Carnegie Institution for Science, 813 Santa Barbara St., Pasadena, CA 91101, USA
\and
Sydney Institute for Astronomy, School of Physics, A28, The University of Sydney, NSW, 2006, Australia
\and
Department of Astrophysical Sciences, Princeton University, 4 Ivy Lane, Princeton, NJ 08544-1001, USA
\and
Yale Center for Astronomy and Astrophysics, and Physics Department, Yale University, PO Box 2018120, New Haven, CT 06520-8120
}

\authorrunning{Husemann et al.}
\titlerunning{A massive multi-phase outflow impacting the edge-on galaxy HE~1353$-$1917}
\abstract{Galaxy-wide outflows driven by star formation and/or an active galactic nucleus (AGN) are thought to play a crucial rule in the evolution of galaxies and the metal enrichment of the inter-galactic medium. Direct measurements of these processes are still scarce and new observations are needed to reveal the nature of outflows in the majority of the galaxy population.}{We combine extensive spatially-resolved multi-wavelength observations, taken as part of the Close AGN Reference Survey (CARS), for the edge-on disc galaxy HE~1353$-$1917 to characterize the impact of the AGN on its host galaxy via outflows and radiation.}{Multi-color broad-band photometry is combined with spatially-resolved optical, NIR and sub-mm and radio observations taken with VLT/MUSE, Gemini-N/NIFS, ALMA and the VLA to map the physical properties and kinematics of the multi-phase inter-stellar medium (ISM).}{We detect a biconical extended narrow-line region (ENLR) ionized by the luminous AGN oriented nearly parallel to the galaxy disc, extending out to at least 25\,kpc. The extra-planar gas originates from galactic fountains initiated by star formation processes in the disc, rather than an AGN outflow, as shown by the kinematics and the metallicity of the gas. Nevertheless, a fast multi-phase AGN-driven outflow with speeds up to 1000\,km/s is detected close to the nucleus at 1\,kpc distance. A radio jet, in connection with the AGN radiation field, is likely responsible for driving the outflow as confirmed by the energetics and the spatial alignment of the jet and multi-phase outflow. Evidence for negative AGN feedback suppressing the star formation rate (SFR) is mild and restricted to the central kpc. But while any SFR suppression must have happened recently, the outflow has the potential to greatly impact the future evolution of the galaxy disc due to its geometrical orientation.}{Our observations reveal that low-power radio jets can play a major role in driving fast multi-phase galaxy-scale outflows even in radio-quiet AGN. Since the outflow energetics for HE~1353$-$1917 are consistent with literature scaling relations of AGN-driven outflows, the contribution of radio jets as the driving mechanisms still needs to be systematically explored.  }
\maketitle

\section{Introduction}
Quantifying the direct impact of active galactic nuclei (AGN) for galaxy evolution across cosmic time remains one of the big observational challenges. The enormous energy released by AGN has long been thought to suppress star formation through dissipation of their energy in the interstellar medium of their host galaxies \citep[e.g.][]{Silk:1998,King:2003,Gaspari:2017b}. Indeed, state-of-the-art numerical simulations and analytic models require some sort of energetic AGN feedback mechanisms in order to reproduce observed galaxy properties at the high-mass end \citep[e.g.][]{Benson:2003,Bower:2006,Cattaneo:2006,Croton:2006,Somerville:2008,Dave:2011,Genel:2014,Crain:2015}. In those models, AGN feedback is often separated into a ``radio mode'' (also known as ``maintenance mode") and a ''quasar-mode`` (''ejective mode"). The radio-mode AGN feedback describes the dissipation of mechanical energy from powerful radio jets. The energy heats the hot gas in halos of galaxies or galaxy groups/clusters such that condensation of gas is prevented and only a low cold gas content is maintained, limiting the ability for ongoing star formation. This mode of AGN feedback has been well established observationally in massive galaxy clusters and groups that host radio galaxies at their centers \citep[see reviews by][]{Fabian:2012,Gaspari:2013}. The quasar-mode feedback instead is thought to remove a significant fraction of the gas from host galaxies of luminous AGN through fast radiatively-driven winds released from the accretion disc \citep[e.g.][]{Feruglio:2015,Tombesi:2015,Bieri:2017}. 

Evidence for fast winds exceeding 10\,000\,$\mathrm{km\,s}^{-1}$ have been observed close to the AGN accretion disc through X-ray and UV spectroscopy \citep[e.g][]{Pounds:2003,Gibson:2009,Tombesi:2010,Tombesi:2010b}. Whether those radiatively-driven AGN winds are energetic enough to expand out to galaxy scale is still a matter of great debate and intensive ongoing investigation. In particular, the optical emission lines of the ionized gas phase have been extensively explored because they systematically show blue-shifted broad wings in their line shapes which are most pronounced in the [\ion{O}{iii}] $\lambda\lambda 4960,5007$ optical emission lines \citep[e.g.][]{Heckman:1981,Whittle:1985a,Mullaney:2013,Zakamska:2016,Woo:2016,Harrison:2016b,Bischetti:2017}. These asymmetric optical line shapes have been interpreted as a common signature of fast bi-conical AGN-driven outflows where the receding side is obscured by dust from the host galaxy \citep[e.g.][]{Pogge:1988a,Bae:2016}. Such biconical outflows have indeed been spatially resolved in several nearby AGN host galaxies with \textit{Hubble} \citep[e.g.][]{Schmitt:2003b,Crenshaw:2000,Das:2006,Fischer:2013,Revalski:2018} and ground-based spectroscopy \citep[e.g.][]{Pogge:1988a,Storchi-Bergmann:1992,Riffel:2013,Cresci:2015b,Mingozzi:2019}. 

An increasing number of AGN have been targeted either with long-slit or integral-field unit (IFU) spectrographs \citep[e.g.][]{Humphrey:2010,Greene:2011,Husemann:2013a,Liu:2013b,Harrison:2014,Husemann:2014,Liu:2014,Carniani:2015,McElroy:2015,Perna:2015,Carniani:2016,Kakkad:2016,Karouzos:2016,Leung:2017,Rupke:2017,Sun:2017,Kang:2018} to spatially map the ionized gas kinematics using the rest-frame optical emission lines. Although there is a clear trend emerging that the [\ion{O}{iii}] outflow velocity is systematically increasing with AGN luminosity \citep{Fiore:2017,Perna:2017,Woo:2017,DiPompeo:2018}, the associated ionized gas kinetic energies can be uncertain by a few orders of magnitude because electron densities $n_\mathrm{e}$ and detailed outflow geometries are usually unconstrained \citep[][and references therein]{Harrison:2018}. Another uncertain aspect of ionized gas outflows is their physical size. Some studies reported ubiquitous outflow sizes of several kpc scales \citep[e.g.][]{Harrison:2014,McElroy:2015,Liu:2013b,Liu:2013}, but recently much lower outflow sizes  of $<$1\,kpc were reported in several studies \citep{Husemann:2016c,Villar-Martin:2016,Tadhunter:2018,Baron:2019}. The difference may partially be explained due to the effect of beam smearing in seeing-limited observations \citep{Husemann:2016c}. 

Furthermore, it has been realized that the cold gas phase may actually dominate the total mass outflow rate budget \citep[e.g.][]{Morganti:2005,Feruglio:2010,Cicone:2014,Morganti:2015,Fiore:2017,Veilleux:2017}. While many observations have been naturally biased towards gas-rich starburst galaxies hosting AGN \citep[e.g.][]{Sturm:2011,Spoon:2013,Veilleux:2013}, a more systematic search on cold gas outflows are currently being conducted \citep{Fluetsch:2019}. Nevertheless, the investigation of the cold gas phase in AGN host galaxies still lags  behind the ionized gas phase as argued in \citet{Cicone:2018}, because extinsive multi-wavelength observations are required to obtain the full outflow rate budget in the different gas phases. This limits a robust assessment on the nature of AGN-driven outflows and prevents to distinguish different theoretical predictions of energy-driven or momentum-driven outflows \citep[e.g.][]{Ostriker:2010,King:2011,Faucher-Giguere:2012,Costa:2014}.

Despite the radio-quiet classification of most AGN being studied so far, they are not radio silent and may host low-luminosity radio jets on various scales from several tens of pc to a few kpc \citep[e.g.][]{Kukula:1998, Blundell:1998,Ulvestad:2005,Leipski:2006b,Doi:2013}. High turbulent velocities in all gas phases have often been associated even with those low-luminosity jets \citep[e.g.][]{Ferruit:1999,Leipski:2006a,Fu:2009,Husemann:2013a,Tadhunter:2014,Villar-Martin:2014,Cresci:2015,Harrison:2015,Villar-Martin:2017} and numerical simulation have shown that even the highly collimated jets can actually lead to large scale outflows when propagating through a clumpy high-density ISM \citep[e.g.][]{Sutherland:2007,Wagner:2012,Mukherjee:2018}. Usually, outflow energetics are compared to AGN luminosity, jet power, or even star formation rate to determine the powering mechanisms in individual sources. All of these comparisons suffer from systematic measurement uncertainties as summarized by \citet{Wylezalek:2018b}. Another complication is that a fast-moving plasma is necessarily creating radio emission itself \citep{Zakamska:2014,Hwang:2018} which may mimic low-luminosity radio jet characteristics. 

\begin{table*}\centering
\caption{Summary of observational characteristics and depth}\label{table:ObsRef}
 \begin{tabular}{ccccccc}\hline\hline
  Band & Instrument & FoV & Sampling & Beam & $t_\mathrm{exp}$ & $1\sigma$ limit\\\hline
  X-ray & \textit{Chandra}/HRC-I & $30\arcmin\times30\arcmin$ & 0\farcs13 & 0\farcs4 & 661\,min & $1\times10^{-6}$ counts s$^{-1}$ arcsec$^{-2}$ \\ 
  H$\alpha$  & VLT/MUSE  & $1\arcmin\times1\arcmin$ & 0\farcs2 & $0\farcs7\times0\farcs7$ & 15\,min & $5\times10^{-18}\,\mathrm{erg\,s}^{-1}\,\mathrm{cm}^{-2}\mathrm{arcsec}^{-2}$ \\
  $[\ion{O}{iii}]$  & VLT/MUSE  & $1\arcmin\times1\arcmin$ & 0\farcs2 & $0\farcs6\times0\farcs6$ & 15\,min & $7.5\times10^{-18}\,\mathrm{erg\,s}^{-1}\,\mathrm{cm}^{-2}\mathrm{arcsec}^{-2}$\\
         $J$ band & Gemini-N/NIFS  & $3\arcsec\times3\arcsec$ & 0\farcs1 & $0\farcs4\times0\farcs4$ & 60\,min & $0.3\times10^{-16}\,\mathrm{erg\,s}^{-1}\,\mathrm{cm}^{-2}\mathrm{arcsec}^{-2}$\\ 
         $K$ band & Gemini-N/NIFS  & $3\arcsec\times3\arcsec$ & 0\farcs1 & $0\farcs4\times0\farcs4$ & 60\,min & $0.8\times10^{-17}\,\mathrm{erg\,s}^{-1}\,\mathrm{cm}^{-2}\mathrm{arcsec}^{-2}$\\ 
	$B$       & SOAR/SOI       & $5\farcm3\times5\farcm3$ & 0\farcs15 & 1\farcs0 & 20\,min & 21.9\,mag(AB)/pix\\
	$V$       & SOAR/SOI       & $5\farcm3\times5\farcm3$ & 0\farcs15 & 1\farcs0 & 20\,min & 20.9\,mag(AB)/pix\\
	$R$       & SOAR/SOI       & $5\farcm3\times5\farcm3$ & 0\farcs15 & 1\farcs0 & 20\,min & 21.5\,mag(AB)/pix\\ 
	$J$       & CAHA2.2m/PANIC & $15\arcmin\times15\arcmin$ & 0\farcs45 & 1\farcs7 & 76\,min & 25.2\,mag(AB)/pix\\
	$H$       & CAHA2.2m/PANIC & $15\arcmin\times15\arcmin$ & 0\farcs45 & 1\farcs5 & 84\,min & 24.9\,mag(AB)/pix\\
	$K_s$     & CAHA2.2m/PANIC & $15\arcmin\times15\arcmin$ & 0\farcs45 & 1\farcs1 & 75\,min & 24.5\,mag(AB)/pix\\
	CO(1-0)   & ALMA	   & $50\arcsec$	        & 0\farcs2  & $0\farcs64\times0\farcs55$ & 104\,min  & $0.21$\,mJy/beam/50$\mathrm{km\,s}^{-1}$ \\
	L  	  & VLA (CnB config)& $32\arcmin$              & 1\farcs5  & $10\farcs0\times5\farcs4$ & 90\,min & $21.5\,\mu$Jy/beam \\
	\ion{H}{I} & VLA (CnB config) & $32\arcmin$            & 1\farcs5 & $10\farcs0\times5\farcs4$ &  90\,min & $0.4$\,mJy/beam/50$\mathrm{km\,s}^{-1}$ \\
 	X  	  & VLA (A config)& $4.5\arcmin$               & 0\farcs04 & $0\farcs4\times0\farcs2$ & 40\,min & $12.5\,\mu$Jy/beam\\\hline
 \end{tabular}
\end{table*}

It is clear that only extensive multi-wavelength observations of AGN host galaxies will improve our understanding about the nature of AGN-driven outflows and the impact on their host galaxies. We therefore started the {\bf C}lose {\bf A}GN {\bf R}eference {\bf S}urvey \citep[CARS, \texttt{www.cars-survey.org},][]{Husemann:2017b} which aims to map the host galaxies of almost 40 nearby ($0.01 < z < 0.06$) unobscured (type 1) AGN host galaxies at all wavelength from radio to X-rays. The CARS sample of AGN is drawn from the Hamburg/ESO survey \citep[HES,][]{Wisotzki:2000} and focuses on the sub-sample of the most luminous optically-selected AGN at those redshift which has previously targeted for single-dish observations of CO(1-0) as described in \citet{Bertram:2007}. By combining optical and near-IR IFU spectroscopy, radio and sub-mm interferometry and multi-band imaging of the entire galaxies we will characterize many aspects of the AGN feedback paradigm, enabling a better understanding of the conditions of star formation in AGN host galaxies, quenching via AGN-driven outflows, and the galaxy-scale (im)balance of AGN feeding~vs.~feedback as described in the chaotic cold accretion model \citep{Gaspari:2017,Gaspari:2018}.

In this paper we present an in-depth multi-wavelength study of the CARS galaxy HE~1353$-$1917 (\object{ESO 578-G009}, $13^\mathrm{h}56^\mathrm{m}36\fs7$ $-19\degr31\arcmin45\arcsec$) at redshift $z=0.035$. This AGN host galaxy appears as an edge-on disc-like galaxy where the AGN ionization cones are directly intercepting the galaxy disc. It is therefore a unique test case to study the process of AGN feedback in detail and to highlight the power of our multi-wavelength approach in constraining the AGN outflow energetics and its putative connection to its host galaxy. 

Throughout this paper we assume $H_0 = 70$\,km s$^{-1}$ Mpc$^{-1}$, $\Omega_M = 0.3$, and $\Omega_{\Lambda} = 0.7$.  In this cosmology, 1\arcsec\ corresponds to 0.697 kpc at the redshift of HE~1353$-$1917 ($z=0.035$), where the associated luminosity distance is 153.9\,Mpc.

\section{The CARS multi-wavelength data set}
As part of the CARS survey we have acquired a large number of multi-wavelength observations for HE~1353$-$1917 from X-rays to radio which are complemented by archival and literature data. Specifically, we obtained optical integral-field spectroscopy with the Multi Unit Spectroscopic Explorer \citep[MUSE,][]{Bacon:2010,Bacon:2014a} at the Very Large Telescope under programme 095.B-0015(A) (PI: Husemann) and near-IR (NIR) integral-field spectroscopy with the Near-Infrared Integral Field Spectrometer \citep[NIFS,][]{McGregor:2003} at Gemini North as a Fast Turnaround project (Program ID GN-2017A-FT-10, PI: J. Scharw\"achter). Deep optical ($B$, $V$ and $R$ bands) and NIR images ($J$, $H$ and $K_s$ bands) were taken with the 4.1\,m Southern Astrophysical Research (SOAR) Telescope using the SOAR Optical Imager (SOI; \citealt{Walker:2003}) and with the 2.2m telescope at the Calar Alto Observatory using the Panoramic Near-Infrared Camera \citep[PANIC,][]{Baumeister:2008}, respectively, under programmes SOAR/NOAO 2016A-0006 (PI: G.~Tremblay) and CAHA F15-2.2-014 (PI: B.~Husemann). $^{12}$CO(1--0) line mapping was obtained with ALMA for HE~1353$-$1917 in the two configurations C40-5 and C40-2 as part of programme 2016.1.00952.S (PI: G.~Tremblay). Radio observations were taken with the Karl G. Jansky Very Large Array (VLA) at $X$-band (10\,GHz) in A configuration and at $L$-band (1.4\,GHz) in CnB configuration under programmes 16B-084 (PI: P\'erez-Torres) and 16A-102 (PI: P\'erez-Torres), respectively. Finally, a deep X-ray spectrum with 62\,ksec integration time was taken with \textit{XMM-Newton} (ObsID: 0803510201, PI: Krumpe) and a high-resolution X-ray image with \textit{Chandra X-ray Observatory} as part of the Guaranteed Time Observation (GTO) program 19700783 (PI: Kraft).

\begin{figure*}
\centering
 \includegraphics[width=0.99\textwidth]{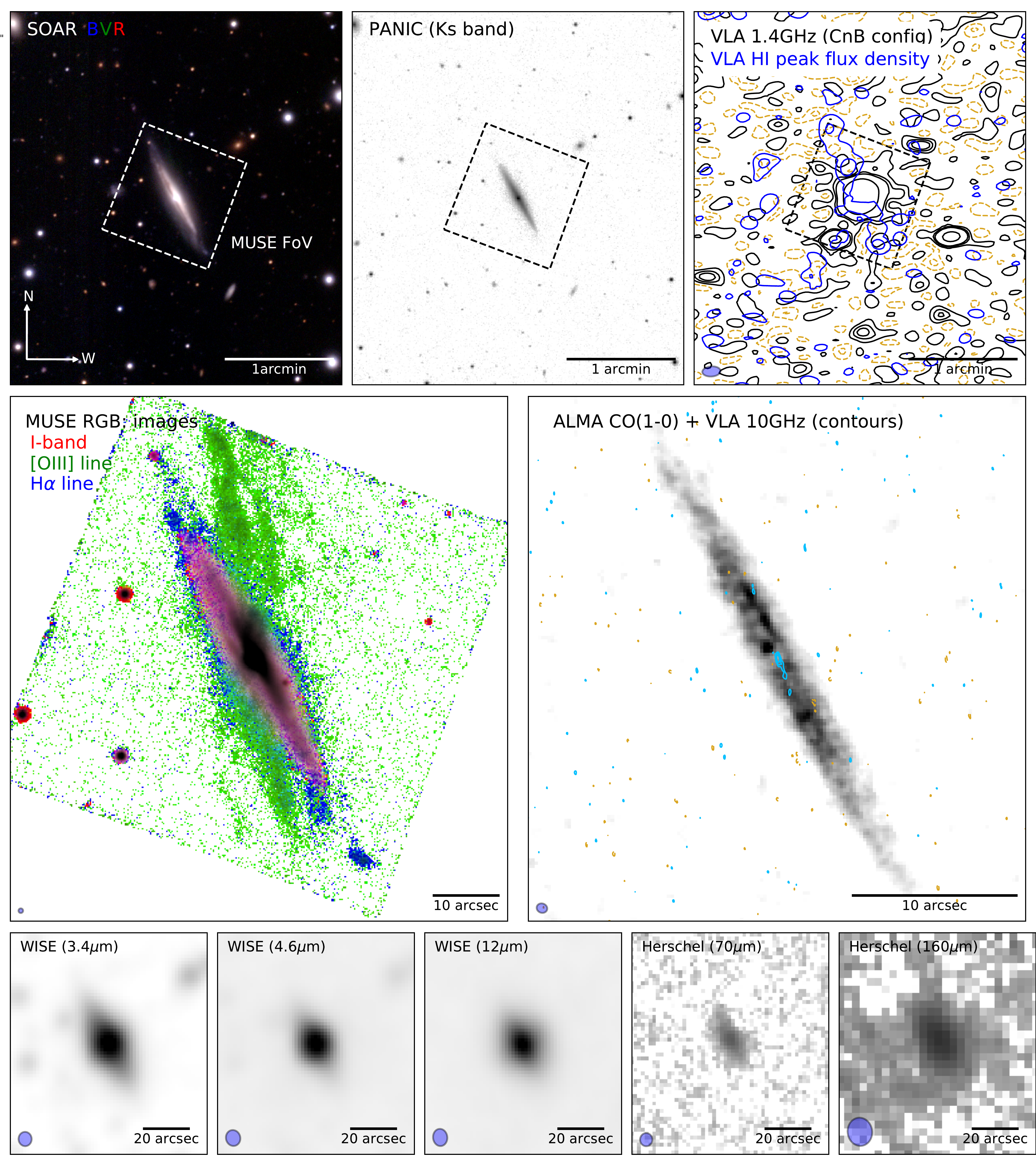}
 \caption{A multi-wavelength overview of HE~1353$-$1917. {\it Top panels:} SOAR optical $BVR$ image composite image, PANIC near-IR $K_s$ band image and a VLA $L$-band (1.4GHz) continuum as contours covering a FoV of $3'\times3'$ ($\sim130~\mathrm{kpc}\times130~\mathrm{kpc}$)  centered on the galaxy. For the $L$-band we display 1, 3, 5, and 25\,$\sigma$ levels as positive (solid black lines) and negative contours (dashed golden lines). We also show the peak flux density in the \ion{H}{i} frequency range at 1.3 and 2\,mJy levels indicating roughly the location of the atomic gas along the galaxy disc. The black dashed box indicates the orientation and position of the MUSE field-of-view (FoV). {\it Middle panels:} A color composite image is shown on the left side based on a narrow-band [\ion{O}{iii}], H$\alpha$ filter and an $I$ band filter reconstructed from the $1\arcmin\times1\arcmin$ MUSE observation. The CO(1-0) flux map observed with ALMA is shown on the right side where the 10\,GHz VLA continuum image is overplotted as 1, 3 and 5$\sigma$ positive (blue lines) and negative (golden lines) contours. {\it Bottom panels:}  Mid- to far-IR archival images from {\it WISE} and {\it Herschel} space observatories for the  3.4\,$\mathrm{\mu}$m to 160\,$\mathrm{\mu}$m wavelength range. Beam sizes are shown when applicable in the lower left corner for guidance in each panel.}
 \label{fig:overview}
\end{figure*}

The basic characteristics of our multi-wavelength observations in terms of field-of-view, spatial resolution and depth are provided in Table~\ref{table:ObsRef}. All details of the observations and the data reduction are fully described in the appendix from Sects.~\ref{sect:MUSE_appendix} to \ref{sect:chandra_appendix}. Furthermore, we collected archival data from the Wide-field Infrared Survey Explorer \citep[{\it WISE}, ][]{Wright:2010} satellite at 3.4, 4.6, 12 and 22\,$\mu$m mid-IR (MIR) bands (see Sect.~\ref{sect:WISE_appendix}), from {\it Herschel} \citep{Pilbratt:2010} with the instruments PACS \citep{Poglitsch:2010} at 70 and 160\,$\mu$m and SPIRE \citep{Griffin:2010} at 250, 350 and 500\,$\mu$m (see Sect.~\ref{sect:Herschel_appendix}), from the Panoramic Survey Telescope and Rapid Response System \citep[Pan-STARRS,][]{Chambers:2016} in the $g,r,i,z,y$ filters (see Sect.~\ref{sect:PNASTARRS_appendix}), and from the Galaxy Evolution Explorer \citep[GALEX,][]{Martin:2005} in the 1350--1750\AA\ (FUV channel) and 1750--2750\AA\ (NUV channel) bands (see Sect.~\ref{sect:GALEX_appendix}). A first visual impression of the processed multi-wavelength data is given in Fig.~\ref{fig:overview}.

\section{Analysis and Results}
\subsection{Broad-band SED fitting}\label{sect:SED}
\begin{figure}
 \resizebox{\hsize}{!}{\includegraphics{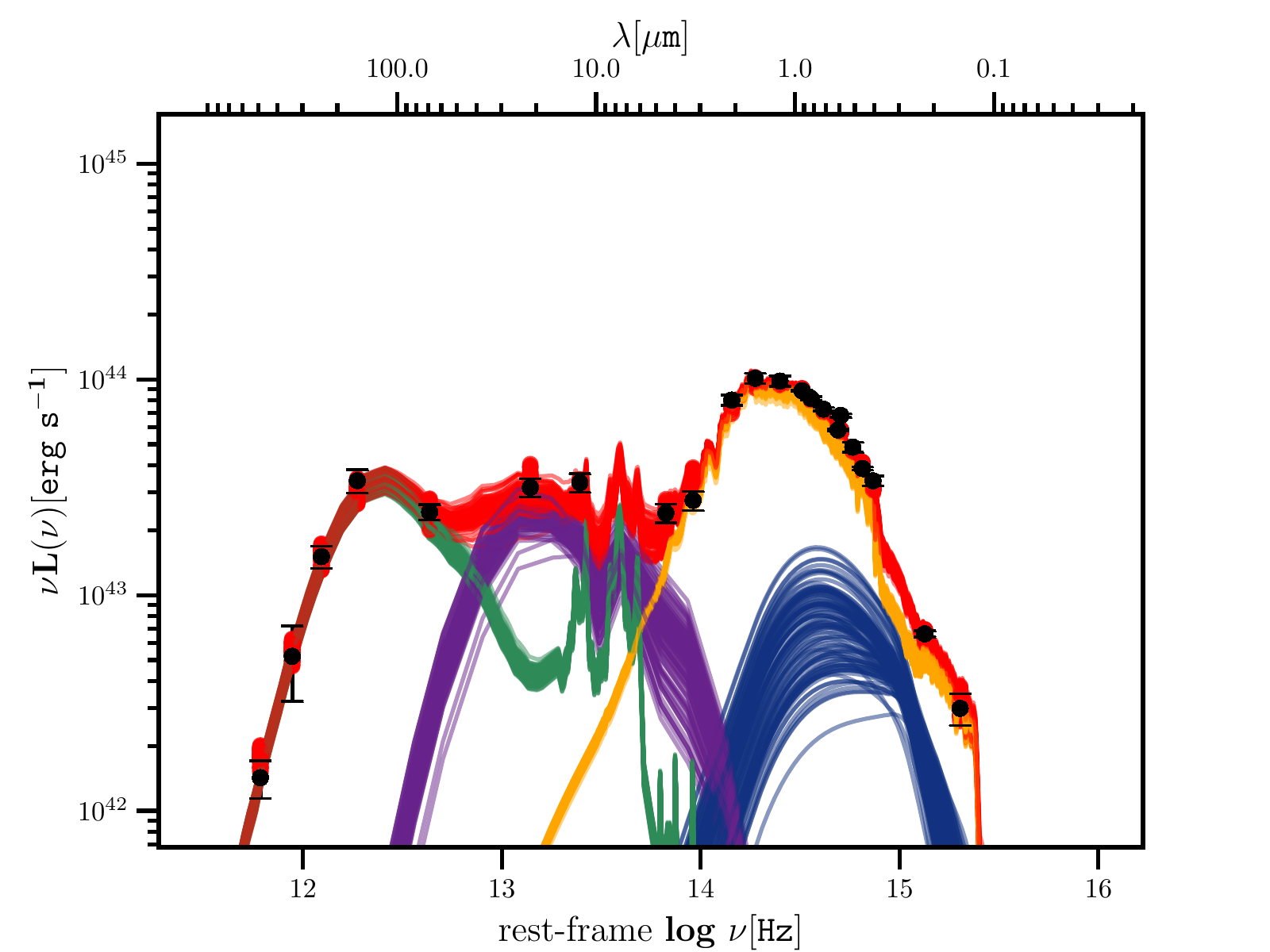}}
 \caption{Broad-band SED of HE~1353$-$1917 and best-fit template models inferred with \textsc{AGNfitter} \citep{CalistroRivera:2016}. The black points represent the observed photometric data  with their associated error bars. One hundred representative template SEDs are overplotted from the MCMC sampling. The total SEDs are shown as red lines which are further split up into their sub-component, 1) the AGN optical-UV emission (blue lines), 2) the stellar continuum (yellow lines), 3) the hot torus emission (purple lines), and 4) the cold+warm dust emission from star formation (green lines).} \label{fig:SED_fit}
\end{figure}

We first measure a reliable stellar mass and star formation rate (SFR) for HE~1353$-$1917 based on our multi-wavelength photometry. The overall spectral-energy distribution from the FIR to the UV  bands is listed in Table~\ref{tab:SED}. We employ a template fitting approach to model the broad-band SED. A Bayesian approach is used by\textsc{AGNfitter} \citep{CalistroRivera:2016} to decompose the total SED into the stellar, AGN and dust emission contribution. We ran \textsc{AGNfitter} with 150 walkers, 3 burn-in sets, of 3000 samples followed by 12000 MCMC samples to infer the distribution of parameters. The broad-band SED and 100 representative MCMC samples are shown in Fig.~\ref{fig:SED_fit}.

From the posterior probability distributions derived by \mbox{\textsc{AGNfitter}} we infer a stellar mass of $M_* = (2.5\pm0.5)\times10^{10}M_\sun$  and a star formation rate of $\mathrm{SFR}=2.3\pm0.1 M_\sun\,\mathrm{yr}^{-1}$. Due to the strong hot dust component from the AGN torus, \mbox{\textsc{AGNfitter}} infers the SFR  by integrating only the starburst component over the 8-1000$\mu$m FIR band as described in \citet{CalistroRivera:2016}. Considering the inferred stellar mass we compute a specific star formation rate (sSFR) of $\log(\mathrm{sSFR}/[\mathrm{yr}^{-1}])=-10.0\pm0.1$. Given the molecular gas content, the derived SFR leads to a gas depletion time scale of $t_\mathrm{dep}=(1.6\pm0.2)\times 10^9$\,yr as expected from normal star forming galaxies \citep[e.g.][]{Leroy:2008,Bigiel:2011}.

\begin{table}
 \caption{Spatially-integrated multi-band SED of HE~1353-1917}\label{tab:SED}
 \centering
 \begin{tabular}{ccc}\hline\hline
  Band & Instrument &  $f_\mathrm{tot}$ [mJy] \\\hline
  FUV  & GALEX      & $0.054\pm0.009$ \\
  NUV  & GALEX      & $0.180\pm0.006$ \\ 
  B   &  SOAR/SOI    & $1.68\pm0.09$  \\ 
  g   & Pan-STARRS    & $2.17\pm0.04$  \\
  V   & SOAR/SOI     & $3.04\pm0.16$  \\
  R   & SOAR/SOI     & $4.33\pm0.06$  \\
  r   & Pan-STARRS    & $4.90\pm0.1$  \\
  i   & Pan-STARRS    & $6.39\pm0.13$  \\
  z   & Pan-STARRS    & $8.34\pm0.17$  \\
  y   & Pan-STARRS    & $10.0\pm0.20$  \\ 
  J   &  CAHA/PANIC  & $14.3\pm0.8$  \\ 
  H   &  CAHA/PANIC  & $19.6\pm1.1$  \\ 
  Ks  &  CAHA/PANIC  & $20.4\pm1.1$ \\  
  W1  &   WISE       &  $11.0\pm1.1$ \\ 
  W2  &   WISE       &  $13.1\pm1.3$ \\ 
  W3  &   WISE       &  $49.0\pm4.9$ \\ 
  W4  &   WISE       &  $82.7\pm8.3$ \\ 
  PACS/70$\mu$m &   Herschel       &  $204\pm17$  \\ 
  PACS/160$\mu$m &   Herschel       & $659\pm82$  \\ 
  SPIRE/250$\mu$m &   Herschel       & $444\pm52$  \\ 
  SPIRE/350$\mu$m &   Herschel       & $215\pm82$  \\ 
  SPIRE/500$\mu$m &   Herschel       &  $85\pm15$ \\ 
  \hline
 \end{tabular}
\end{table}

\subsection{Deblending of AGN and host galaxy light}
\subsubsection{IFU spectroscopy}
We also have to decompose the optical and NIR IFU spectroscopic data from MUSE and NIFS into an AGN and a host galaxy component. To do that we use the fact that the emission from the AGN appears as a point source at all wavelength while the host galaxy necessarily is an extended component. Thus, the first step in the decomposition process is to estimate the PSF for the data. While MUSE covers a large wavelength range leading to wavelength dependent PSF, the PSF for the separate $J$ and $K$ band NIFS observations will be assumed to be constant for the respective band. Following the prescription of \citet{Jahnke:2004b} we empirically obtain the PSF from the intensity distribution of the broad AGN emission lines using our own software \QDeb\ \citep{Husemann:2013a,Husemann:2014}. We reconstruct the PSF in the optical MUSE data from the broad H$\beta$, H$\alpha$ and the \ion{O}{i} $\lambda 4448$, which we interpolate in wavelength pixel-by-pixel using a 2 order polynomial after normalizing the brightest pixel to 1.  The $J$ and $K$ band observations with NIFS cover the prominent broad Pa$\beta$ and Br$\gamma$ emission, respectively, which are used to construct the PSF for each band without the need to interpolate. 

Following the iterative AGN-host galaxy deblending scheme presented in \citet{Husemann:2013a} we initially construct an AGN cube from the brightest AGN spectrum and the wavelength depended PSF. After subtracting the AGN from the original cube, the residual cube contains the host galaxy signal. However, even the brightest AGN spectrum in the initial cubes has a host galaxy contribution which causes an AGN over-subtraction at the center of the galaxy. We mitigate this effect by iteratively subtracting the residual host galaxy spectrum from the initial AGN spectrum taking into account the surface brightness distribution. Given the large field-of-view (FoV) and relatively low spatial resolution of the MUSE data, we use the measured surface brightness profile from the broad-band photometry to interpolate the brightness of the  galaxy spectrum towards the nucleus, but assume a constant surface brightness of the host on the very small scales probed by the NIFS data.

\begin{figure*}
 \includegraphics[width=\textwidth]{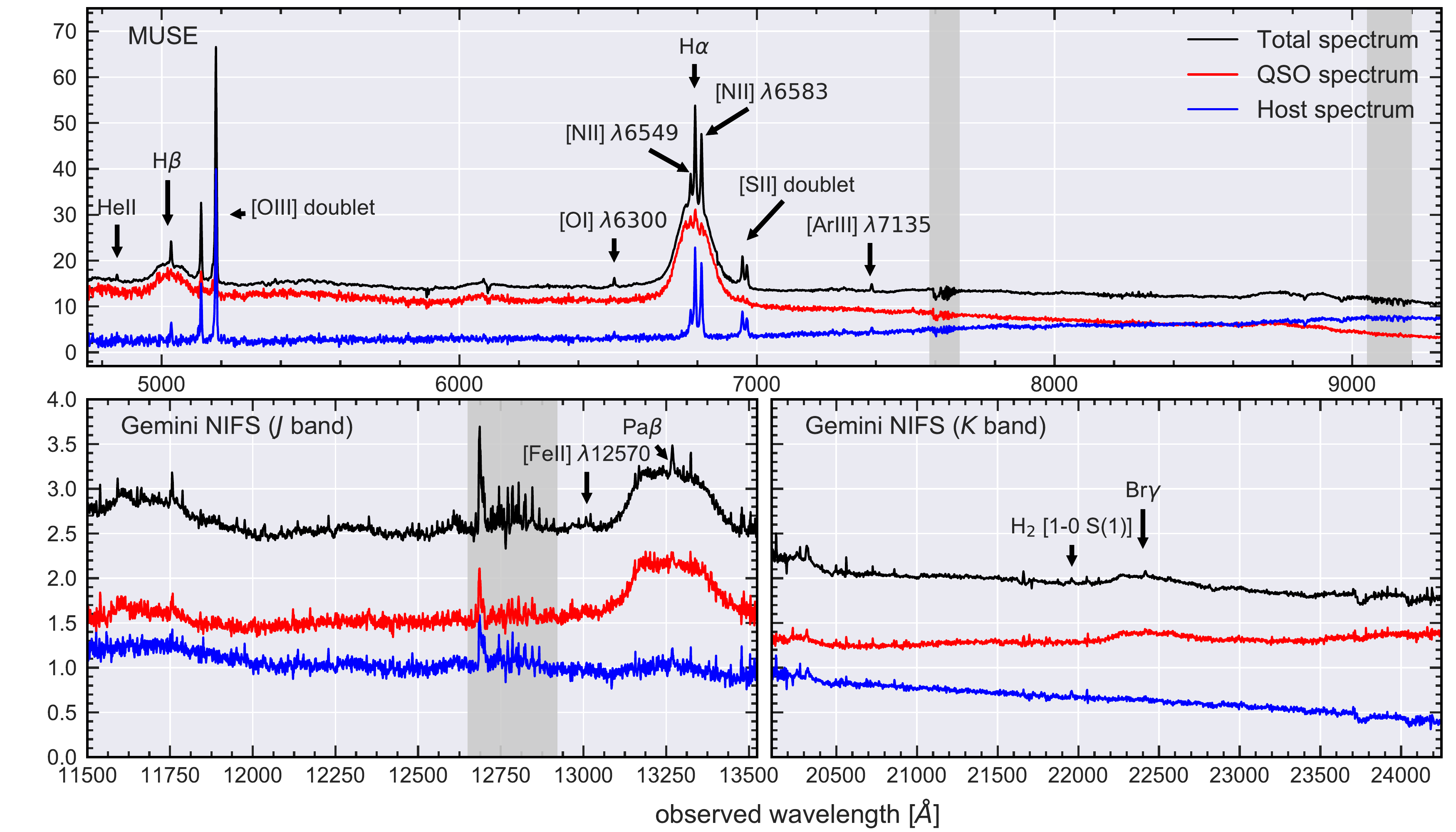}
 \caption{The total AGN and host galaxy spectrum within a 3\arcsec\ aperture centred on the AGN after applying the AGN-host galaxy deblending procedure as described in the main text. Important emission lines are clearly marked and wavelength regions prone to residuals from telluric absorption correction are  highlighted with the dark gray shaded areas.}
\label{fig:decomp}
\end{figure*}

We stop the iterative process after 6 and 4 iterations for MUSE and NIFS, respectively, at which the changes are less than 1\%. In Fig.~\ref{fig:decomp} we show the decomposed spectra from MUSE and NIFS within an aperture of 3\arcsec\ and 1\arcsec, respectively, which confirm that the host galaxy spectra are free from any remaining broad line components. Based on the recovered PSF we estimate a spatial resolution due to the seeing of $\sim$0\farcs7 full-width-at-half-maximum (FWHM) at the H$\alpha$ wavelength for the MUSE observations and $\sim$0\farc4 for the NIFS observations.

\subsection{AGN parameters}
\label{section:AGNparameter}
\begin{figure*}
\centering
 \includegraphics[width=0.9\textwidth]{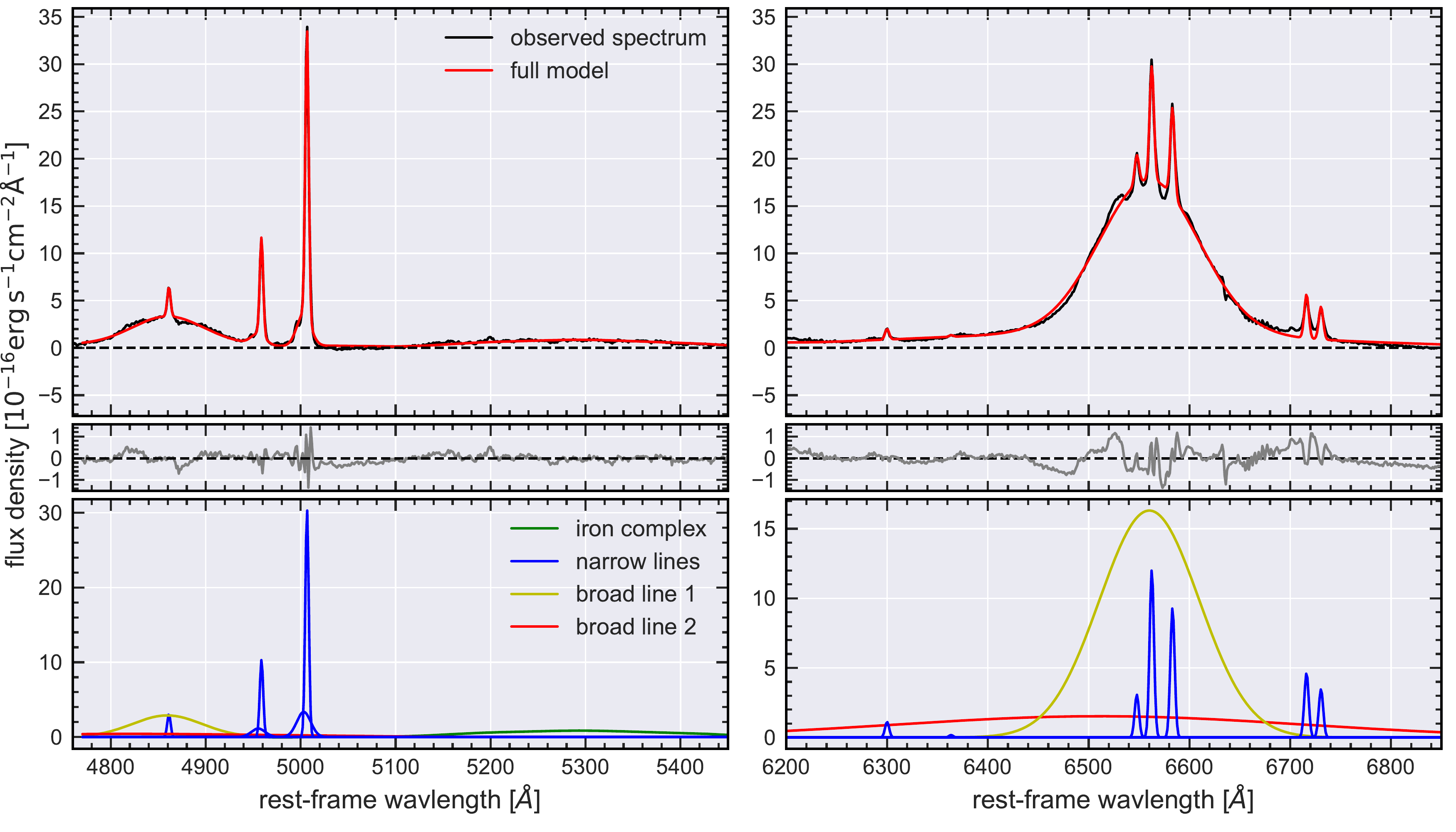}
 \caption{Fit to the nuclear spectrum of HE~1353-1917 in the wavelength regions around H$\beta$ and H$\alpha$ after continuum subtraction. The upper panels show an overlay of the nuclear spectrum (black) and the total fit (red). The residuals from the fit are indicated in gray (offset from the nuclear spectrum for better visibility).
The lower panels show the individual Gaussian components fitted to the nuclear spectrum together with an \ion{Fe}{ii} template from \citet{Kovacevic:2010}. All components shown with the same color were kinematically tied to each other during the fit. This includes the narrow components (H$\beta_\mathrm{n}$, [\ion{O}{iii}]~$\lambda 4959$ and $\lambda 5007$, [\ion{O}{i}]~$\lambda 6300$ and $\lambda 6364$, [\ion{N}{ii}]~$\lambda 6548$ and $\lambda 6583$, and H$\alpha_\mathrm{n}$ and [\ion{S}{ii}]~$\lambda\lambda6717,6731$) shown in blue, the broad components (H$\beta_\mathrm{b}$ and H$\alpha_\mathrm{b}$) shown in yellow, and the very broad components (H$\beta_\mathrm{vb}$ and H$\alpha_\mathrm{vb}$) shown in red.}
 \label{fig:AGN_fit}
\end{figure*}

Basic AGN parameters are determined by fitting the emission lines in the nuclear spectrum of HE~1353$-$1917 (Fig.~\ref{fig:AGN_fit}). The nuclear spectrum was obtained from the MUSE datacube by integrating an aperture of 2\arcsec\ radius centered on the AGN position. A stellar continuum datacube was subtracted beforehand which we constructed by fitting stellar population synthesis model spectra to the QSO-deblended datacube as described below in Sect.~\ref{sect:line_measurements}. The remaining pure AGN continuum was subtracted by fitting a power-law to the continuum sampled at four different rest wavelength ranges (4700--4732\,\AA, 5070--5100\,\AA, 5630--5650\,\AA, and 6810--6870\,\AA). Emission lines in the nuclear spectrum were modeled by simultaneously fitting the two wavelength regions around H$\beta$ and H$\alpha$ with a set of Gaussian line profiles together with an \ion{Fe}{ii} template from \citet{Kovacevic:2010}. During the fit, all narrow permitted and forbidden lines were kinematically tied to each other in velocity and line width, assuming instrumental spectral resolutions of $R_\mathrm{S}=1800$ and 2500 in the wavelength regions around H$\beta$ and H$\alpha$, respectively.

The broad H$\beta$ and H$\alpha$ lines were modeled using the superposition of two broad Gaussian components. Two components are needed to capture the extended blue wings of the broad H$\alpha$ line. Both components for the broad H$\beta$ and H$\alpha$ line were forced to share the same kinematics. A two component model can, however, not capture the full line asymmetry of the broad line as seen in the residuals of Fig.~\ref{fig:AGN_fit}. Those deviations do however not affect the FWHM estimates of the full line profile beyond the systematic errors. For \ion{Fe}{ii}, the line width and velocity shift were tied to the broad Balmer lines during the fit. We used the narrowest \ion{Fe}{ii} templates from \citet{Kovacevic:2010} and convolved these with Gaussian kernels to adjust the line width during the fit. To further constrain the fit, we used fixed flux ratios of 3.05 and 2.948 for [\ion{N}{ii}]~$\lambda 6583$/[\ion{N}{ii}]~$\lambda 6548$ and [\ion{O}{iii}]~$\lambda 5007$/[\ion{O}{iii}]~$\lambda 4959$, respectively (see discussion in \citet{Schirmer:2013}).

From the best-fit model we infer a broad Balmer line width of $\mathrm{FWHM}_\mathrm{BLR}=(5820\pm80)\,\mathrm{km\,s}^{-1}$ and $\sigma_{BLR}=(3990\pm180)\,\mathrm{km\,s}^{-1}$. The measured continuum flux at the rest-frame wavelength of 5100\AA\ is $f_{5100}=(10.5\pm0.1)\times10^{-16}\,\mathrm{erg}\,\mathrm{s}^{-1}\,\mathrm{cm}^{-2}\,\mathrm{\AA}^{-1}$ and the broad Balmer emission-line fluxes are $f_{\mathrm{H}\beta}=(4.1\pm0.2)\times10^{-14}\,\mathrm{erg}\,\mathrm{s}^{-1}\,\mathrm{cm}^{-2}$ and $f_{\mathrm{H}\alpha}=(29.5\pm0.2)\times10^{-14}\,\mathrm{erg}\,\mathrm{s}^{-1}\,\mathrm{cm}^{-2}$. Here, we estimate the single-epoch BH mass using the broad H$\alpha$ line luminosity and FWHM as calibrated by \citet{Woo:2015}
\begin{equation}
 M_\mathrm{BH}=1.12\left(\frac{\mathrm{FWHM}_{\mathrm{H}\alpha}}{1000\,\mathrm{km\,s}^{-1}}\right)^{2.06}\left(\frac{L_{\mathrm{H}\alpha}}{10^{42}\,\mathrm{erg\,s}^{-1}}\right)^{0.46}10^{6.544}M_\odot
\end{equation}
This yields a BH mass for HE1353-1917 of $M_\mathrm{BH}=(1.36\pm0.04)\times10^8\,M_\odot$ based on our measurements. However, the uncertainties on single-epoch BH masses are dominated by systematics so that a typical error of 0.4\,dex is assumed \citep[e.g.][]{Vestergaard:2006,Denney:2009} for high S/N measurements as in this case.

The bolometric AGN luminosities can be estimated from various indicators. Adopting a bolometric correction factor of $\sim10$ for the continuum luminosity at 5100\AA\ \citep{Richards:2006} and the conversion from the broad H$\alpha$ to continuum luminosity reported by \citet{Greene:2005} we estimate a bolometric luminosity of $L_\mathrm{bol}=(2.0\pm0.2)\times10^{44}\mathrm{erg}\,\mathrm{s}^{-1}$. This estimate from the optical line emission of the BLR is fully consistent with the estimated bolometric luminosity from our deep X-ray observations. Considering our BH mass estimate we computed an Eddington ratio of $\log(L_\mathrm{bol}/L_\mathrm{Edd})=-1.9\pm0.4$, so the BH is accreting close to $\sim$1\% of its Eddington limit. \\

\begin{figure*}
 \includegraphics[width=\textwidth]{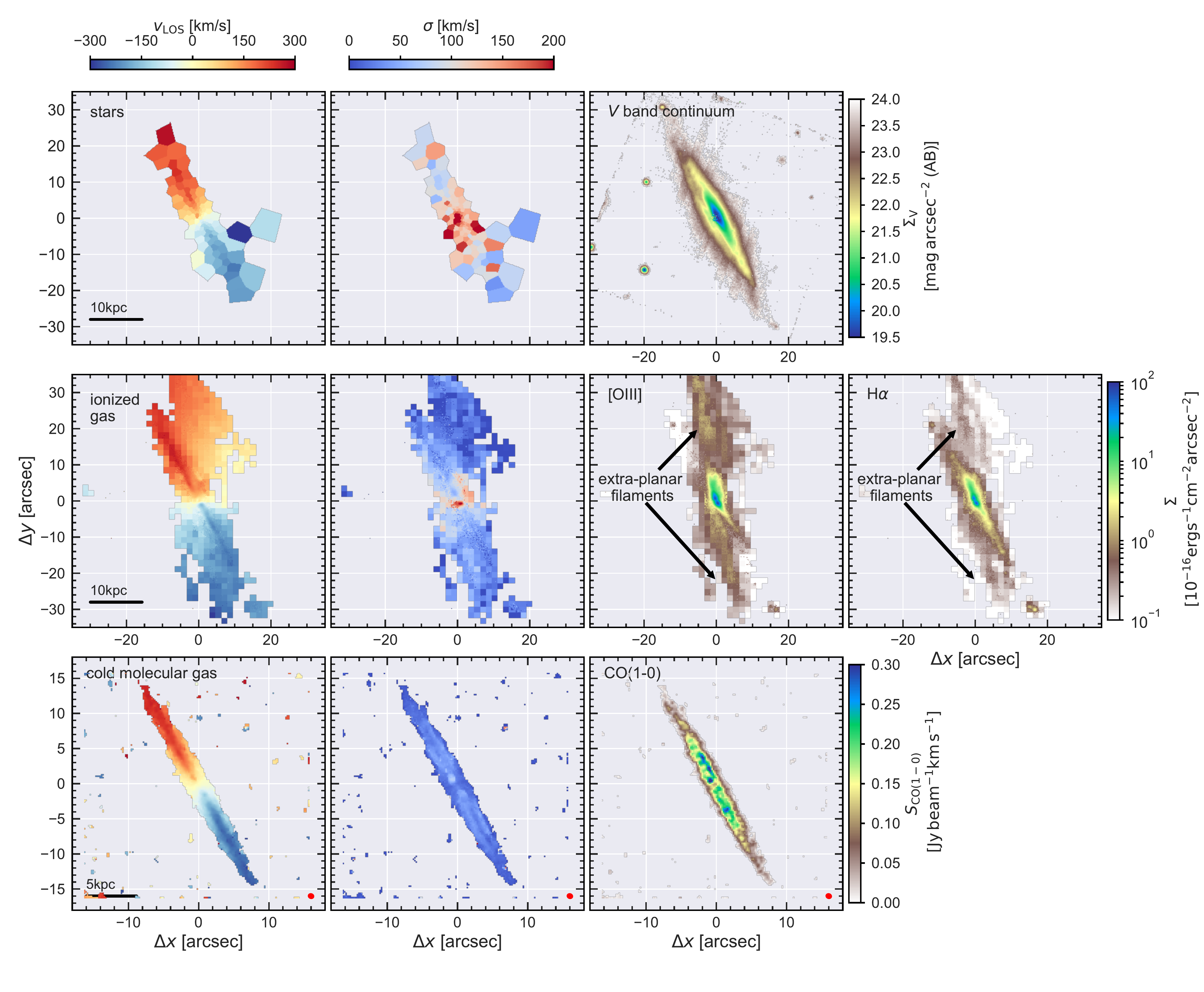}
 \caption{Maps of the stellar (first row), ionized gas (second row) and cold molecular gas (third row) kinematics and flux distribution. The first column shows the line-of-sight LOS velocity with respect to the systemic velocity of the galaxy, the second column is the velocity dispersion and third/forth column highlights the surface brightness of the respective tracer. The beam size for the ALMA observations is indicated as a red symbol in the respective panels.}
\label{fig:maps}
\end{figure*}

\subsection{Spatially-resolved multi-phase emission-line measurements}\label{sect:line_measurements}
\subsubsection{Ionized gas phase}
To obtain clean measurements of the ionized-gas emission lines in the optical MUSE IFU spectra, we first have to subtract the stellar continuum properly avoiding any influence of absorption lines on emission-line fluxes. Modelling the stellar continuum and subsequent emission-line analysis is implemented in \textsc{PyParadise} \citep[see][]{Husemann:2016a,Weaver:2018}, which is a Python version of the stellar population synthesis fitting code \textsc{Paradise} \citep{Walcher:2015} to which emission-line fitting capabilities are added.  The stellar and emission-line modelling is done in two subsequent steps. First, \textsc{PyParadise} fits the continuum spectrum as a linear super-position of a template library of  stellar or stellar population spectra after convolving them with a Gaussian line-of-sight velocity distribution (LOSVD). Second, the best-fit stellar continuum models are  subtracted from the original spectra and the emission-lines are modelled as Gaussian functions. 

A key feature of \textsc{PyParadise} is that the input spectra and the template library spectra are initially normalized by a running mean where emission lines or strong sky line residuals are masked and linearly interpolated. In this way, we minimize systematic effects caused by wavelength-dependent flux calibration issues or low-frequency residuals after the AGN subtraction. Below we briefly describe the underlying algorithm and its application to our MUSE data after AGN-host galaxy deblending.

\textsc{PyParadise} determines the best-fit LOSVD and linear combination of normalized spectral templates independently based on an MCMC algorithm and a non-negative linear least square fitting, respectively, in an iterative scheme. First, the algorithm infers an initial guess on the LOSVD based on a single spectrum drawn from the template library. Second, the previously determined best-fit LOSVD is applied to all spectra in the template library to obtain the best-fit non-negative linear combination of template spectra excluding masked spectral regions contaminated by emission lines or sky line residuals. This best-fit linear combination of spectra is used as the  input spectrum for further iterations and the two step process of estimating LOSVD and linear combination is repeated. The final best-fit spectrum is then denormalized and subtracted from the original spectrum. 

Afterwards emission lines are then measured subsequently in the residual spectrum. Given the spectral coverage of MUSE and redshift of HE~1353$-$1917, we focus on modelling the H$\beta$, [\ion{O}{iii}] $\lambda\lambda 4960,5007$, [\ion{O}{ii}] $\lambda 6300$, H$\alpha$, [\ion{N}{ii}] $\lambda\lambda 6545,6583$ and [\ion{S}{ii}] $\lambda\lambda 6717,6731$ emission lines. All the emission lines are modelled as a system of Gaussians with a common LOS velocity and velocity dispersion taking into account the wavelength-dependent spectral resolution of MUSE as derived by \citet{Bacon:2017}. In addition, the line ratios of the [\ion{O}{iii}] $\lambda\lambda 4960,5007$ and [\ion{N}{ii}] doublets $\lambda\lambda 6545,6583$ are fixed to their theoretical values \citep{Storey:2000}. 

\begin{figure*}
 \includegraphics[width=0.95\textwidth]{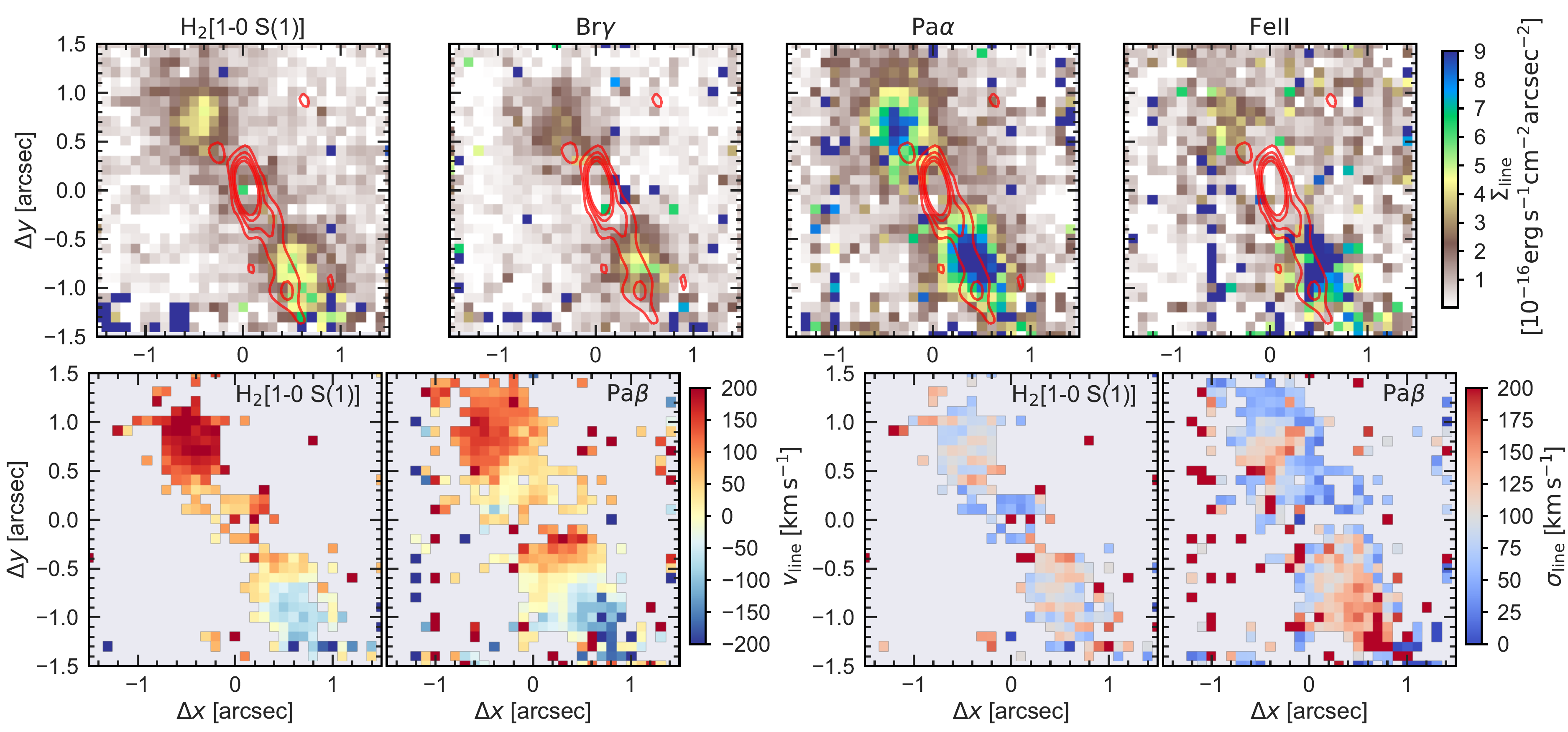}
 \caption{\textit{Top panels:} Surface brightness maps for the H$_2$ [1-0 S(1)] $\lambda$21218\AA, Br$\gamma$, Pa$\beta$ and [\ion{Fe}{ii}] $\lambda$12570\AA\ emission lines in our Gemini-NIFS NIR observations after subtracting the point-like AGN contribution. The red contours correspond to the distribution of 10GHz radio continuum emission as resolved with our VLA radio interferometric observations. \textit{Lower panels:} Radial velocity and velocity dispersion for the two brightest lines, H$_2$ and Pa$\beta$ respectively. }
\label{fig:NIFS_maps}
\end{figure*}

Since the S/N of the stellar continuum is much lower than that of the emission-lines in the MUSE cube of HE~1353$-$1917, we apply a Voronoi binning \citep{Cappellari:2003} to increase the S/N to a minimum of 20 in the stellar continuum in the wavelength range  $7000\AA<\lambda<7100\AA$. The emission lines are then modelled again per spaxel after repeating the spectral synthesis modelling with \textsc{PyParadise} but now fixing the previously estimated LOSVD parameters of the corresponding Voronoi bin used for the continuum fitting. Given that some of the important emission lines, e.g. H$\beta$, are often too weak in a single spaxel, we also perform the emission-line fitting after binning the original cube by $8\times8$ spaxel  ($1\farcs6\times1\farcs6$). Uncertainties for all parameters are estimated using a Monte-Carlo approach where the stellar continuum and emission lines fitting are repeated 40 times after each spectrum is modulated randomly within the error of each spectral pixel. 

The resulting LOS velocity $v_\mathrm{LOS}$, velocity dispersion $\sigma$ and flux maps are shown in the first two rows of Fig.~\ref{fig:maps}. As expected the stellar and ionized gas kinematics reveal a clear rotational pattern along the major axis of the nearly edge-on disc galaxy. The unexpected features for this galaxy are the quite extended filamentary structures of ionized gas outside of the disc. The LOS velocity of the filaments shows that the gas rotates in the same sense as the galaxy disc, which already suggests that the ionized gas is linked to the galaxy disc. Furthermore, we detect exceptionally high velocity dispersion in the ionized gas about 1\farcs3 (900\,pc) south-west of the AGN position. Understanding the origin of both features is the main topic and driver for the following analysis of this paper.

\subsubsection{Cold molecular and atomic gas phase}\label{sect:coldphase}
From the cleaned ALMA datacube we created a zeroth moment (integrated intensity) map, a first moment (mean velocity) map, and a second moment (velocity dispersion) map of the detected CO(1-0) line emission using a masked moment technique \citep{Dame:2011}. A copy of the clean data cube was Gaussian-smoothed spatially (with a FWHM equal to that of the synthesized beam), and then smoothed (using a 4 channel FWHM Gaussian) in velocity. A three-dimensional mask was then defined by selecting all pixels above a fixed flux threshold of 3 times the RMS noise in this new cube.  The moment maps were then created using the original un-smoothed cubes within the masked regions only, without any threshold.

The resulting moment maps are shown in the third row of Fig.~\ref{fig:maps}. Strong molecular gas emission is detected from the disc of the galaxy. The velocity dispersion (second moment) map of this source also shows the typical X-shaped structure of a beam-smeared gas rotating regularly, apart from two hot spots in the south-west and north-east of the nucleus of the galaxy, at the same location as that found in the ionised gas. The origin of these features will be discussed more in Sect.~\ref{sect:gas_kin}.

After integrating all the CO(1-0) emission, we obtain a cold gas mass of $M_\mathrm{H2}=(3.6\pm0.3)\times10^9\mathrm{M}_\odot$ adopting a Galactic conversion factor of $\alpha_\mathrm{CO}=4.35\,\mathrm{M}_\odot(\mathrm{K\,km\,s}^{-1}\,\mathrm{pc}^{-2})^{-1}$. Our estimate is slightly higher than the single-dish measurements from the IRAM-30m telescope of $M_\mathrm{H2}=(2.7\pm0.2)\times10^9\mathrm{M}_\odot$ reported by \citet{Bertram:2007}. The discrepancy is mainly caused by the large extent of the source with respect to the IRAM beam ($\sim$22\arcsec) so that the integrated signal would need to be corrected by the primary beam response. This correction can only be applied to the ALMA data where the CO(1-0) flux distribution is properly resolved with two different baselines to cover the diffuse extended emission with maximum recoverable scale of $20\arcsec$.

Based on the integrated \ion{H}{i} line flux (see Sect.~\ref{sect:VLA_HI}) we estimate a at total atomic gas mass of $M_\mathrm{\ion{H}{i}}=(1.3\pm0.2)\times10^{10}M_\sun$ following the prescription of \citet{Koenig:2009}. The inferred atomic gas mass is fully consistent with the upper limit previously reported by \citet{Koenig:2009} for this source using the single-dish Effelsberg telescope. Unfortunately, the S/N is not sufficient to make proper moment maps as in the case of the CO data, but at least we present the peak flux morphology in Fig.~\ref{fig:overview} which confirms that the atomic gas is aligned with disc.

\subsubsection{Warm-molecular gas phase}
Warm-molecular gas emission in the NIFS  $\sim 3\arcsec \times 3\arcsec$ FoV is traced via the near-infrared rovibrational H$_2$ [1-0 S(1)] $\lambda$21218\AA\ line. Based on the QSO-deblended NIFS data cubes, we derived emission-line maps for the H$_2$ [1-0 S(1)] and narrow Br$\gamma$ component in the $K$-band  as well as Pa$\beta$ and [\ion{Fe}{ii}] $\lambda$12570\AA\ in the $J$-band as shown in Fig.~\ref{fig:NIFS_maps}. Those emission lines were fitted with a Gaussian line profile together with a constant continuum. We also convolved the line widths with the instrumental resolutions of $R_\mathrm{S} \sim 5290$ and $R_\mathrm{S} \sim 6040$ in the $K$- and $J$-bands, respectively. 

Fig.~\ref{fig:NIFS_maps} shows that all emission lines in the central $\sim 3\arcsec \times 3\arcsec$ of HE1353-1917 originate mainly from two emission regions located in the galaxy disc at about 0\farcs8 ($\sim 0.6$~kpc) to the north-east and south-west of the nucleus. The emission from the narrow component of the hydrogen recombination lines (i.e. Br$\gamma$ in the $K$-band and Pa$\beta$ in the $J$-band) is found to be co-spatial with the H$_2$ emission. The NIFS data also reveal a deficiency of H$_2$ or narrow-line hydrogen recombination emission within a 0\farcs3 (0.2 kpc) region (radius) around the nucleus. The nuclear region is instead dominated by the PSF from the broad Br$\gamma$ and Pa$\beta$ emission associated with the AGN broad-line region (see Fig.~\ref{fig:decomp}) and decomposed during the QSO-host deblending step. 

Interestingly, a comparison with the radio continuum map shows that the southern emission peak is nearly co-spatial with the putative radio jet. Also the kinematics reveal a velocity gradient along the jet axis which is not following a pure rotational pattern of the galaxy disc  (Fig.~\ref{fig:NIFS_maps}). Hence, the excitation of H$_2$ and [\ion{Fe}{ii}] together with potentially non-gravitational motions of the gas may be directly linked to the impact of an AGN-driven radio jet on the cold gas disc. The enhanced radio emission on the southern approaching side may be explained either by Doppler boosting of a relativistic jet \citep[e.g.][]{Laing:1988,Cohen:2007} or by an intrinsic asymmetry of a bi-polar jet.

\subsection{Emission-line diagnostics}
\subsubsection{Ongoing star formation and extended NLR}
\begin{figure*}
  \includegraphics[width=\textwidth]{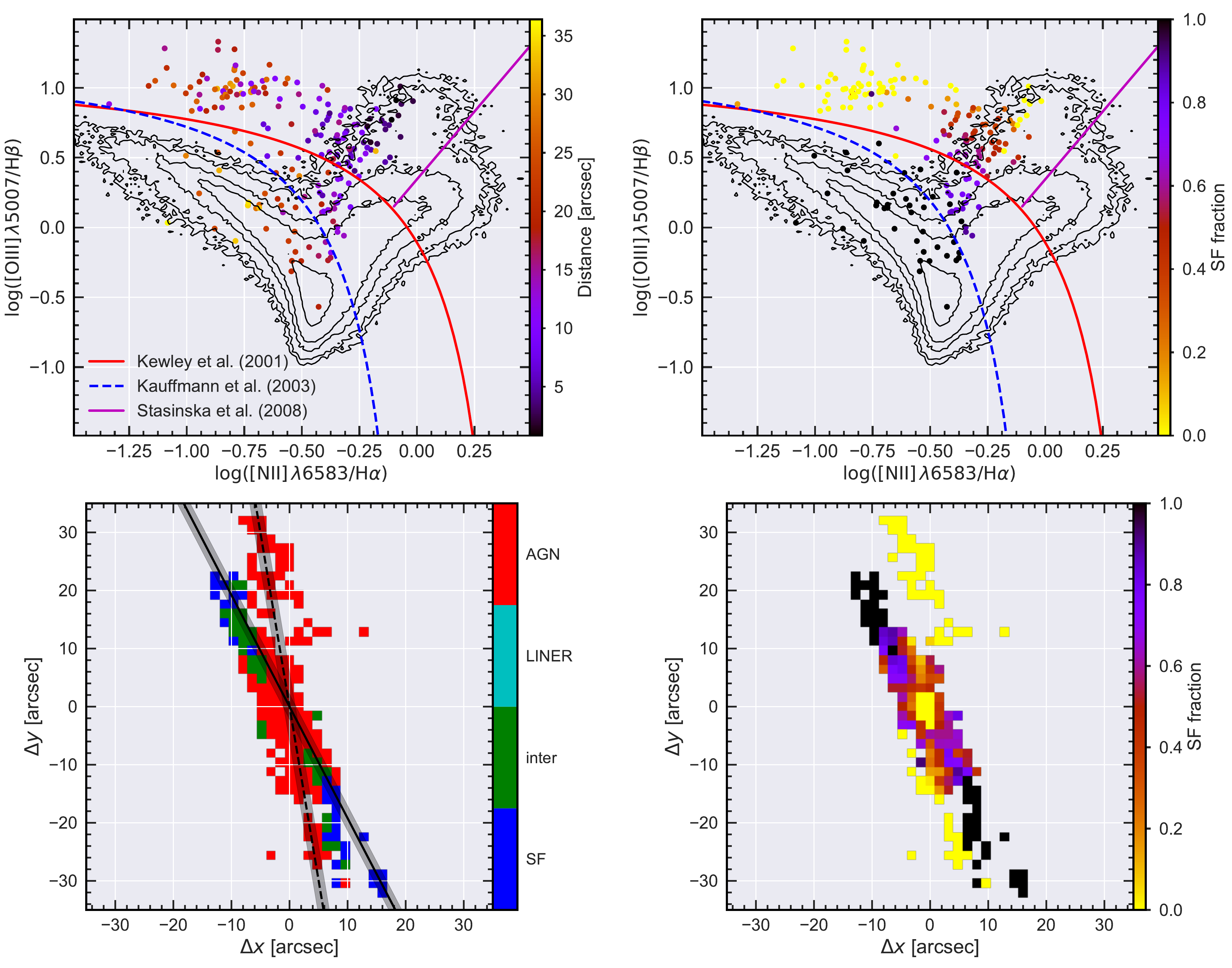}
  \caption{\textit{Top left panel:} Classical BPT [\ion{O}{iii}]/H$\beta$ vs. [\ion{N}{ii}]/H$\alpha$ diagnostic plot after binning the MUSE data to $1\farcs2$ spatial sampling. The distance from the nucleus is color coded and the distribution of line ratios from the SDSS MPA/JHU catalog is shown as contours for comparison. Demarcation curves from \citet{Kauffmann:2003} (dashed blue line), \citet{Kewley:2002} (red solid line) and \citet{Stasinska:2008} (solid magenta line) are displayed for reference.  \textit{Top right panel:} Same as the left panel where the color coding represent the SF fraction to the emission line measured as described in the main text. The SF fraction is set to zero for emission regions above the galaxy disc and set to one for regions below the \citeauthor{Kauffmann:2003} demarcation line. \textit{Bottom left panel:} 2D excitation classification map based on the boundaries set by the demarcation lines. The solid and dashed line represent the primary disc galaxy axis ($\sim28\degr$ with respect to North) and an ENLR axis along ($\sim10\degr$ with respect to North) the strongest emission, respectively. The black shaded area around the axis indicates the 2\arcsec-wide pseudo long-slit extraction region for the metallicity gradients shown in Fig.~\ref{fig:N2S2_metallicity}. \textit{Bottom right panel:} Spatial distribution of the SF fraction as defined from the emission-line ratios.}
  \label{fig:BPT_diagram}
 \end{figure*}
A key analysis from the IFU spectroscopy is to discriminate various ionization sources of the ionized gas throughout the galaxy and beyond. In the case of AGN host galaxies, we do expect a mix of ionization from ongoing star formation, AGN photoionization and potentially from shocks due to AGN-driven or starburst-driven outflows. Various optical emission-line ratios are usually employed to discriminate between different ionization mechanisms \citep[e.g.][]{Baldwin:1981,Veilleux:1987,Kewley:2006,Stasinska:2008}. We show the [\ion{O}{iii}]/H$\beta$ vs. [\ion{N}{ii}]/H$\alpha$ diagnostic diagram for HE~1353$-$1917 in Fig.~\ref{fig:BPT_diagram} (top left panel) with the distance from the nucleus indicated as a color gradient. While the QSO-host deblending algorithm already takes out artificial mixing of light from the bright nucleus with the surrounding galaxy, there is an additional mixing between ionization from star formation and the extended NLR. 

To disentangle the varying contribution of ionization by young stars and the AGN we use a modified version of the algorithm presented in \citet{Davies:2016}. The basic idea is that all measured extinction-corrected emission-line fluxes can be modelled as a linear superposition of a characteristic emission-line basis spectrum for the AGN photoionized and classical \ion{H}{ii} regions. While two basis spectra are sufficient to disentangle the ionization close to the nucleus, the line ratios change across the galaxy due to a radial gradient in gas-phase metallicity. Hence, we expand the method of \citet{Davies:2016} by considering a sample of  \ion{H}{II} basis spectra across the \ion{H}{ii} sequence from SDSS and an empirical set of basis spectra for the AGN ionization taken from the MUSE data itself. The optimal combination for each spaxel is then determined through a Markov Chain Monte Carlo approach which also provides estimates on the uncertainty on the SF fraction due to the possible combination of basis spectra combined with the intrinsic errors on the emission-line ratios. 

In Fig.~\ref{fig:BPT_diagram} (top right)  we present the estimated SF fraction across the classical BPT diagram. While we compute the AGN fraction for the mixing region within the disc of galaxy ranging from 0 to 1, we set all spaxel regions to an AGN faction of  1 that are well outside the edge-on disc galaxy area. Those spaxels correspond to the AGN ionization cones which are illuminating gas above the main body of the galaxy where a contribution from \ion{H}{ii} region can be neglected. Those regions indeed keep a high [\ion{O}{iii}]/H$\beta$ ratio while the [\ion{N}{ii}]/H$\alpha$ ratio is decreasing. This is characteristic for a decreasing metallicity \citep[e.g.][]{Groves:2006,Husemann:2011} rather than a mixing with \ion{H}{ii} regions. 

Combining the spatially-resolved extinction-corrected H$\alpha$ flux with the SF fraction estimates, we obtain a total H$\alpha$ flux of $f_{\mathrm{H}\alpha}=(5.5\pm0.1)\times10^{-14}\mathrm{erg}\,\mathrm{s}^{-1}\mathrm{cm}^{-2}$. This H$\alpha$ flux corresponds to a luminosity of $L_{\mathrm{H}\alpha}=(1.55\pm0.03)\times10^{41}\,\mathrm{erg}\,\mathrm{s}^{-1}$ which leads to a H$\alpha$-based SFR estimate of $\mathrm{SFR}_{\mathrm{H}\alpha}=1.23\pm0.03\,\mathrm{M}_\sun\mathrm{yr}^{-1}$. Similarly, we estimate an integrated extinction-corrected [\ion{O}{iii}] luminosity from the ENLR as $L_{[\ion{O}{iii}]}=(1.15\pm0.01)\times10^{42}\,\mathrm{erg}\,\mathrm{s}^{-1}$, with an extension that seems to be limited by the MUSE FoV and exceeds 35\arcsec\ or 25\,kpc in projection. Hence, the AGN phase has already been  lasting for at least 0.1\,Myr given the light travel time across the entire ENLR.

\subsubsection{Gas-phase metallicity along the galaxy disc and extended NLR}
The gas-phase metallicity is an important diagnostic as it provides a record of the metal enrichment history of galaxies. Since the weak auroral [\ion{O}{iii}] $\lambda4363$ emission line is outside the wavelength range covered by MUSE we can only use strong-line methods to infer the gas-phase metallicity of the ionized gas. The N2 index (N2=$\log$([\ion{N}{ii}]\,$\lambda6583$/H$\alpha$ flux ratio) or the O3N2 index (O3N2=$\log$\{(\ion{O}{iii}$\lambda5007$/H$\beta$)/(\ion{N}{ii}$\lambda6583$/H$\alpha$)\} flux ratio) have been empirically calibrated as a proxy for the oxygen abundance of \ion{H}{ii} regions in star-forming galaxies \citep[e.g.][]{Pettini:2004,Marino:2013}. In the case of HE~1353$-$1917 we cannot directly apply those metallicity indicators, because a significant part of the galaxy is contaminated by strong AGN ionization due to the orientation of the ionization cone. Although variations in the [\ion{N}{ii}]\,$\lambda6583$/H$\alpha$ line ratio have been attributed to changes in the metallicity even for AGN-ionized gas clouds based on photo-ionization models \citep[e.g.][]{Storchi-Bergmann:1998, Groves:2006}, there may also be a dependence on AGN luminosity as highlighted by \citet{Stern:2013}.

An alternative metallicity diagnostic is the N2S2 index (N2S2=$\log$([\ion{N}{ii}]$\,\lambda6583$/[\ion{S}{ii}]\,$\lambda\lambda6717,6731$)) which is a proxy for the N/O abundance \citep[e.g][]{Sabbadin:1977,Viironen:2007} which in turn is also correlated with the O/H abundance \citep[e.g.][]{Alloin:1979,Perez-Montero:2009}. The advantage of the N2S2 index is that it correlates well with the metallicity even for the AGN photo-ionized ENLR, similar to the N2 index, but does not exhibit a strong dependence on the AGN luminosity as discussed by \citet{Stern:2013}. While \citet{Stern:2013} provides a calibration of N2S2 to the oxygen abundance scale established by \citet{Tremonti:2004}, we independently correlate the N2S2 index with the oxygen abundance as inferred by the O3N2 index of \citet{Pettini:2004} based on the line ratios of 2300 star-forming galaxies from the SDSS DR7 value-added catalog \citep{Brinchmann:2004},
\begin{equation}
 12+\log{\mathrm{O/H}} = 8.658 + 0.34\times \mathrm{N2S2}  - 0.312\times \mathrm{N2S2}^2
\end{equation}
This exercise is necessary to compare our results for HE~1353$-$1917 with those of \citet{Sanchez:2014} as the absolute calibration of the O/H abundance is significantly dependent on the used empirical calibration  \citep{Kewley:2008}.

\begin{figure}
 \resizebox{\hsize}{!}{\includegraphics{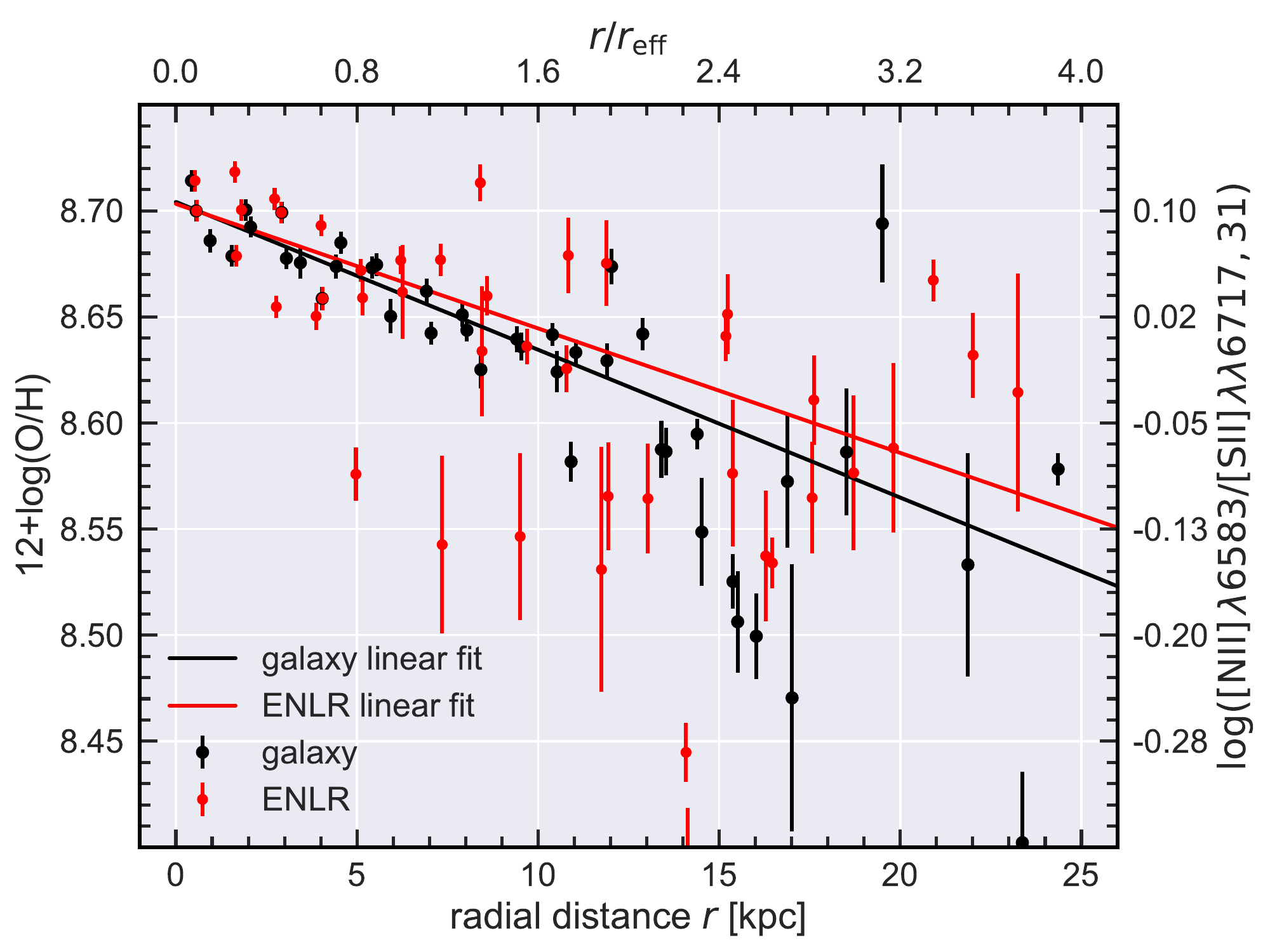}}
 \caption{Radial metallicity gradient as a function of radius based on the N2S2 index for 2\arcsec-wide pseudo long-slits along galaxy disc (black points) and along the ENLR bicone (red points) as defined in Fig.~\ref{fig:BPT_diagram}. The best-fit linear relations are shown as solid lines in the respective color.}
 \label{fig:N2S2_metallicity}
\end{figure}

In Fig.~\ref{fig:N2S2_metallicity} we show the O/H abundance based on the N2S2 index as a function of galacto-centric distance in units of kpc and normalized to the effective radius of the galaxy disc. We extract pseudo long-slits along the major axis of the galaxy and the ENLR of the putative AGN-ionization cone (see Fig.~\ref{fig:BPT_diagram}), which allow us to measure the gas-phase metallicity out to 15\,kpc or 2.5 effective radii. We find an error-weighted slope of $\alpha_\mathrm{[O/H]}=-0.043\pm0.006\,\mathrm{dex}\,r_\mathrm{eff}^{-1}$ and $\alpha_\mathrm{[O/H]}=-0.036\pm0.011\,\mathrm{dex}\,r_\mathrm{eff}^{-1}$ for the O/H gradients along the galaxy disc and the ENLR, respectively, with an absolute zero-point of $12+\log([\mathrm{O/H}])=8.703\pm0.006$ in both cases. The errors are derived by bootstrapping where we fit the relation 1000 times for a random selection of half the data points and then compute the standard deviations for the slope and zero-point. Due to the narrow width of the pseudo-long slit the metallicity measurements along the disc and the ENLR are totally independent already after 5\arcsec\ (3.5\,kpc) away from the nucleus. The measured gradients are consistent with the characteristic O/H gradients of undisturbed disc galaxies which are $\alpha_\mathrm{[O/H]}=-0.1\pm0.09\mathrm{dex}\,r_\mathrm{eff}^{-1}$ \citep{Sanchez:2014}.  The central metallicity as expected from the mass--metallicity relation of galaxies is $12+\log(\mathrm{[O/H]})_\mathrm{PP04\,O3N2}=8.72\pm0.10$ for our measured stellar mass of HE~1353$-$1917 following the calibration of \citet{Kewley:2008}. 

The consistency checks above shows that our metallicity estimates based on the N2S2 index provide reliable measurements even in the case of dominating AGN-ionization. Most importantly, it highlights the fact that the gas illuminated by the AGN above the galaxy's disc plane shares the same radial metallicity dependence seen in the ionized gas within the galaxy's disc. This is an important piece of evidence which suggests that the gas illuminated by the AGN has been driven out \emph{vertically} from the disc plane likely by stellar feedback through stellar mass loss and supernova explosions rather than originating from the galaxy centre through AGN outflows.

\begin{figure*}
\includegraphics[width=\textwidth]{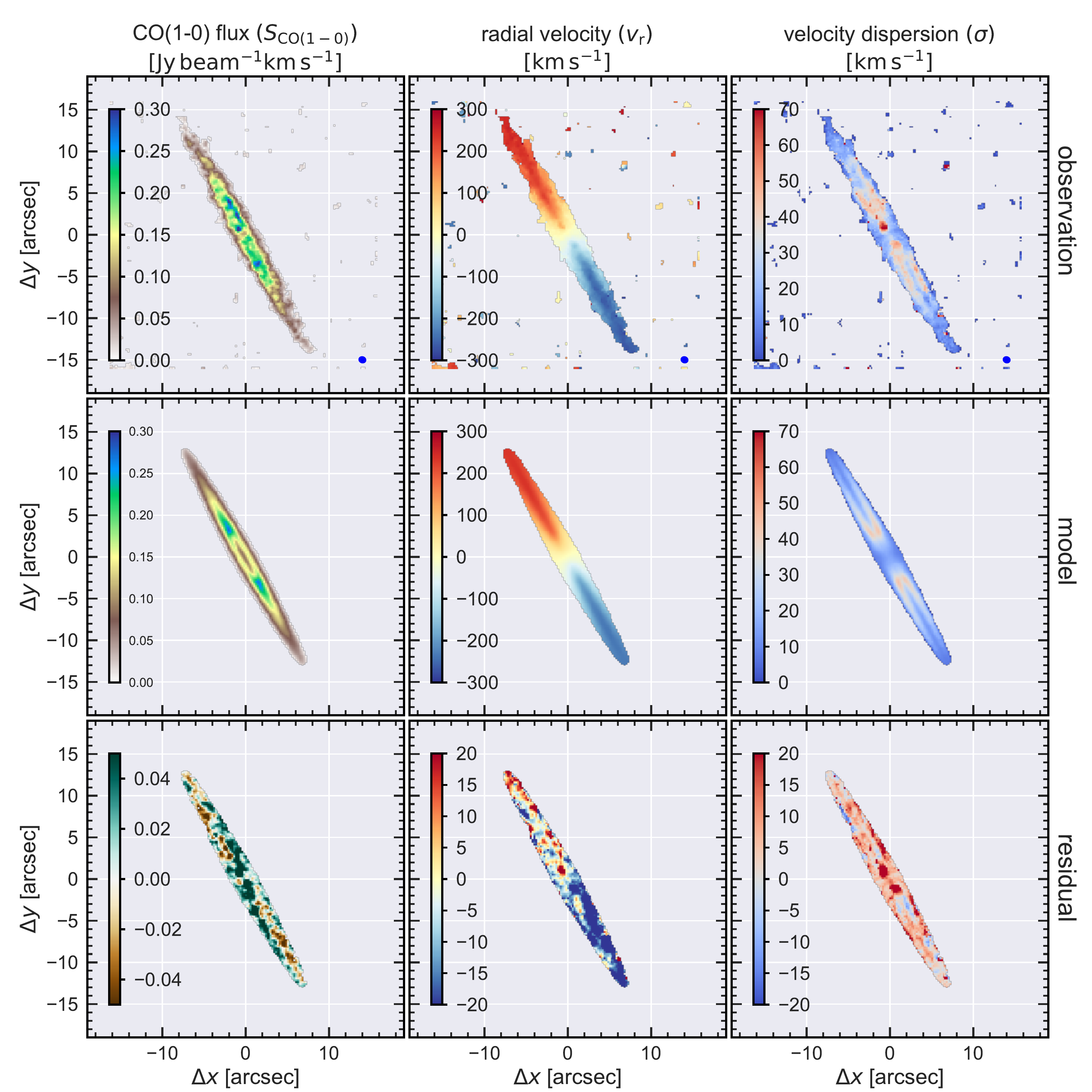}
\caption{Kinematic modelling of the molecular gas traced by CO(1-0). \textit{Upper panels:} Moment maps of the CO(1-0) line from the ALMA observations. The beam size is indicated in all panels as the filled blue elliptical in the lower right for comparison. \textit{Middle panels:} Best-fit kinematic model determined with the KinMS software tool \citep{Davis:2013}. \textit{Lower panels:} Residuals of the best-fit model maps compared to the original data.  }
\label{fig:CO_model}
\end{figure*}

\subsection{Multi-phase gas kinematics}\label{sect:gas_kin}
\subsubsection{CO(1-0) kinematic modelling and dynamical mass}
The cold gas kinematics as shown in Fig.~\ref{fig:maps} seems to be dominated by classical rotational motion of a disc galaxy. Hence, we model the ALMA CO observation directly with the KINematic Molecular Simulation \citep[KinMS,][]{Davis:2013,Davis:2013b} software tool\footnote{freely available at \url{https://github.com/TimothyADavis/kinms}} taking the beam size and velocity binning into account. As an input for the model we perform a multi-Gaussian expansion \citep[MGE,][]{Emsellem:1994, Cappellari:2002} to our $K_s$ band image observed with PANIC to determine the deprojected mass model assuming a linear mass-to-light ratio gradient (M/L) which is a free parameter for the fitting. The MGE parameters are listed in Table \ref{mgetable}. We neglect the unresolved MGE component at the galaxy centre, which arises from the AGN. We assume a truncated exponential disc for the CO surface brightness distribution with the scale radius ($r_\mathrm{exp}$) and the inner truncation radius ($r_\mathrm{trunc}$) as free parameters together with the position angle ($\mathrm{PA}_\mathrm{disc}$), inclination ($i_\mathrm{disc}$), the intrinsic velocity dispersion ($\sigma_\mathrm{cold}$), systemic velocity ($v_\mathrm{sys}$) and the kinematic center ($x_\mathrm{kin},y_\mathrm{kin}$) of the disc.

\begin{figure*}
 \includegraphics[width=\textwidth]{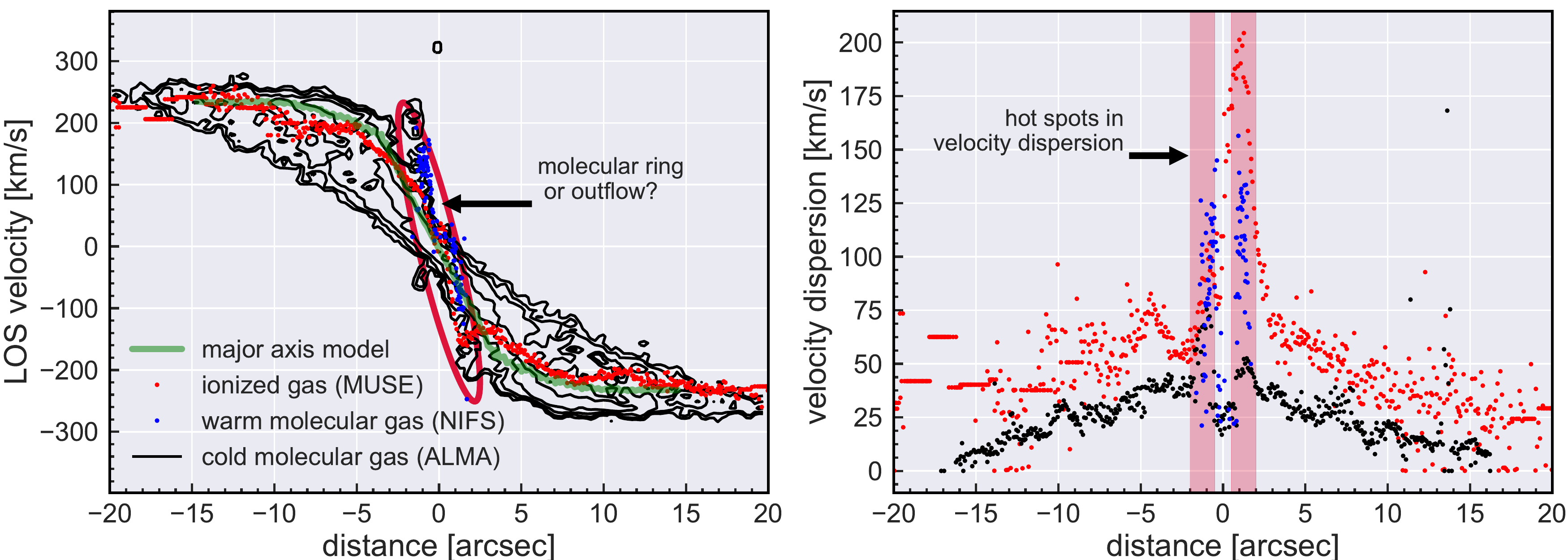}
 \caption{Position-velocity (left panel) and position-dispersion (right panel) diagrams along the major axis of HE1353-1917. The black contours are for the cold molecular gas and the colored data points are the curves for the ionized and the warm molecular phase where the latter is limited by the much smaller NIFS FoV. The maximum velocity curve from the best-fit MGE model prediction is shown in the PV diagram on the left. Within the central 3\arcsec\ we detect significant deviation from the PVD model (left) which may be related either to a galaxy ring structure of a bipolar outflow. ``Hot spots'' with an excess in velocity dispersion (right) are found on both sides of the nucleus in the same region.}
 \label{fig:pV-diagram}
\end{figure*}

\begin{table}
\caption{MGE parametrization of the galaxies $K_s$ band light profile.}
\begin{center}
\begin{tabular}{ccc}\hline\hline
$\log_{10}$ I$_j$\tablefootmark{a} & $\log_{10}$  $\sigma_j$\tablefootmark{b} & $q_j$\tablefootmark{c} \\
$[L_{\odot,K_s}$ pc$^{-2}]$ & [arcsec] & \\
\hline
214091.0\tablefootmark{*} & 0.25 &    0.529\\
4156.4  &    1.15  &   0.669\\
886.5  &    3.80  &   0.390\\
1752.9   &   6.51  &   0.100\\
109.9   &   10.13  &   0.950\\
977.7   &   10.22  &   0.129\\
 \hline
\end{tabular}
\tablefoot{
\tablefoottext{a}{$K_s$ surface brightness of Gaussian component.}
\tablefoottext{b}{Standard deviation (width) of Gaussian component.}
\tablefoottext{c}{Axis ratio of Gaussian components}
\tablefoottext{*}{The central unresolved Gaussian is dominated by AGN light, and is thus subtracted to minimize the effect of the AGN on our kinematic fitting.}
}
\end{center}
\label{mgetable}
\end{table}%

\begin{table}
\centering
 \caption{Best-fit CO(1-0) kinematic model parameters and 1$\sigma$ errors}\label{tab:kin_parameters}
 \begin{tabular}{cc}\hline\hline
 Parameter & Value\\\hline
  $\mathrm{PA}_\mathrm{disc}$ & $29.1\degr\pm0.5$\\
  $i_\mathrm{disc}$ & $85.4\degr\ _{-0.4}^{+0.1}$\\
  $\sigma_\mathrm{cold}$ & $7.0\pm1.8\,\mathrm{km\,s}^{-1}$\\
  M/L$_{Ks}(R=0\arcsec)$  &  $1.18\pm0.07$\\
  M/L$_{Ks}(R=15\arcsec)$  &  $1.45\pm0.08$\\
  $v_\mathrm{sys,opt}$ & $ 10449\pm3\,\mathrm{km\,s}^{-1}$\\
  $r_\mathrm{exp}$ & $5\farcs8\pm0.\farcs$\\
  $r_\mathrm{trunc}$ & $3\farcs1\pm0\farcs1$\\\hline
 \end{tabular}
\end{table}

In Fig.~\ref{fig:CO_model} we show the CO input maps, model maps and residuals for the best-fit model parameters listed in Table~\ref{tab:kin_parameters}. Our model is indeed a very good description of the data where the residual in the flux distribution naturally originates from the clumpiness of the gas in the disc. This gas is not distributed with a simple exponentially declining profile, instead shows a central flux deficit within a truncation radius of $r_\mathrm{trunc}=3\farcs1\pm0\farcs1$. The residuals in velocity along the minor axis are caused by the high inclination and corresponding line-of-sight superposition related to the thickness of the disc and the beam size. With a maximum projected LOS velocity of $v_\mathrm{LOS}=240\,\pm\,5\,\mathrm{km\,s}^{-1}$ at $R=17.2$\arcsec\ (12.65\,kpc) away from the galaxy centre, we obtain a dynamical mass of $M_\mathrm{dyn}(R<=12.65\,\mathrm{kpc})= 0.6\times(v_\mathrm{rad}/\sin(i))^2R/G=(1.0\pm0.3)\times10^{11}M_\sun$ following the prescription of \citet{Lequeux:1983} for disc galaxies. Compared to a stellar mass of $M_*=(2.5\pm0.5)\times10^{10}M_\sun$, a molecular gas mass of $M_\mathrm{H_2}=(3.6\pm0.3)\times10^9\mathrm{M}_\odot$ and the atomic gas $M_\mathrm{HI}=(1.3\pm0.2)\times10^{10}M_\sun$, we estimate a dark matter fraction of $M_\mathrm{DM}/M_\mathrm{dyn}$ of $\approx$60\% within 2.1 effective radii. This dark matter fraction is fairly typical for spiral galaxies of this mass \citep[e.g.][]{Salucci:1991,Martinsson:2013}.

\begin{figure*}
 \includegraphics[width=\textwidth]{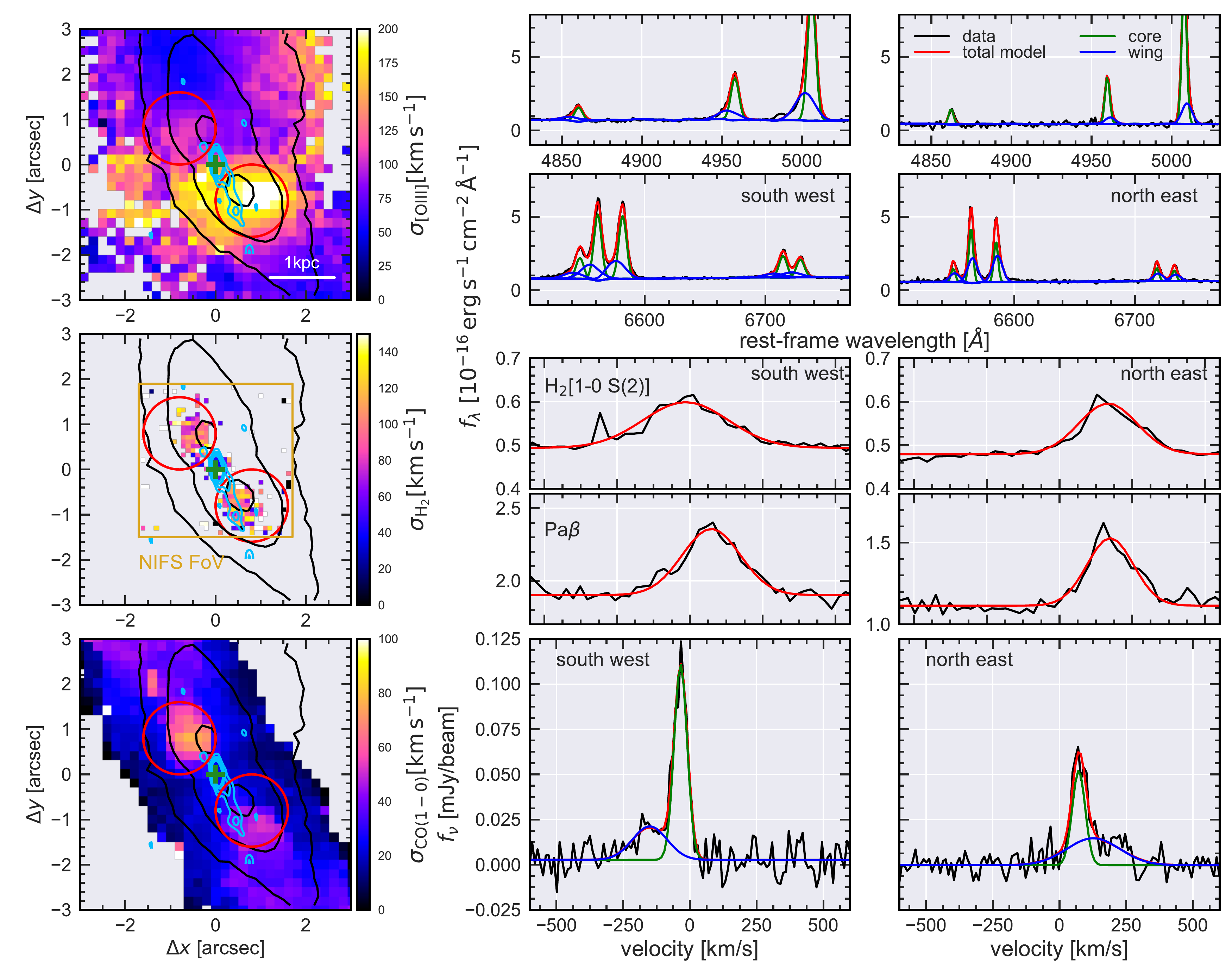}
 \caption{{\it Left panels:} The velocity dispersion maps for the [\ion{O}{iii}], $\mathrm{H}_2$[1-0\ S(2)], and CO(1-0) emission lines (assuming a single Gaussian LOSVD) in the central $5''\times5''$ ($\sim3.5\,\mathrm{kpc}\times3.5\,\mathrm{kpc}$) for [\ion{O}{iii}], $\mathrm{H}_2$ and CO(1-0). The black contours indicate the \Ox\ surface brightness distribution as a reference in all maps. The green cross marks the position of the AGN and the light blue contours represent the distribution of 10\,GHz radio continuum emission as resolved with our VLA radio interferometric observations. The red circles indicates a circular aperture of 0.8\arcsec\ in radius which we use to extract spectra covering the two high dispersion region located south west and north east of the nucleus.  {\it Middle and right panels:} Spectra of the high dispersion regions in various emission lines from the optical, NIR and sub-mm observations for the south-west (middle) and north-east region (right). Most emission lines reveal an asymmetry towards the blue or red side that we model with two Gaussian line components which we refer to as the core and wing component. Only Pa$\beta$ and $\mathrm{H}_2$[1-0\ S(2)] can be well-described by only a single Gaussian component.}
\label{fig:outflow_maps}
\end{figure*}

A striking feature in the cold gas kinematics are two well-defined regions on both sides of the nucleus which exhibit a significantly higher velocity dispersion than our kinematic model predicts. To understand the nature and properties of this region we reconstruct the multi-phase position-velocity diagram (PVD) along the major axis of the galaxy (Fig.~\ref{fig:pV-diagram}) based on the CO(1-0), H${}_2[1-0 \mathrm{S}(1)]$ and ionized gas kinematic maps. We find that the LOS velocity profile of the ionized and molecular gas matches exactly along the disc at nearly all radii. This shows that also the ionized gas kinematics are determined pre-dominantly by gravitational motions. However, the LOS velocity and, in particular, the velocity dispersion most strongly deviates from the kinematical model \emph{in all gas phases} about 1\farcs5--2\farcs0 away from the center on both sides. The warm molecular gas, which typically arises in strong shocks, is only detected at the location of those two ``hot spots'' supporting the notion that it is indeed a special region within the galaxy. Most importantly, we detect an extra-ordinary high velocity dispersion in the ionized gas phase at the same location, but only on one side of the galaxy centre. Given the high inclination and dust content of the disc,  we assume that we cannot see the optical emission lines of the ionized gas on the other side of the nucleus due to the high obscuration. 

It might be possible to explain the elevated velocity dispersion in the molecular gas by adding an additional point-like mass in the centre or a circum-nuclear bar. However, the significantly stronger deviations in the warm  and ionized gas phase, as clearly visible in the PVD and velocity dispersion diagram (Fig.~\ref{fig:pV-diagram}) and the kinematic maps (Fig.~\ref{fig:maps}), cannot be explained by gravitational motion and appear symmetrically offset from the nucleus. Hence, we disfavor a simple gravitational induced scenario and conclude that we are detecting the signature of a fast bipolar multi-phase outflow driven by the AGN.

\subsubsection{Multi-phase kinematics of the bipolar outflow}
The velocity dispersion maps shown in Fig.~\ref{fig:outflow_maps} (left panels) reveal the exact location of regions with a high velocity dispersion. Those kinematic features have already been seen in PVD along the major axis of the disc (Fig.~\ref{fig:pV-diagram}). All the different gas phases show a higher dispersion about $\sim$1\farcs2 or $\sim$900\,pc south-west of the nucleus. A similar region is on the opposite side north-east of the nucleus. It appears much weaker in the optical \Ox\ than in the H$\alpha$ emission line, because it is the receding side of the outflow and obscured by the foreground dust screen of the disc. We extract a co-added spectrum  within an aperture of 0.8\arcsec radius for both high dispersion regions. The spectra for the south-west and north-east region are shown in the right panels of Fig.~\ref{fig:outflow_maps}. All optical lines as well as the CO(1-0) line exhibit a prominent broad wing on the blue and red side, respectively. The lack of a broad wing in the NIR lines is likely caused by the limited sensitivity given the peak flux ratio of the narrow-to-broad components seen in the CO(1-0) lines.

\begin{table*}
\caption{Emission-line measurements for the southern outflow region}\label{tab:outflow}
\centering
\hspace*{-8mm}
\input{table1.tex}
\end{table*}

\begin{figure*}
 \includegraphics[width=\textwidth]{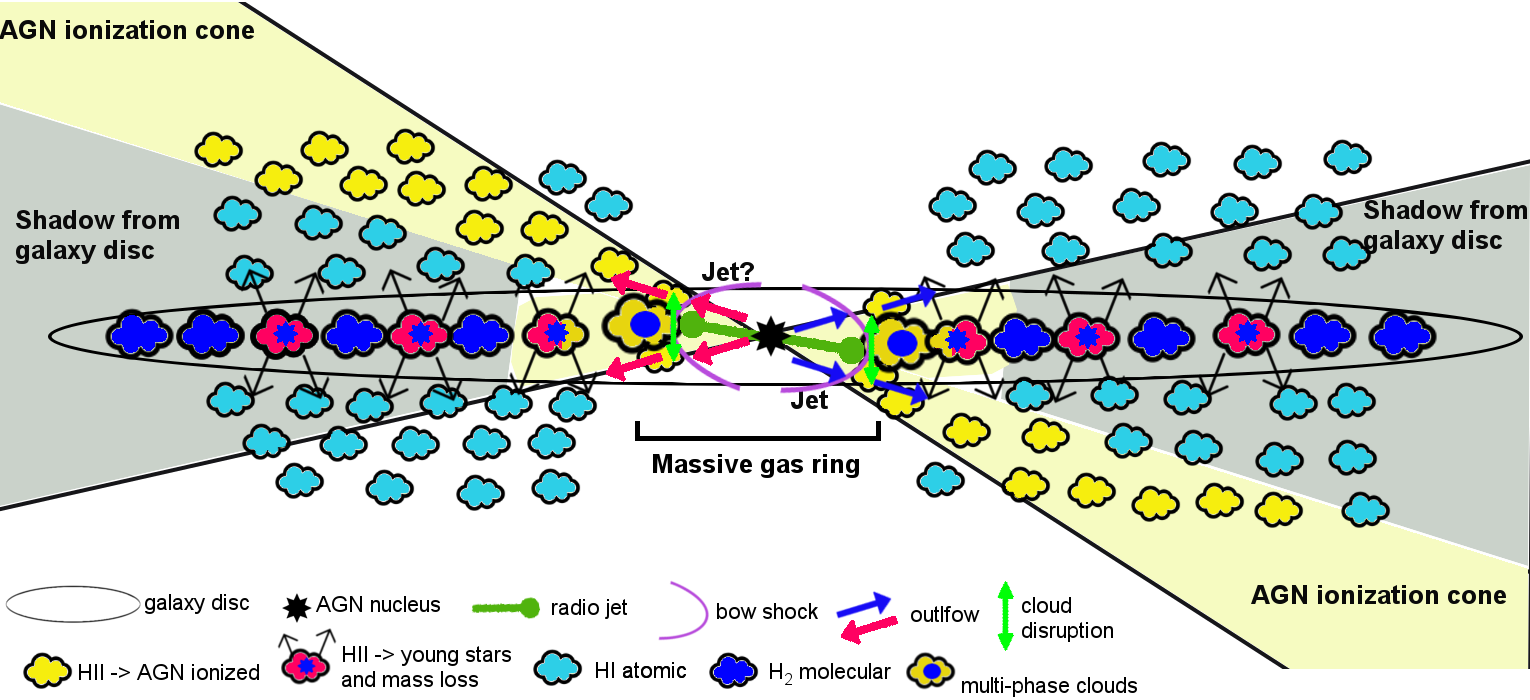}
 \caption{A simple cartoon like diagram to highlight the interaction of the AGN with its host galaxy in HE~1353$-$1917 seen fully edge-on but perpendicular to the ionization cones. In this orientation our observing position in reality would be about $5^\circ$ above the disc plane and an unknown azimuth angle to the right so that we can look down the ionization cone into the nucleus. The AGN ionization cones and radio jet are nearly perpendicular to the rotation axis of this galaxy disc in our proposed scenario. The AGN ionization is then able to illuminate  extra-planar material above the disc plane. This material was likely expelled from the disc due to mass-loss from star formation and supernova explosions. The gas becomes visible only on one side of the galaxy due the shadowing of the ionization cone given the intrinsic thickness of the dusty galaxy disc. A jet is hitting the inner wall of a molecular gas ring which significantly enhances the cloud turbulence and disrupts the clouds in the vertical direction. The fast outflow in the ionized gas is then accelerated, either by the radiation pressure or the hot outflow from the jet, and moving around the dense material at the inner edge of the dense gas ring in the disc plane.}\label{fig:Cartoon}
\end{figure*}

We modelled the emission lines as a superposition of two Gaussian profiles when applicable. We force doublet lines to be fixed in their theoretical flux ratios and coupled the velocity dispersion of all lines for the two distinct kinematic systems to avoid degeneracies in the solution. The parameters for the best-fit model are listed in Table~\ref{tab:outflow} for the south-west and north-east nucleus. The velocity of the core component is in all cases very close to the corresponding LOS velocity of the stars that we used as a reference. The velocity dispersion of the core component varies between $\sigma_\mathrm{core}\sim$60--140$\,\mathrm{km\,s}^{-1}$ (corrected for spectral resolution) for the ionized and warm-molecular gas phases, but much lower for the cold molecular gas as expected. The wing components are  systematically blueshifted by about $-300\,\mathrm{km\,s}^{-1}$ and $-160\,\mathrm{km\,s}^{-1}$ on the south-west side and redshifted by about $140\,\mathrm{km\,s}^{-1}$ and $100\,\mathrm{km\,s}^{-1}$ on the north-east side for the ionized and molecular gas, respectively. The ionized gas velocity dispersion of $365\pm27\,\mathrm{km\,s}^{-1}$ on the south-west side is significantly higher than the stellar velocity dispersion of $\sigma_*=239\pm19\,\mathrm{km\,s}^{-1}$ (in the same aperture) and the broad component of cold molecular gas with $67\pm25\,\mathrm{km\,s}^{-1}$. Those extreme ionized gas kinematics cannot be easily explained solely by gravitational motions and are a rather clear signature of a kpc-scale AGN-driven outflow.

Strikingly, the different gas phases are not exactly co-spatial after properly co-registering point-like AGN continuum emission of the various observations as seen in Fig.~\ref{fig:outflow_maps}. The warm-molecular gas as probed by H$_2$ [1-0 S(1)] is mainly excited at the location of the radio emission of the jet in the south-west region and the putative counter-jet in the north-east region. The high-dispersion molecular gas peaks are located at a slightly larger distance from the nucleus whereas the very fast ionized gas appears closer to the nucleus and slightly offset vertically from the disc plane. This suggests that the AGN-driven outflow is pushing on the cold gas in the disc with a stratification of excitation and speed as the shock front is passing the cold gas clouds which disrupts the front side of the clouds as sketched in Fig.~\ref{fig:Cartoon}. The gas inside the shock front is much warmer than the cold gas outside where the ionized gas is flowing around the thin and dense cold gas disc and diverted outside of the plane of the galaxy following the path of less resistance. In the following we will estimate the outflow energetics of the different gas phases to quantify the driving mechanisms of the outflow and to predict the potential damage to the galaxy.

\begin{table*}
\caption{Mass outflow rate, momentum injection rate and kinetic energy injection at a distance of 1\,kpc from the nucleus}\label{tab:energetics}
\centering
\input{table2.tex}
\end{table*}

\subsubsection{Outflow energetics and mass outflow rates}\label{sect:outflow_results}
As a first step we compute the outflowing ionized gas mass from the extinction-corrected H$\alpha$ luminosity ($L_{\mathrm{H}\alpha}$) and the electron density $(n_\mathrm{e})$ with
\begin{equation}
 M_\mathrm{ion} = 3.4\times10^6\left(\frac{100\mathrm{cm}^{-2}}{n_\mathrm{e}}\right)\times\left(\frac{L_{\mathrm{H}\alpha}}{10^{41}\mathrm{erg\,s}^{-1}}\right) M_\sun
\end{equation}
as described in \citet{Husemann:2016a} following \citet{Osterbrock:2006}. From the H$\alpha$/H$\beta$ Balmer decrement we infer an optical LOS attenuation, adopting a Milky Way-like attenuation curve \citep{Cardelli:1989}, of $A_{V,\mathrm{core}}=2.3\pm0.1$\,mag $A_{V,\mathrm{wing}}=1.8\pm0.2$\,mag in the south-west side for the core and outflow component, respectively. The corresponding values for the north-east side are $A_{V,\mathrm{core}}=1.2\pm0.3$\,mag $A_{V,\mathrm{wing}}>3.5$\,mag, which highlights the severe extinction along the LOS for the broad component at optical wavelengths and weak signature in the \Ox\ line on the north-east side of the nucleus. Based on the measured attenuations,  we obtained extinction-corrected H$\alpha$ luminosities of $\log(L_{\mathrm{H}\alpha,\mathrm{core}}/[\mathrm{erg\,s}^{-1}])=40.79\pm0.04$ and $\log(L_{\mathrm{H}\alpha,\mathrm{wing}}/[\mathrm{erg\,s}^{-1}])=40.34\pm0.06$ on the south-west side and $\log(L_{\mathrm{H}\alpha,\mathrm{core}}/[\mathrm{erg\,s}^{-1}])=40.05\pm0.08$ and $\log(L_{\mathrm{H}\alpha,\mathrm{wing}}/[\mathrm{erg\,s}^{-1}])>40.85$ on the north-east side. Here, we face the issue that the luminosity of the wing component is highly uncertain on the North side due to the severe extinction. The electron density can be estimated from the [\ion{S}{ii}] $\lambda\lambda$6717,6731 doublet line \citep{Osterbrock:2006}, which is often the greatest obstacle to robustly measure the outflowing ionized gas mass as discussed in \citet{Harrison:2018}, \citet{Kakkad:2018} and \citet{Rose:2018}.  Here, we are able to robustly deblend the [\ion{S}{ii}] doublet into a narrow and broad component for both the south-west and north-east region. We measure an $\mathrm{S2}$=[\ion{S}{ii}]$\lambda$6717/[\ion{S}{ii}]$\lambda$6731 ratio of $\mathrm{S2}_\mathrm{core}=1.21\pm0.04$ and $\mathrm{S2}_\mathrm{wing}=0.73\pm0.12$ for the core and wing components  in the south-west region, respectively, and $\mathrm{S2}_\mathrm{core}=1.20\pm0.14$ and $\mathrm{S2}_\mathrm{wing}=1.19\pm0.15$ for the north-east region. Assuming an electron temperature of $T_e\sim10\,000$\,K, we estimate electron densities of $n_\mathrm{e,core}=198_{-41}^{+45}\mathrm{cm}^{-2}$  and $n_\mathrm{e,wing}=1500_{-670}^{+1700}\mathrm{cm}^{-2}$ for the south-west region and $n_\mathrm{e,core}=185_{-111}^{+172}\mathrm{cm}^{-2}$  and $n_\mathrm{e,wing}=214_{-130}^{+175}\mathrm{cm}^{-2}$ for the north-east region.  Hence, we obtain ionized gas masses of $M_\mathrm{ion,core}=\left(1.1_{-0.2}^{+0.35}\right)\times10^{6}M_\sun$ and $M_\mathrm{ion,wing}=\left(0.5_{-0.3}^{+0.4}\right)\times10^{5}M_\sun$ in the south-west region  and $M_\mathrm{ion,core}=\left(1.9_{-0.9}^{+2.5}\right)\times10^{5}M_\sun$ in the north-east region.  Since the wing component of H$\beta$  remains undetected due to extreme obscuration towards the north-east region, we can only compute a lower limit on the ionized gas mass of $M_\mathrm{ion,wing}>1.1\times10^{6}M_\sun$ by applying the same extinction correction to the H$\alpha$ wing component as for the south-west region. 

\begin{figure*}
 \includegraphics[width=\textwidth]{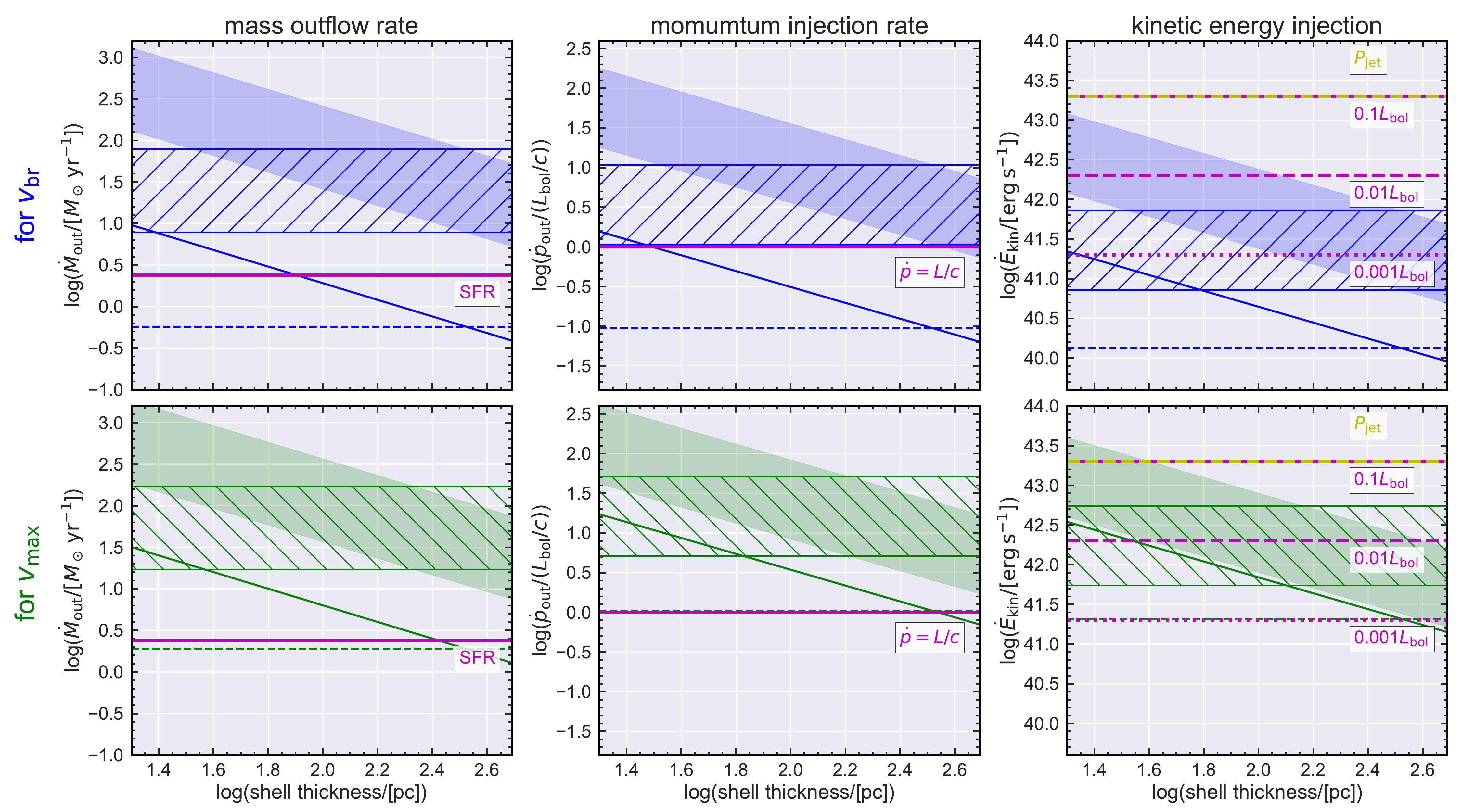}
 \caption{Mass outflow rate (left panel), momentum injection rate (middle panel) and kinetic energy injection rate (right panel) through a sphere at 1\,kpc radius as a function of shell thickness. Measurements based on $v_\mathrm{br}$ and $v_\mathrm{max}$ are shown in top panels  and bottom panels, respectively. Combined molecular+ionized gas measurements are shown as shaded (shell assumption) or hatched (cone assumption) areas highlighting the one order of magnitude uncertainty of $X_\mathrm{CO}$, while pure ionized gas mass computations results are represented by solid lines (shell assumption) or dashed lines (cone assumption). For comparison, the magenta lines highlight the SFR of the entire galaxy from the FIR measurements in the left panel, the point where $L_\mathrm{bol}/c=\dot p_\mathrm{out}$ in the middle panel, and the fraction of $L_\mathrm{bol}$ at 10\% (solid line), 1\% (dashed line) and 0.1\% (dotted line) as well as the jet power $P_\mathrm{jet}$ (yellow dashed line) in the right panel.}\label{fig:energetics}
\end{figure*}

Based on the ro-vibrational H$_2$ [1-0 S(1)] NIR emission line we estimate the warm molecular gas mass following the prescription of \citet{Mezcua:2015}, which assumes a temperature of 2000\,K, a transition probability of $A=3.47\times10^{-7}\,\mathrm{s}^{-1}$ and population fraction of $f=0.0122$. The leads to a mass conversion of
\begin{equation}
 M_\mathrm{warm} = 5.088\times10^{13}\left(\frac{D_L}{\mathrm{Mpc}}\right)^2\left(\frac{f_{\mathrm{H}_2\mathrm{\,[1-0 S(1)]}}}{\mathrm{erg\,s\,cm}^{-2}}\right)M_\sun
\end{equation}
Although the H$_2$ [1-0 S(1)] shows only one kinematic component in our NIFS observations, it is most likely associated with the outflow as the kinematics are close to those of the molecular gas wing component. We estimate warm-molecular gas outflow masses of $M_\mathrm{warm}=370\pm37 M_\sun$ for the south-west region and $M_\mathrm{warm}=250\pm25 M_\sun$ under the assumption stated above. Since those gas masses are orders of magnitude less than the outflowing ionized and molecular gas masses, the warm-molecular gas does not significantly contribute to outflow energetics and mass outflow rate budgets. We therefore ignore this gas-phase for the discussion of the outflow energy budget in the following.

We estimate the cold molecular gas from the CO luminosity following the prescription of \citet{Davis:2015}:
\begin{equation}
 M_\mathrm{mol} = 3.93\times10^{-17}X_\mathrm{CO}\left(\frac{D_L}{\mathrm{Mpc}}\right)^2\left(\frac{f_\mathrm{CO(1-0)}}{\mathrm{Jy\,km\,s}^{-1}}\right)
\end{equation}
for which we choose a conversion factor of  $X_{\rm CO}=2\times10^{20}\,\mathrm{cm}^{-2}(\mathrm{K\,km\,s}^{-1}\mathrm{pc}^2)^{-1}$ as a typical value for massive spiral galaxies like our Milky Way in the nearby Universe. This yields molecular gas masses of $M_\mathrm{mol,core}=(1.8\pm0.11)\times10^8M_\odot$ and $M_\mathrm{mol,winge}=(0.84\pm0.25)\times10^8M_\odot$ for the south-west region and $M_\mathrm{mol,core}=(0.9\pm0.2)\times10^8M_\odot$ and $M_\mathrm{mol,wing}=(1.0\pm0.3)\times10^8M_\odot$ for the north-east region. However, the conversion factor is the greatest source of systematic uncertainties and the conditions of the gas in the fast moving clouds is still matter of debate with a lower limit provided by optically-thin gas at a temperature of T$_{ex}=30$\,K corresponding to  X$_{\rm CO}$= 1.6$\times 10^{19}\,\mathrm{cm}^{-2}(\mathrm{K\,km\,s}^{-1}\mathrm{pc}^2)^{-1}$. While several studies have found significantly higher conversion factors for the outflowing molecular gas \citep{Cicone:2018b, Walter:2017,Zschaechner:2018} also optically-thin conditions have been reported \citep{Dasyra:2016,Oosterloo:2017}. Hence, our molecular gas estimates for the putative outflows could be at most an order of magnitude lower and represent a hard lower limit.

Mass outflow rates have often be computed using the assumption of a homogeneously filled cone \citep[e.g.][]{Fiore:2017} with a size $R_\mathrm{out}$ which leads to 
\begin{equation}
 \dot M_\mathrm{out,cone} = 3\times \left(\frac{v_\mathrm{out}}{100\,\mathrm{km\,s}^{-1}}\right)\left(\frac{M_\mathrm{out}}{10^7M_\sun}\right)\left(\frac{1\,\mathrm{kpc}}{R_\mathrm{out}}\right) \,M_\sun\,\mathrm{yr}^{-1}
\end{equation}
where $v_\mathrm{out}$ is the outflow velocity for which different measurements have been assumed in the literature, which strongly depends on which part of the line shape is assumed to be participating in the outflow. A maximum velocity was defined by \citet{Rupke:2013} to be $v_\mathrm{max} = v_\mathrm{broad}+2\times\sigma_\mathrm{broad}$ instead of $v_\mathrm{broad}$ alone for the bulk motion of the broad component. Alternatively, non-parametric line shape measurements of the entire line shape have been used to define $v_{10}$, the velocity containing 10\% of the line flux, and $W_{80}$, the width of the line between 10 and 90\% of the line flux. We list all those outflow velocity parameters in Table~\ref{tab:kin_parameters}. While for $v_\mathrm{max}$ and $v_\mathrm{broad}$ only the mass in the broad component is considered, $v_{10}$ and $W_{80}$ are often used with the gas mass inferred from the entire line shape.

In the specific case of HE~1353$-$1917 it seems that a filled cone geometry is not a good assumption because our spatially-resolved kinematic mapping suggests an expanding shell-like shock front scenario that is propagating through the galaxy. In this case the mass outflow rate can be described as 
\begin{equation}
 \dot M_\mathrm{out,shell} = \left(\frac{v_\mathrm{out}}{100\,\mathrm{km\,s}^{-1}}\right)\left(\frac{M_\mathrm{out}}{10^6M_\sun}\right)\left(\frac{100\,\mathrm{pc}}{\Delta R}\right) \,M_\sun\,\mathrm{yr}^{-1}
\end{equation}
which implicitly represents the localized outflow rate at the shell distance  from the nucleus independent of the outflow opening angle. The shell-thickness $\Delta R$ cannot be directly measured from our observations of HE~1353-1917 at our given spatial resolution. For the outflow energy in Table~\ref{tab:energetics} we adopt a thickness of $\Delta R=200$\,pc, but parametrize the thickness within a range of 20 and 500\,pc in Fig.~\ref{fig:energetics}. In addition to the mass outflow rate we also compute the momentum injection rate $\dot p_\mathrm{out}=v_\mathrm{out}\dot M_\mathrm{out}$ and kinetic energy injection rate as $\dot E_\mathrm{kin}=0.5v^2_\mathrm{out}\dot M_\mathrm{out}$. The turbulent motion is implicitly considered in all outflow velocity definitions except for $v_\mathrm{br}$ for which we add $\sigma_\mathrm{br}$ in quadrature to be comparable.

We visualize the results in Fig.~\ref{fig:energetics} for $\dot M_\mathrm{out}$, $\dot p_\mathrm{out}$ and $\dot E_\mathrm{kin}$ for the shell and cone geometry, various outflow velocities and the ionized and total gas phases. Since there is a one order of magnitude systematic uncertainty on the molecular gas mass depending on the actual $X_\mathrm{CO}$ factor in the outflow we show them as bands in all panels.  To separate the ionized gas and total gas mass is important for various reasons, 1) most observations of AGN are only able to measure the ionized gas energetics, 2) the molecular and ionized gas have different kinematics, and 3) it is unclear what fraction of the broad component in the molecular gas is caused by an outflow or governed by internal galaxy structure, such as a circumnuclear ring. This already tells us that even with our high-quality measurements computation of outflow energetics is hard and strongly model dependent. While we have taken out the uncertainty on the electron density for the ionized gas, the definition of outflow velocity, geometry and $X_\mathrm{CO}$ still leads to order of magnitude uncertainties.

Nevertheless, we can draw some solid conclusions from the data. The total mass outflow rate already exceeds the total SFR independent of $X_\mathrm{CO}$ or geometry. The total mass-loading factor $\eta=\dot M/\mathrm{SFR}$ therefore ranges between $\sim$1--30, if the molecular gas is indeed outflowing. Given that the outflow is still confined to the central kpc, the mass-loading factor would further increase by at least an order of magnitude if the SFR would be taken from the same region. Compared to the ionized gas the molecular phase clearly dominates $\dot M_\mathrm{out}$ even at much lower speed, which implies  $\eta<1$ (for the total SFR) for the ionized gas phase alone for all cases when the shell thickness is $\Delta R>100$\,pc. The difference in the momentum injection rate $\dot p_\mathrm{out}$ and the kinetic energy injection $\dot E_\mathrm{out}$ is significantly reduced between the ionized and molecular outflow due to the higher outflow velocity of the ionized gas phase. Our observations are consistent with $\dot p_\mathrm{out}\sim L_\mathrm{bol}/c$ only for the cone geometry, if $X_\mathrm{CO}$ is on the low side, and for the ionized gas phase if $\Delta R>300$\,pc. For the shell geometry we find $\dot p_\mathrm{out}> L_\mathrm{bol}/c$ if $\Delta R<100\,pc$\, independent of $X_\mathrm{CO}$. A caveat here is that the covering factor of the spherical AGN luminosity is significantly smaller than 1. Assuming a vertical height of $\pm250$\,pc for the outflow at 1\,kpc distance means that the covering factor is only 25\%, which would imply a momentum boost for radiation pressure only of $>1$ for the total $\dot p_\mathrm{out}$ in all cases. 

Only $\sim 1\%$ of the total bolometric luminosity is required to explain the estimated kinetic energy injection rates $\dot E_\mathrm{kin}$ in the outflow for all cases where $\Delta R>300$\,pc. Taking into account the covering factor would bring this to 4\%. Hence, the AGN radiation may indeed be able to drive the outflow if such a fraction of the luminosity can be efficiently transferred to the gas in the disc. However, we also see a radio jet impinging on the gas disc. We convert the inferred radio luminosity at 1.4\,GHz from our VLA observations at 10\,GHz to a kinetic jet power of $\log(P_\mathrm{jet}/[\mathrm{erg\,s}^{-1}])=43.3$ based on the calibration by \citet{Cavagnolo:2010}, which is nearly equal to $0.1L_\mathrm{bol}$. The radio jet alone would therefore also be able to mechanically drive the outflow with just 10\% of coupling efficiency as seen in Fig.~\ref{fig:energetics}.

\section{Discussion}\label{sect:discussion}
\subsection{The origin of the large extended narrow-line region}
One of the most prominent feature seen in the MUSE data is the extended narrow-line region (ENLR) where AGN photoionized gas can be traced out several tens of kpc to the limits of the MUSE FoV. The two-side geometry combined with a prior on the AGN orientation through the type 1 AGN classification implies that the geometry of the ENLR is driven by the opening angle and orientation of the AGN ionization cone fully consistent in line with the  unification model of AGN \citep{Antonucci:1993,Urry:1995}. However, various ENLR studies  usually find that the ENLR is bound to the galaxy in the majority of cases as the AGN illuminates primarily the ISM \citep[e.g.][]{Bennert:2002,Schmitt:2003b,Netzer:2004,Husemann:2013a,Hainline:2013,Husemann:2014}. There is only a minority of cases reported where the ENLR significantly reaches beyond the galaxy which have been explained by the illumination of gas accretion flows from the environment \citep[e.g.][]{Husemann:2011,Johnson:2018}, gas debris or filaments due to past or ongoing galaxy interaction \citep[e.g.][]{Villar-Martin:2010,daSilva:2011,Husemann:2016a,Storchi-Bergmann:2018,Villar-Martin:2018}, or large-scale gas outflows \citep[e.g.][]{Greene:2012,Liu:2013}. 

For HE~1353$-$1917 we have presented a number of measurements that let us draw a conclusive picture for the origin of the ionized gas a few kpc above the galaxy disc. The gas kinematics show that the ENLR shares the same direction of the angular momentum vector as the gas and stars of the disc without any signatures of counter-rotation. The LOS velocity amplitude on both sides of the ENLR is slightly smaller than that of the gas in the disc at the same radial distance along the plane. This could in principle be a signature of an additional velocity component (like an outflow) or simply an inclination effect since the ionization cone cannot be fully perpendicular to our line-of-sight given that we observe an unobscured AGN nucleus.

Additionally, the metallicity gradients along the galaxy disc and along the ENLR ionization cone axis are in good agreement. This is evidence against an outflow scenario because the metal enriched gas from the center would be ejected so that we would expect a significant increase of metals along the ENLR with respect to the galaxy disc. That the ENLR follows the metallicity gradient of the disc strongly suggests that the material in the ENLR was expelled nearly vertically from the disc and thereby retains a similar metallicity pattern as well as the same angular momentum as a function of radius. Hence, the AGN-illuminated gas above the plane is tightly connected to the disc properties as a function of radius. Stellar mass loss and supernova-driven winds is a natural explanation for the coupled similar kinematics and metal enrichment. However, the velocity dispersion in the extra-planar gas is much smaller than in typical starburst-driven winds \citep[e.g.][]{Ho:2014,Ho:2016}. This means that the outflow is significantly weaker or that the gas is already falling back after a short burst in star formation a few Myr ago. Both scenarios agree with the small amount of illuminated ionized gas $10^5M_\sun$ and the unusual star formation activity in the disc.

\subsection{Origin of the compact ring-like gas structure}
Another striking feature in our data is the enhancement of velocity dispersion in the cold gas and shocked-heated H$_2$ gas in two symmetric regions around the nucleus (Fig.~\ref{fig:outflow_maps}), which also resembles a ring-like structure seen edge-on. Such a ring-like structure is tentatively supported by a deficit of ionized gas (Br$\gamma$, Pa$\alpha$, or [FeII]) seen in the high-angular resolution Gemini observations (Fig.~\ref{fig:NIFS_maps}).  This is not affacted by the AGN subtraction as the PSF is significantly narrower than the apparent diameter of the ring. Given the orientation of the ionized gas outflow and the edge-on nature of this galaxy, there are several scenarios that may cause such a ring structure.

HE~1353$-$1917 is one of the few cases, such as NGC~1068 \citep{Garcia-Burillo:2014,Gallimore:2016} or NGC~5643 \citep{Alonso-Herrero:2018, Cresci:2015b}, where the ionization cones and jet axis are directly intercepting the cold gas of the galaxy. Hence, one potential scenario is that the AGN-driven outflow has been pushing the gas outwards for some time. This scenario is attractive to support the impact of AGN outflow, since a multi-phase outflow and a radio jet within the depleted gas disc is clearly detected. However, this scenario assumes that the cold gas disc was initially homogenous and reaching down to the cricumnuclear scales of a few pc.

Alternatively, the outflow is impinging on a pre-existing nuclear ring with a diameter of 2\farcs3 or 1.6\,kpc seen edge-on. The position-velocity diagram of Fig.~\ref{fig:pV-diagram} indeed can be interpreted as a typical signature for a nuclear ring rather than a pure outflow. The velocity signature of a ring is a straight and steep line in the center, 
as can be seen in our own Milky-Way central molecular zone for example \citep[e.g.][]{Yusef-Zadeh:2009}, with a ring of 200\,pc radius. Other examples of edge-on rings are M82 \citep{Neininger:1998} or NGC~4565 \citep{Yim:2014}. Obviously, it is more easy to identify nuclear rings when not edge-on, like in NGC~1433 \citep{Combes:2013}, NGC~1068 \citep{Garcia-Burillo:2014} or NGC~1097 \citep{Martin:2015}.

These rings are thought to be due to the gravity torques of a bar, driving gas toward the centre, where it accumulates transiently into a ring at the inner Lindbald resonance of the bar \citep[e.g.][]{Buta:1996}. The ring is then massive, and produces some kinematic decoupling with a characteristic feature in PVDs. When the barred galaxies are associated to a low-luminosity AGN (Seyfert, LINER), it is frequent to see also a molecular outflow, as revealed in the references above. Weak outflows are usually not the origin of such ring structure, which pre-existed the nuclear activity. However,  we cannot detect a bar due to the inclination of the galaxy so we cannot test if the ring is matching with the expected Lindbald resonance. Alternatively, the turbulent regime at the galaxy centre with a  turbulent Taylor number of ${\rm Ta_t} \equiv v_{\rm rot}/\sigma_v < 1$ will lead to inelastic collisions of gas clouds enhancing their condensation fostering rapid feeding of the BH \citep[e.g.][]{Gaspari:2017}. Whether the molecular gas in the ring has already being pushed outwards by the outflow or acts as a stiff barrier for the outflow given its mass and density is still unclear at this point. In the following section we assume nevertheless that most of the high dispersion in the molecular phase is associated with an outflow to compute the associated outflow energetics.

\subsection{The powering source of the multi-phase gas outflow}
In contrast to the large ENLR on tens of kpc scales, the prominent multi-phase gas outflow in HE~1353$-$1917 is confined to the inner $\sim$1\,kpc from the nucleus. This is a large size for the outflow compared to what we find in other CARS targets (Singha et al. in prep.). HE~1353$-$1917 is therefore an ideal laboratory to understand the driving mechanisms for this outflow as we could make relatively precise estimations of the outflow energetics from our multi-wavelength observations. One big question is whether these outflows are radiatively driven by the AGN radiation or mechanically driven by a jet even in radio-quiet AGN \citep{Wylezalek:2018b}. 

\begin{figure}
 \resizebox{\hsize}{!}{\includegraphics{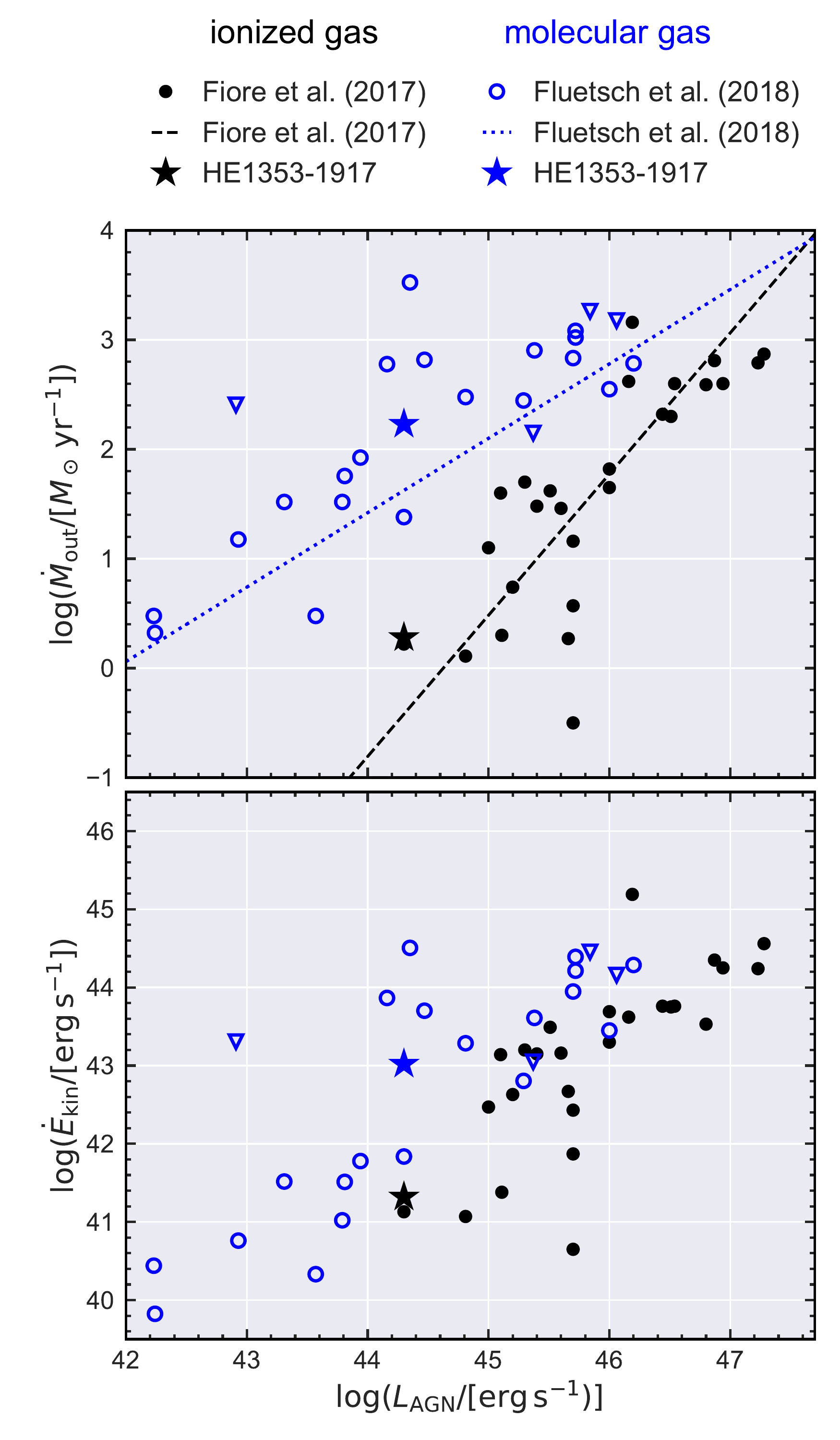}}
 \caption{Comparison of mass outflow rate (upper panel) and kinetic power (lower panel) with AGN bolometric luminosity as computed for a conical outflow geometry as described in the text. Literature compilations of AGN outflow in the ionized gas phase (black symbols) and the molecular gas phase (blue symbols) are taken from \citet{Fiore:2017} and \citet{Fluetsch:2019}, respectively. The measurements for HE~1353$-$1917 are denoted by the star symbol. Best-fit correlations for the literature compilation are shown as dashed lines.} \label{fig:outflows_literature}
\end{figure}

Considering the energetics of the outflow in comparison to the AGN bolometric luminosity and the jet power as shown in Fig.~\ref{fig:energetics} it seems that both the AGN and the radio jet are capable of powering the outflow. Yet, the jet power is already highly directional and the size of the jet exactly matches with the location of the warm molecular gas and the turbulent molecular gas. Since only a fraction of the AGN luminosity would impact the disc, the dissipation of the radiation into kinetic energy would need to be much more efficient than 1\% to power the entire multi-phase outflow.  Indeed, the orientation of the jet axis such that it almost exactly intercepts the gas disc is uncommon which makes HE~1353$-$1917 a rare case. However, similar examples have been reported before such as IC~5063 \citep[e.g.][]{Kulkarni:1998,Morganti:2013, Tadhunter:2014, Morganti:2015, Dasyra:2015, Dasyra:2016, Oosterloo:2017}, 3C~293 \citep[e.g.][]{Emonts:2005,Mahony:2013,Mahony:2016} or NGC~3079 \citep[e.g.][]{Duric:1988,Irwin:1992,Veilleux:1994,Cecil:2001,Middelberg:2007}. In all cases, high turbulent velocities have been found at the hot spots of radio jets. 

A spatial coincidence of radio jet morphology and velocity dispersion of the ionized gas has already been reported for spatially-resolved spectroscopy of more luminous radio-quiet AGN \citep[e.g.][]{Husemann:2013a,Villar-Martin:2017} and powerful compact radio sources \citep[e.g.][]{Roche:2016}, but it has been correctly proposed that the fast moving plasma itself can lead to radio emission that mimics jet activity \citep{Zakamska:2014,Hwang:2018}. In the case of HE~1353$-$1917 we can rule out that the ionized plasma is creating the radio emission because the high-velocity ionized gas traced by [\ion{O}{iii}] is significantly displaced compared to the observed jet-like radio emission. Hence, we think that the radio jet is transferring its energy and momentum to the ambient medium through an extended shock front which creates turbulence in a dense clumpy interstellar medium. Such a great impact of the radio jet has been observationally shown in many cases \citep[e.g.][]{Villar-Martin:1999b,ODea:2002,Nesvadba:2006,Holt:2008,Guillard:2012,Villar-Martin:2014,Harrison:2015,Santoro:2018,Tremblay:2018} and theoretically supported through detailed hydrodynamic simulations \citep[e.g.][]{Krause:2007,Sutherland:2007,Wagner:2011,Wagner:2012,Cielo:2018,Mukherjee:2018}. As we discussed in Sect.~\ref{sect:gas_kin}, the jet power alone is sufficient to energetically drive the outflow because only a small fraction of the AGN luminosity would impact the thin disc of the galaxy implying conversion efficiencies of more than 10\% of $L_\mathrm{bol}$. \citet{Hopkins:2010b} proposed a two-stage process for efficient radiation-driven outflows. They describe a scenario in which an initial weak wind in the hot gas phase, possibly initiated by an accretion disc wind or a radio jet, creates additional turbulence in the surrounding medium so that massive gas clouds will subsequently expand and disperse. This expansion of gas clouds would significantly increase their apparent cross-section with respect to incident radiation field of the AGN. Such a two-stage process may increase the coupling efficiency by an order of magnitude. While we cannot directly confirm this process with our observations, the close alignment of the jet axis and the ionization cone greatly suggest that the outflow is driven jointly by both mechanical and radiative energy with an unknown ratio of the two. The open question is whether the same powerful outflow could have developed without the fast radio jet impacting the cold gas directly given its unique orientation.

Recent compilations of ionized \citep{Fiore:2017} and molecular \citep{Fluetsch:2019} outflows found a significant correlation with outflow kinetic power and AGN luminosity. \citet{Fiore:2017} assumed a conical outflow geometry with $v_\mathrm{max}$ as the outflow velocity, while \citet{Fluetsch:2019} used the average thin shell approximation that is systematically smaller by a factor of 3. In Fig.~\ref{fig:outflows_literature} we compare the literature measurements re-computed for a conical outflow geometry with  our conical outflow measurements for HE~1353$-$1917 using $v_\mathrm{max}$ as in Table~\ref{tab:energetics}. We find that the mass outflow rates and kinetic power of the outflow measured for HE~1353$-$1917 is fully consistent with the clear trends between outflow energetics and AGN luminosity as found in previous studies. While we argue that a conical geometry is an inappropriate description for the outflow it is still valid for a homogenous relative comparison between different sample measurements where spatial information is limited. Although this could be taken as evidence that the outflow must be driven by the AGN luminosity, the fact that the radio jet clearly contributes to the AGN outflow cautions against this simple interpretation. Since the radio luminosity correlates with optical AGN luminosity even for radio-quiet AGN the relative contribution to the outflow is not clear and is not sufficiently probed in the current literature. Statistically, \citet{Mullaney:2013} and \citet{Villar-Martin:2014} argued for a significant role of radio jets in driving ionized gas outflows whereas \citet{Woo:2016} attributes this to simple sample selection effects. Only spatially-resolved mapping of the multi-phase outflows in relation to radio jet morphology and AGN ionization cones as presented here will be able to disentangle the relative importance of both mechanisms as discussed in \citet{Wylezalek:2018b}. 

\begin{figure}
 \resizebox{\hsize}{!}{\includegraphics{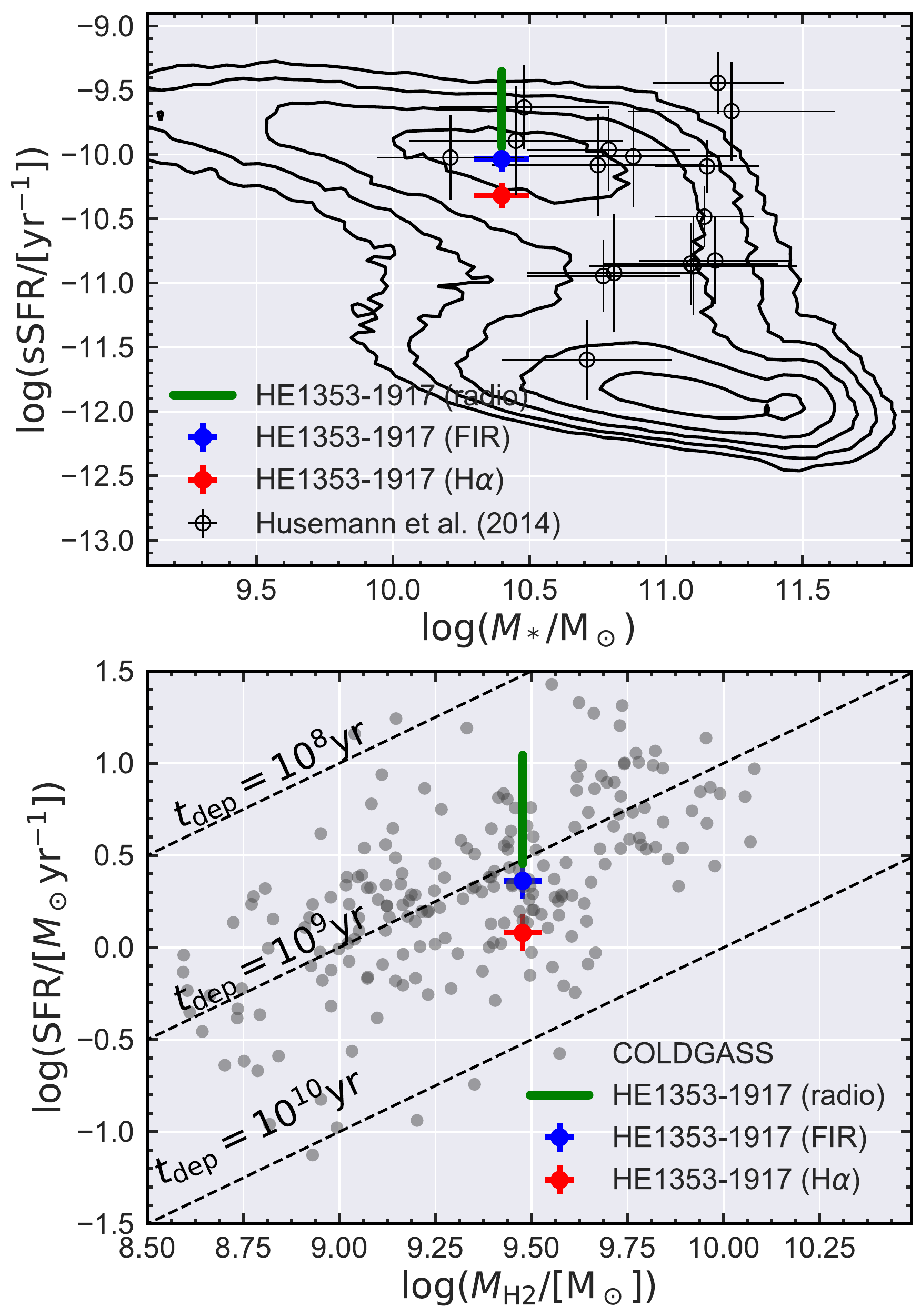}}
 \caption{\textit{Upper panel:} Specific star formation rate against stellar mass. Contours highlight the distribution of local galaxies as probed by SDSS DR7 \citep{Brinchmann:2004}. The position of HE~1353$-$1917 is denoted by the filled symbols, where the SFR is traced by the FIR continuum (blue symbol) and by the H$\alpha$ emission lines (red symbol) after deblending with the ENLR contribution as discussed in the text. The open symbols represent a local sample of luminous AGN with H$\alpha$-based SFR measurements from \citet{Husemann:2014} for comparison. \textit{Lower panel:} Star formation rate (SFR) against molecular gas mass ($M_{\mathrm{H}_2}$). The gray symbols represent the properties of a local galaxy sample from the COLDGASS survey \citep{Saintonge:2011}. The dashed diagonal lines represent constant value of the gas depletion time. The colored symbols are the measurements for HE~1353$-$1917 as in the upper panel.} \label{fig:mass_SFR}
\end{figure}

\subsection{Implications for AGN feedback}
The direct suppression of star formation caused by large-scale AGN-driven winds is one potential mechanism to regulate the growth of massive galaxies. While the cold gas in disc galaxies is hard to destroy when the outflow expands perpendicular to the disc \citep[e.g.][]{Gabor:2014}, the outflow in HE~1353$-$1917 is orientated such that it is directly impacting the cold gas disc. Hence, we would expect the efficiency of negative AGN feedback, i.e. suppression of star formation, to be maximized at the given outflow energy. In Fig.~\ref{fig:mass_SFR} we compare the 
sSFR and the associated cold gas depletion time ($t_\mathrm{dep}$) of HE~1353$-$1917 with the overall distribution of local non-AGN galaxies from SDSS \citep{Brinchmann:2004} and the COLDGASS survey \citep{Saintonge:2011}. We find that HE~1353$-$1917 is nearly matching galaxies in the star-forming main sequence pointing against negative AGN feedback caused by the outflow. 

However, we also find a significant difference between the sSFR based on the FIR and H$\alpha$-based SFR tracers where the FIR-based sSFR is closer to the star-forming main sequence than the H$\alpha$-based SFR. The difference is a factor of 2, which is significant considering the measurement errors. As pointed out by \citet{Hayward:2014}, the FIR-based SFRs may over-predict the SFR in galaxies undergoing star formation suppression, because the FIR emission is excited by stars with life times of up to $\sim$100\,Myr and may not react on SFR changes on much smaller timescales. Here, we also compute the SFR based on the AGN core subtracted 1.4GHz radio luminosity and the radio-SFR calibration by \citet{Bell:2003}. The SFR range between 2.8\,$M_\sun\,\mathrm{yr}^{-1}$ and 10\,$M_\sun\,\mathrm{yr}^{-1}$ for an assumed range in spectral index of $-1.2<\alpha<-0.5$. This suggest that the SFR has remained nearly constant till a few tens of Myr as probed by radio emission. Since the H$\alpha$ line can only be excited by stars with life times of up to $\sim$Myr it is much more sensitive to more rapid SFR variations than the FIR or the radio. This timescale effect has been detected in merging galaxies where the SFR is enhanced on small timescale at close galaxy separation which is only recovered in H$\alpha$ and not in the longer timescales SFR tracers \citep{Davies:2015}. Considering a time scale of at most a few Myr years for the AGN phase in HE~1353$-$1917 and the associated impact on the surrounding medium, it is plausible that the difference in the SFR is caused by the same timescale effect. On the contrary, the H$\alpha$ luminosity may be particularly prone to the systematics of dust extinction given the edge-on view we have on HE1353$-$1917, which may also cause the discrepancy in the SFRs. 

In any case, the impact of the current powerful AGN-driven outflow in HE~1353$-$1917 on \emph{the total} SFR and star formation efficiency is small and nearly indistinguishable from the parent population of non-AGN galaxies. This is in agreement with several works finding that AGN host galaxies have similar star formation properties as their non-AGN counterparts \citep[e.g.][]{Harrison:2012b,Husemann:2014,Xu:2015,Balmaverde:2016,Vaddi:2016,Zhang:2016,Suh:2017,Rosario:2018,Shangguan:2018}, while others report a significant decrease in the SFR \citep[e.g.][]{Farrah:2012,Mullaney:2015,Shimizu:2015,Wylezalek:2016} for different kind of AGN samples.

\section{Summary and conclusions}\label{sect:conclusion}
In this paper we presented an extensive spatially-resolved multi-wavelength investigation of the edge-on disc galaxy HE~1353$-$1917 hosting a luminous AGN. The observations offer unique insights into the interaction of an AGN with its host galaxy. Our main results can be summarized as follows:
\begin{itemize}
 \item HE~1353$-$1817 reveals a large biconical ENLR which extends about 25\,kpc nearly parallel to the galaxy disc. Considering the kinematics and the metallicity of the ionized gas we conclude that the gas likely originated from the galaxy disc and was vertically lifted by stellar mass loss and supernovae on several Myr timescales. Without the illumination by the AGN such low-density diffuse gas would be a neutral gas phase and undetectable at optical wavelengths.
 \item A fast multi-phase outflow about $\sim$1\,kpc away from the nucleus is detected in the ionized gas ($v_\mathrm{max}\sim1000$km/s) and molecular gas ($v_\mathrm{max}\sim300$km/s). This region of the outflow appears as a distinct feature in the PVD and is coincident with jet-like radio emission at 10\,GHz as detected with the VLA. Warm molecular emission is also directly associated with the radio jet location. This co-location implies that the ISM is heated directly through a fast moving shock, which is hitting the dense, cold gas disc of the galaxy. The cold gas has possibly formed a ring structure dynamically at this radius, created through a bar or other internal mechanisms, such as chaotic cold accretion. 
 \item Despite the great uncertainties in computing outflow energetics, we can confirm that the molecular gas phase would dominate the mass outflow rate if a ring structure is not dominating the observed cold gas kinematics. In the most pessimistic case the mass outflow rate is $\dot M>2M_\odot\,\mathrm{yr}^{-1}$ and may be as high as $\dot M\sim200M_\odot\,\mathrm{yr}^{-1}$ implying a mass loading factor greater than unity. 
 \item It is unclear whether the momentum injection rate of the outflow is consistent with $\dot p = L_\mathrm{bol}/c$ or requires a momentum boost as suggested for many other AGN-driven outflows. However, the power in the radio jet is sufficient to drive the outflow alone, so that the outflow is not necessarily radiatively driven. While indeed 5\% of the total AGN luminosity would be sufficient to power the outflow as well, the geometric covering factor of the AGN radiation field intercepting the gas disc is almost an order of magnitude lower which implies that the radiation alone may not be able to power the outflow.
 \item The ionized and molecular mass outflow rates and the associated kinetic outflow power for HE~1353$-$1917 is fully in line with previous trends with AGN luminosity as found in larger samples of AGN host galaxies. 
 \item The host galaxy of HE~1353$-$1917 is still on the star-forming main sequence based on FIR and radio continuum diagnostics which would not imply strong negative or positive feedback by the outflow. However, we detect a mild reduction in the SFR by a factor of 2 when estimated from the extinction-corrected and ionization-corrected H$\alpha$ luminosity. It is unclear at this point if this implies a recent reduction of the SFR in the last 5\,Myr (H$\alpha$) compared to 100\,Myr (FIR) or if a large fraction of the H$\alpha$ luminosity is missed by the high attenuation of dust due to edge-on orientation of the galaxy.
\end{itemize}

Our study of HE~1353$-$1917 demonstrates the unique potential of CARS to unravel the details of multi-phase AGN-driven outflows, their driving mechanism and impact on the host galaxies. This single-object study shows how the different observations can be combined to provide a complete picture of the galaxy characteristics. While it is difficult to make general conclusions for the overall AGN population from a single object study, there are a few important points to be drawn from this initial investigation, which will drive further studies from the overall CARS sample.

A striking feature is the orientation of the AGN ionization cones and the radio jet axis which directly intercept the cold gas disc of this galaxy. This orientation effect certainly maximizes the efficiency of the AGN to drive a massive outflow that already reaches out to distance of 1\,kpc from the nucleus. The direct spatial coincidence of the radio jet with the fast outflow suggests a dominant role of the jet power in driving the outflow which is confirmed based on the outflow energetics. Nevertheless, the impact of the outflow on the SFR is still very mild and may at most affect the central kpc at very recent times. Whether AGN feedback can significantly reduce the molecular gas content and SFR of the entire disc on longer timescales is unclear at this point.

Considering that the outflow rates and energetics of HE~1353$-$1917 are in agreement with estimates for larger AGN samples from the literature, we may pose the question whether the previous trends with AGN luminosity can be considered as unambiguous evidence for radiatively-driven AGN outflows. Systematic investigations of the connection between small-scale low-power radio jets and the multi-phase outflows are still lacking for many existing samples. One of the primary goals of CARS is therefore to systematically explore the relative role of jets and radiation as the powering mechanisms of AGN-driven outflows and its impact on star formation.

\begin{acknowledgements}
We thank the referee for providing very valuable comments which significantly improved the quality of the manuscript. MK acknowledges support from DLR grant 50OR1802. GRT acknowledges support from the NASA through Einstein Postdoctoral Fellowship Award Number PF-150128, issued by the Chandra X-ray Observatory Center, which is operated by the Smithsonian Astrophysical Observatory for and on behalf of NASA under contract NAS8-03060. MG is supported by the Lyman Spitzer Jr. Fellowship (Princeton University) and by NASA Chandra grants GO7-18121X/GO8-19104X. SMC acknowledges support from the Australian Research Council (DP190102714). We thank Alex Markowitz for helpful discussions on the RGS data in the context of warm absorbers.  The work of SAB, CPO and MS was supported by a generous grant from the Natural Sciences and Engineering Research Council of Canada.

Based on observations collected at the European Organization for Astronomical Research in the Southern Hemisphere under ESO programme 095.B-0015(A).Based on observations obtained at the Gemini Observatory, which is operated by the Association of Universities for Research in Astronomy, Inc., under a cooperative agreement with the NSF on behalf of the Gemini partnership: the National Science Foundation (United States), National Research Council (Canada), CONICYT (Chile), Ministerio de Ciencia, Tecnolog\'{i}a e Innovaci\'{o}n Productiva (Argentina), Minist\'{e}rio da Ci\^{e}ncia, Tecnologia e Inova\c{c}\~{a}o (Brazil), and Korea Astronomy and Space Science Institute (Republic of Korea).
Based on observations obtained at the Southern Astrophysical Research (SOAR) telescope, which is a joint project of the Minist\'{e}rio da Ci\^{e}ncia, Tecnologia, Inova\c{c}\~{o}es e Comunica\c{c}\~{o}es (MCTIC) do Brasil, the U.S. National Optical Astronomy Observatory (NOAO), the University of North Carolina at Chapel Hill (UNC), and Michigan State University (MSU). Based on observations collected at the German-Spanish Astronomical Center, Calar Alto, jointly operated by the Max-Planck-Institut für Astronomie Heidelberg and the Instituto de Astrofísica de Andalucía (CSIC). This paper makes use of the following ALMA data: ADS/JAO.ALMA\#2016.1.00952.S. ALMA is a partnership of ESO (representing its member states), NSF (USA) and NINS (Japan), together with NRC (Canada), MOST and ASIAA (Taiwan), and KASI (Republic of Korea), in cooperation with the Republic of Chile. The Joint ALMA Observatory is operated by ESO, AUI/NRAO and NAOJ. This work is based on observations obtained with XMM-Newton, an ESA science mission with instruments and contributions directly funded by ESA Member States and NASA. The VLA is operated by the National Radio Astronomy Observatory, a facility of the National Science Foundation operated under cooperative agreement by Associated Universities, Inc. {\it Herschel} is an ESA space observatory with science instruments provided by European-led Principal Investigator consortia and with important participation from NASA. This publication makes use of data products from the Wide-field Infrared Survey Explorer, which is a joint project of the University of California, Los Angeles, and the Jet Propulsion Laboratory/California Institute of Technology, funded by the National Aeronautics and Space Administration. This research has made use of the NASA/IPAC Infrared Science Archive, which is operated by the Jet Propulsion Laboratory, California Institute of Technology, under contract with the National Aeronautics and Space Administration.
The Pan-STARRS1 Surveys (PS1) and the PS1 public science archive have been made possible through contributions by the Institute for Astronomy, the University of Hawaii, the Pan-STARRS Project Office, the Max-Planck Society and its participating institutes, the Max Planck Institute for Astronomy, Heidelberg and the Max Planck Institute for Extraterrestrial Physics, Garching, The Johns Hopkins University, Durham University, the University of Edinburgh, the Queen's University Belfast, the Harvard-Smithsonian Center for Astrophysics, the Las Cumbres Observatory Global Telescope Network Incorporated, the National Central University of Taiwan, the Space Telescope Science Institute, the National Aeronautics and Space Administration under Grant No. NNX08AR22G issued through the Planetary Science Division of the NASA Science Mission Directorate, the National Science Foundation Grant No. AST-1238877, the University of Maryland, Eotvos Lorand University (ELTE), the Los Alamos National Laboratory, and the Gordon and Betty Moore Foundation. This work is based in part on observations made with the Galaxy Evolution Explorer (GALEX). GALEX is a NASA Small Explorer, whose mission was developed in cooperation with the Centre National d’Etudes Spatiales (CNES) of France and the Korean Ministry of Science and Technology. GALEX is operated for NASA by the California Institute of Technology under NASA contract NAS5-98034.
\end{acknowledgements}

\bibliographystyle{aa}
\bibliography{references}

\begin{appendix}
\section{Multi-wavelength data processing}
\subsection{MUSE observations}\label{sect:MUSE_appendix}
HE~1353$-$1917 was observed with MUSE in the WFM-NOAO-N configuration during the night of June 20, 2015. The MUSE $1\arcmin\times1\arcmin$ field-of-view (FoV) is large enough to cover the entire optical extent of this edge-on disc-like galaxy. Three individual exposures with 300\,s on-source integration time were taken including small dither offsets and 90$\degr$ field rotation to reduce systematics. The observations were carried out with thin cirrus clouds passing and excellent seeing conditions with $0.6''$ (FWHM) as measured by the slow guiding system (SGS) within MUSE. As part of the standard calibration plan, a sequence of bias frames, arc lamp and continuum lamp exposures were taken at the end of the night and the standard star EG~153 was observed at the beginning of the night. In addition, a snapshot continuum lamp exposure was taken within an hour of the target observation to improve the temperature-dependent illumination correction across the MUSE FoV. A set of twilight exposures are available from the beginning of the previous night.

We reduced the data with standard MUSE pipeline \citep[version 1.6.0,][]{Weilbacher:2012,Weilbacher:2014}, which performs all the reduction steps till the creation of the flux-calibrated science cube. The process includes bias subtraction, wavelength calibration, flat-field correction, illumination correction across the field, sky subtraction, flux calibration and correction for differential atmospheric refraction before the reconstruction of the cube. Sky subtraction is performed based on the collapsed spectrum within the blank sky areas of the FoV. We further suppress the residuals of sky lines adopting a PCA analysis of the individual sky spectra in the blank sky area. Our custom-made PCA processing is similar but not identical to the PCA sky subtraction algorithm \textsc{ZAP} \citep{Soto:2016} and will be released with the CARS survey presentation article (Husemann et al. in prep.). The final cube has a spatial sampling of $0\farcs2$ and covers the wavelength range $4750\AA<\lambda<9300\AA$ at a spectral resolution of $\sim$2.5\AA.

\subsection{Gemini NIFS observations}
Near-infrared IFU data of the central $\sim 3\arcsec\times 3\arcsec$ were obtained with NIFS at Gemini-North as a Fast Turnaround project on 2017 April 6 and 7. The NIFS data were taken under natural seeing and non-photometric (CC70) conditions. The spectra cover the near-infrared $J$- and $K$-bands at spectral resolutions of $\sim$ 6040 and 5290, respectively. The data for each band were observed using 6 dithered on-source exposures of 600\,s. Telluric standard stars (of type A0V) were observed at a similar airmass immediately before and after the science exposures, respectively.

he data for HE1353-1917 and the telluric calibration stars were reduced using standard tasks included in the Gemini IRAF Package (Version 1.14). The object exposures were interleaved with off-target sky exposures for sky subtraction. The object and sky exposures were flat-fielded and rectified spatially and spectrally. These data products were used as input to perform the sky subtraction using {\tt Skycorr} \citep{Noll:2014}. In order to correct for telluric absorption features, each sky-subtracted on-source exposure was divided by the telluric standard star spectrum observed closest in time and normalized by a black body spectrum of $T=9480$\,K corresponding to the A0V stellar type. Hydrogen absorption lines, intrinsic to A0V stars, were deblended from the telluric spectra beforehand. The science data were flux calibrated based on the telluric standard stars by assuming their corresponding 2MASS $J$- and $K$-band fluxes. The six calibrated science data cubes per band were aligned by using the galaxy nucleus as a reference, median-combined, and binned to $0\farc1 \times 0\farc1$ pixels.

Given the non-photometric conditions, we compared the reconstructed $J$ and $K$ broad-band flux within a diameter of $3''$ from the NIFS datacubes with the corresponding flux in our PANIC NIR images (see Sect.~\ref{sect:PANIC}). While the $K$ band flux agrees within the errors, we find a systemic offset in the $J$ band flux of 0.76. We multiplied the $J$ band NIFS data with this factor to correct for this empirically measured offset with respect to the 2MASS photometric system.

\subsection{SOAR optical imaging observations}
We observed HE~1353$-$1917 with SOI \citep{Walker:2003} at the SOAR telescope using its set of Bessel $B$, $V$, and $R$ filters on 18 February 2016 as part of a follow-up imaging run for CARS. We integrated for 1200\,s per filter over a four-point dither pattern. The data were reduced and combined with the \texttt{mscred} package in {\sc iraf}.  The three-color composite is shown in the top left panel of Fig.~\ref{fig:overview}.

From the reduced images we infer integrated magnitudes for HE~1353$-$1917 within an elliptical aperture ($r_\mathrm{major}=21.1\arcsec$ and $r_\mathrm{minor}=5.6\arcsec$) covering the entire galaxy. The resulting brightnesses are $m_B=15.88\pm0.05$\,mag(AB), $m_V=15.2\pm0.01$\,mag(AB), and $m_R=14.81\pm0.01$\,mag(AB).  The errors represent the zero-point uncertainties inferred from the standard star fields.

\subsection{PANIC Near-IR imaging observations}\label{sect:PANIC}
Deep NIR $JHKs$ images of HE~1353$-$1917 were obtained during the nights of 6-9 March 2017 with the PANIC at the 2.2m telescope of the Calar Alto Observatory. The target galaxy was positioned in the center of quadrant 1 of PANIC which offers the best cosmetics of the detector mosaic. A sequence of  16 dithered exposure with offsets of 1.5\arcmin\ in between were obtained to aid background subtraction. The effective exposure times are 4600\,s, 5000\,s and 4500\,s for the $J$, $H$ and $Ks$ filter, respectively, after removing bad frames due to clouds or satellite stripes.  The observing conditions were overall good with a seeing in the range of 1--2\arcsec. The data was reduced with the dedicated Pipeline for PANIC \citep[PAPI,][]{Ibanez:2012} which performs dark subtraction, flat-fielding, background subtraction based on dithered exposures, astrometry based on star catalogs and co-addition of frames. We then use the software {PhotometryPipeline} \citep{Mommert:2017} to determine the zero-points for each co-added images based on cross-match with the 2MASS catalog \citep{Skrutskie:2006}. We infer total brightnesses of $m_J=12.60\pm0.06$\,mag(Vega), $m_H=11.82\pm0.06$\,mag(Vega), and $m_{Ks}=11.30\pm0.06$\,mag(Vega) within the same elliptical aperture as used for the SOAR images.

\subsection{ALMA observations}
The  $^{12}$CO(1--0) line in HE~1353$-$1917 was observed with ALMA on both the 3rd November and the 24th of December 2016 for 70.56 and 34.23 minutes in configurations C40-5 and C40-2, respectively. Combining these observations gives us sensitivity to emission on scales up to 22\farc6. An 1850 MHz correlator window was placed over the $^{12}$CO(1--0) line, yielding a continuous velocity coverage of $\approx$3000 km s$^{-1}$ with a raw velocity resolution of $\approx$1 km s$^{-1}$. Three 2 GHz wide low-resolution correlator windows were simultaneously used to detect continuum emission.

The raw ALMA data were calibrated using the standard ALMA pipeline, as provided by the ALMA regional centre staff. Additional flagging was carried out where necessary to improve the data quality. Amplitude and bandpass calibration were performed using Ganymede and the quasar J1337-1257.  J1400-1858 was used as the phase calibrator.  We then used the \texttt{Common Astronomy Software Applications} \citep[{\tt CASA},][]{McMullin:2007} package to combine and image the visibility files of the two tracks, producing a three-dimensional RA-Dec-velocity data cube (with velocities determined with respect to the rest frequency of the $^{12}$CO(1-0) line). In this work we primarily use data with a channel width of 50\, km s$^{-1}$, and pixels of 0\farc2 were chosen in order to  approximately Nyquist sample the synthesized beam.

The data presented here were imaged using Briggs weighting with a robust parameter of 0.5, yielding a synthesized beam of 0\farc64\,$\times$\,0\farc55 at a position angle of 71$^{\circ}$ (a physical resolution of 476\,$\times$\,409 pc$^2$). 
Continuum emission was detected, measured over the full line-free bandwidth, and then subtracted from the data in the $uv$ plane using the {\tt CASA} task {\tt uvcontsub}. The achieved continuum root-mean square (RMS) noise is 9.3 $\mu$Jy. The continuum-subtracted dirty cubes were cleaned in regions of source emission (identified interactively) to a threshold equal to the RMS noise of the dirty channels.  The clean components were then added back and re-convolved using a Gaussian beam with a FWHM equal to that of the dirty beam.  This produced the final, reduced, and fully calibrated $^{12}$CO(1--0) data cubes, with RMS noise levels of 0.21 mJy beam$^{-1}$ in each 50 km s$^{-1}$ channel.

\subsection{VLA radio 10\,GHz radio continuum}
We observed HE~1353$-$1917  on 22 December 2016 with the VLA at X-band in the A configuration. The observations were centered at 10.0\,GHz, with a synthesized bandwidth of 4 GHz (from 8 up to 12 GHz). We used the source 3C~286 for bandpass and absolute flux density calibration purposes.  For phase calibration purposes, we used the nearby, bright sources J1357$-$1744. The total on-source time was approximately 40\,min. We calibrated and imaged the data using the {\tt CASA}, applying standard procedures. The resulting cleaned image with a spatial resolution of about 0\farcs2 is shown as contours in Fig.~\ref{fig:overview}. It reveals an apparently one-sided jet-like radio morphology with a size of about 1\arcsec. The peak flux density at the synthesized frequency of 10\,GHz is of 241.1\,$\mu$Jy/beam, and the total flux density is of about 728.6\,$\mu$Jy, which are highly significant given an off-source rms of $12.5\,\mu$Jy/beam. The total radio flux at 10\,GHz corresponds to a radio luminosity of $L_{10\mathrm{GHz}}=2.0\times10^{21}\,\mathrm{W Hz}^{-1}$ which corresponds to 1.4\,GHz radio luminosities ranging from $L_{1.4\,\mathrm{GHz}}=5.3\times10^{21}\,\mathrm{W\,Hz}^{-1}$ to $L_{1.4\,\mathrm{GHz}}=2.5\times10^{22}\,\mathrm{W\,Hz}^{-1}$ considering a likely range of $-1.2<\alpha<-0.5$ spectral indices.

\subsection{VLA 1.7\,GHz  radio continuum and \ion{H}{I} line}\label{sect:VLA_HI}
We observed HE~1353$-$1917  on 30 April 2016 with the VLA at L-band in the CnB configuration. The observations consist of 6 continuum sub-bands with $64\times2$\,MHz channels each corresponding to a synthesized bandwidth of 730\,MHz (from 1.2 up to 2.1 GHz with some gaps in between due to RFI).  One sub-band with $512\times15.6$\,kHz was set for line mapping centered on the redshifted \ion{H}{I}. 3C~286 was used again as a bandpass and absolute flux density calibrator while Phase calibration was performed using the bright source J1351$-$1449. The total on-source time was approximately 90\,min. The data was calibrated with {\tt CASA}  applying standard routines. The cleaned continuum and \ion{H}{I} images have a spatial resolution of about $10\farcs0\times5\farcs4$ and are shown in Fig.~\ref{fig:overview}. Furthermore, continuum emission was detected and measured over the full \ion{H}{I} line-free bandwidth and subsequently subtracted from the data in the $uv$ plane using the {\tt CASA} task {\tt uvcontsub}.

\begin{figure}
 \resizebox{\hsize}{!}{\includegraphics{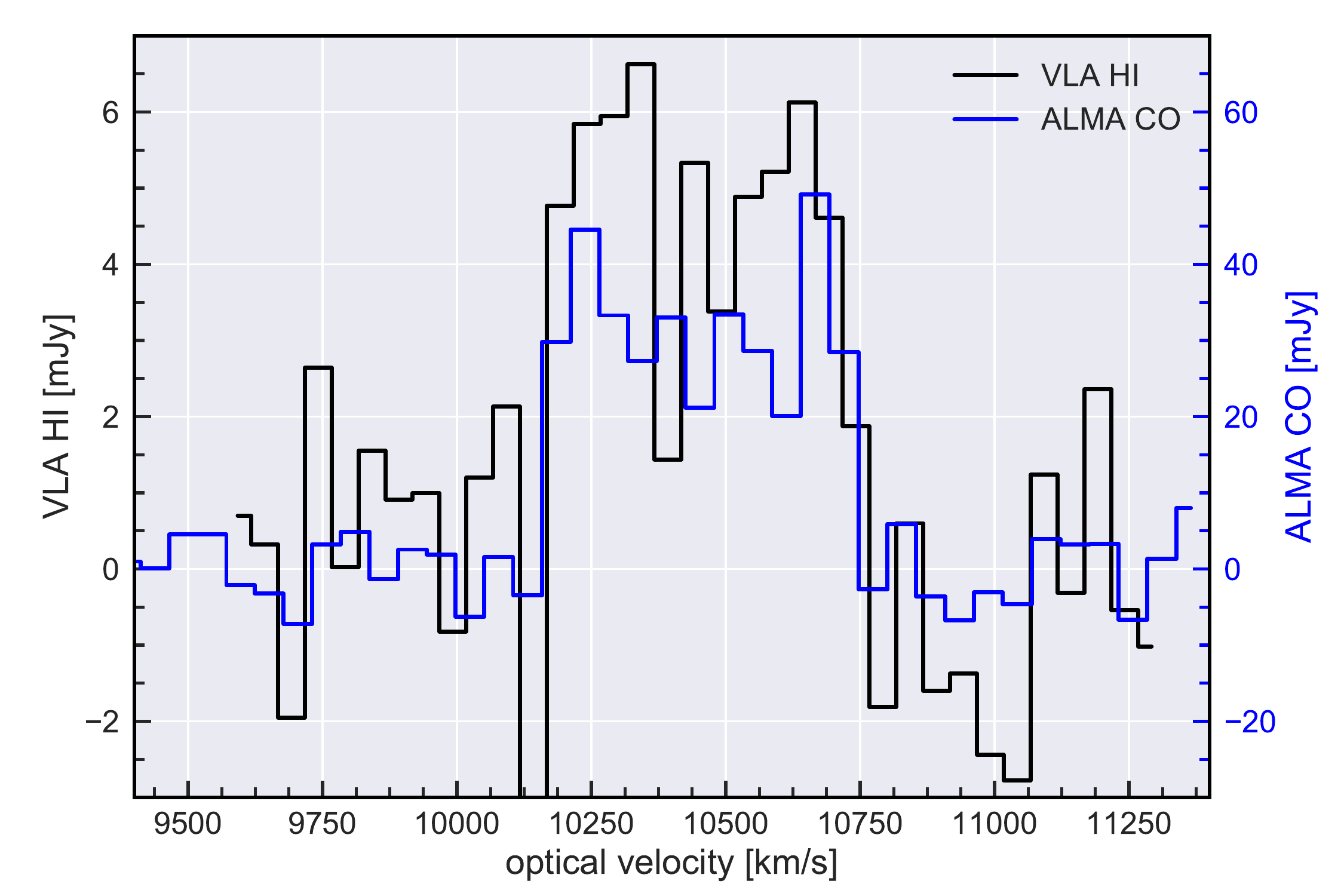}}
 \caption{Comparison of the integrated CO(1-0) and \ion{H}{i} line profiles as measured with ALMA and VLA, respectively.} \label{fig:CO_HI_profile}
\end{figure}

We detect spatially extended continuum emission from the host galaxy with a peak flux density of 3.8\,mJy/beam with an off-source rms if $21.5\,\mu$Jy/beam. The integral flux density is 9.4\,mJy which corresponds to a 1.4\,GHz luminosity of  $L_{1.4\,\mathrm{GHz}}=2.1\times10^{22}\,\mathrm{W\,Hz}^{-1}$ assuming a radio spectral index of $\alpha=-0.8$. Also extended \ion{H}{i} emission was detected associated with the host galaxy of HE1353$-$1917 despite a non-detection with the Effelsberg radio telescope as reported by \citet{Koenig:2009}. The integrated \ion{H}{i} line spectrum over 2\arcmin as obtained from the VLA is shown in Fig.~\ref{fig:CO_HI_profile}. We detect the \ion{H}{i} line at $5\sigma$ significance with the right shape and a total line flux of $f_{\mathrm{\ion{H}{i}}}=2.4\pm0.5\,\mathrm{Jy\,km\,s}^{-1}$.

\subsection{XMM-Newton X-ray observations}\label{sect:XMM_appendix}
HE~1353$-$1917 was observed with {\it XMM-Newton} on 2017 August 8 with the EPIC cameras and RGS for 62\,ksec. The spectra were extracted with the {\tt SAS} package (16.1.0) and the {\tt HEASOFT} (v6.21) using the standard setting for point sources. Since the small window mode was used, pile-up is not affecting the observations. We created a standard source and background spectrum in PN, MOS1, and MOS2. The spectra are grouped with a minimum binning of 20 counts. Standard source reductions procedures are also used for the RGS data.

We fit the X-ray spectra using {\tt Xspec} version 12.9.1e \citep{Arnaud:1996} in the 0.3--12 keV energy range. We use the cosmic abundances of \citet{Wilms:2000} and the photoelectric absorption cross sections provided by \citet{Verner:1996}.
The best-fit model yields Galactic absorption-corrected fluxes of $f_\mathrm{0.5-2\,keV}=(3.40\pm0.02) \times 10^{-12}$ erg cm$^{-2}$ s$^{-1}$ and $f_\mathrm{2-10\,keV}=(5.97\pm0.03) \times 10^{-12}$ erg cm$^{-2}$ s$^{-1}$, which corresponds to rest-frame luminosities of $L_\mathrm{0.5-2\,keV}=(9.50\pm0.06) \times 10^{42}$ erg s$^{-1}$ and $L_\mathrm{2-10\,keV}=(1.69\pm0.01) \times 10^{43}$ erg s$^{-1}$. Using \cite{Marconi:2004} we estimate $L_{\rm bol}/L_{\rm 2-10\,keV}\sim 10$. Thus, we estimate a bolometric luminosity of $L_{\rm bol} = (1.7\pm0.6) \times 10^{44}$ erg s$^{-1}$.

\begin{figure}
 \resizebox{\hsize}{!}{\includegraphics{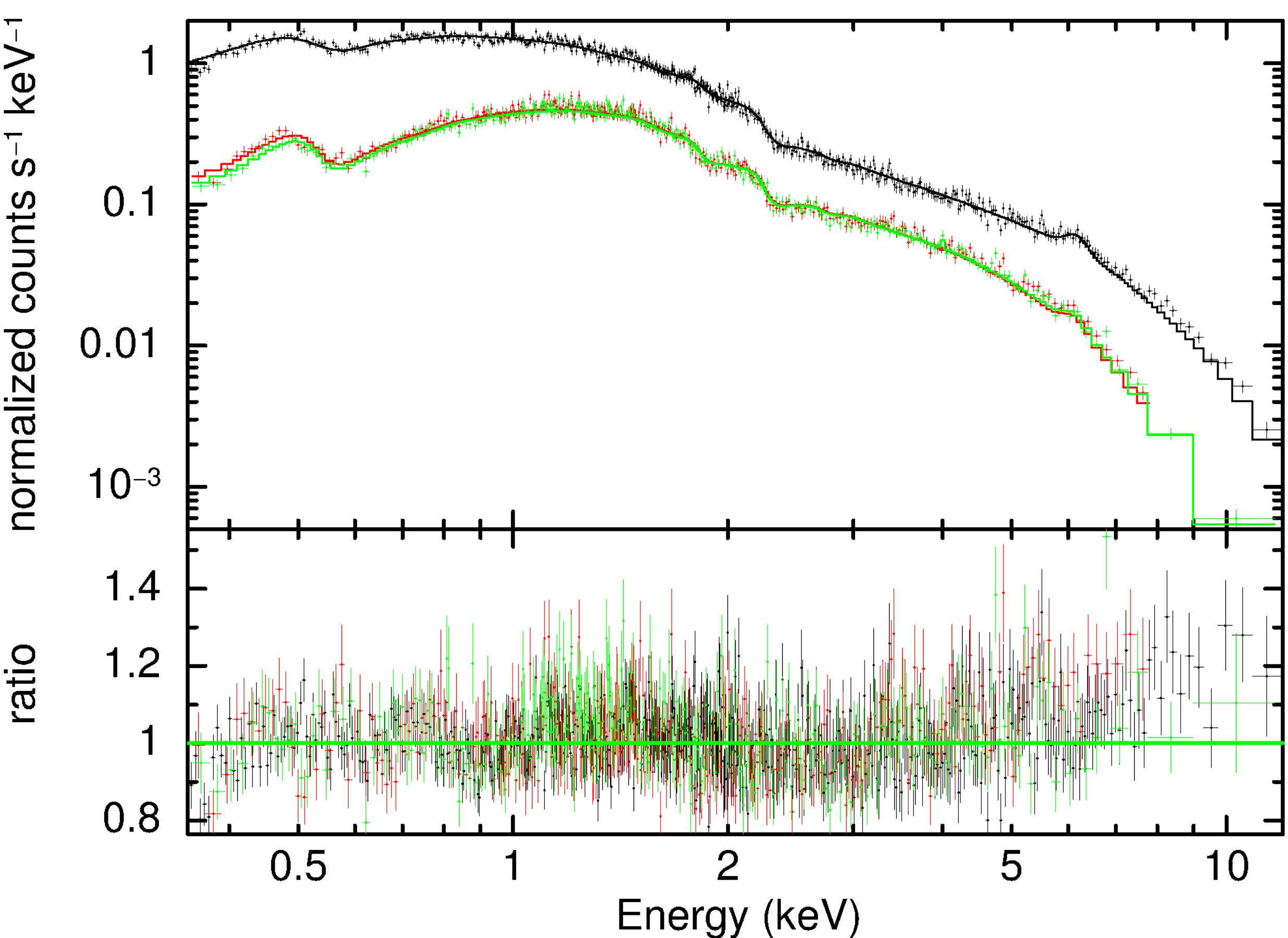}}
\caption{\textit{XMM-Newton} EPIC broad-band spectra of HE~1353$-$1917. 
The best-fit model to the data (top) and the residuals (bottom) are shown. To achieve this best fit, we have to model the Galactic absorption, a power-law component, a 6.4 keV (rest-frame) Fe-line, and two zones of ionized absorption (XSTAR grid model). One absorption zone has a low and the other a high ionization parameter ($\log \xi \sim -1.4$, $\log \xi \sim 2.7$, respectively). The low S/N RGS data confirm the presence of these two absorption systems with $v_{\rm out}<1000$ km s$^{-1}$.}
\label{fig:Xray_spec}
\end{figure}

HE~1353$-$1917 is also detected in the 105-month {\it Swift}-BAT All-sky hard X-ray Survey \citep{Oh:2018} with a total 14-195 keV flux of $f_{14-195~\mathrm{keV}}=1.39^{+0.34}_{-0.19}\times10^{-11}$ erg s$^{-1} $cm$^{-2}$ and a photon index of $\gamma=2.1^{+0.6}_{-0.5}$ (both quoting 90\% confidence levels). This hard X-ray flux corresponds to a luminosity of $\log (L_{14-195~\mathrm{keV}}/[\mathrm{erg\,s}^{-1}]) = 43.5$.

\subsection{\textit{Chandra} X-ray Observatory observations}
\label{sect:chandra_appendix}
\begin{figure}
\resizebox{\hsize}{!}{\includegraphics[width=\textwidth]{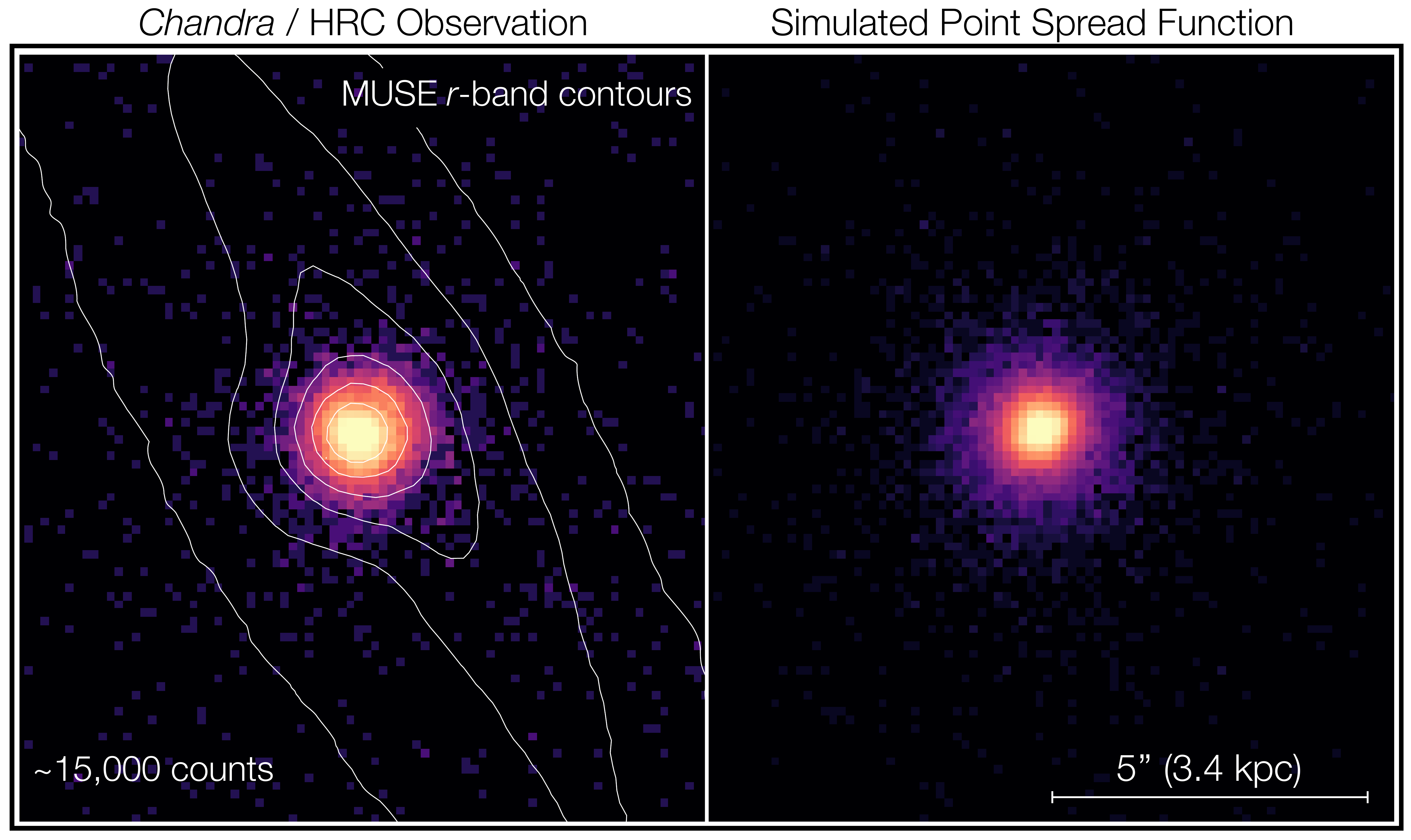}}
\caption{\textit{Left panel:} The $\sim40$ ksec \textit{Chandra} High Resolution Camera (HRC) observation of HE~1353$-$1917, with MUSE $r$-band contours overlaid in white. \textit{Right panel:} The simulated point spread function for the observation. We find no evidence for any extended emission superposed over the PSF. We do, however, make note of a semicircular arc of $\sim$10 excess counts $\sim2\arcsec$ SW of the nucleus and shown more clearly in Fig.~\ref{fig:chandra_excess}.}
\label{fig:chandra}
\end{figure}

We also observed HE~1353$-$1917 with the \textit{Chandra X-ray Observatory}'s High Resolution Camera (HRC) as part of Guaranteed Time Observation (GTO) program 19700783 (PI: Kraft). The galaxy's nucleus was placed on the nominal aim point of the HRC-I detector for the 39.62\,ksec observation, imaging the source with the sharpest point spread function available to \textit{Chandra} ($0\farcs4$ FWHM). Beyond the nuclear point source, which contains $\sim$15\,000 counts in the 0.5--7\,keV band, there is a $\sim$2$\sigma$ enhancement in total counts over the background level that forms a linear streak extending $\sim$6\arcsec southwest of the nucleus (see Fig.~\ref{fig:chandra}, left panel). 
The number of counts in this feature is $\sim$0.4\% of the total counts found in the nuclear PSF. The feature also happens to align exactly along the $U$-axis of the HRC-I detector, and therefore shares many characteristics with the well-known HRC ``ghost image'' artifact \footnote{\url{http://cxc.harvard.edu/proposer/POG/html/chap7.html}}.
Especially in the absence of any supporting multi-wavelength evidence it is most likely that this feature is indeed a detector artifact.

In search of any extended emission superposed on the nuclear point source, we extracted the radial profile of X-ray surface brightness from a series of 15 concentric annuli centered on the observation's photo-centroid. Using the best-fit \textit{XMM-Newton} X-ray spectrum (see Sect.~\ref{sect:XMM_appendix})
as an input, we simulated the observation's PSF using the Chandra Ray Tracer (\texttt{ChaRT}) and \texttt{marx} tools\footnote{\url{http://cxc.harvard.edu/ciao/threads/psf.html}}. We compared the radial profile extracted from the data with that from the modeled point spread function (Fig.~\ref{fig:chandra}, right panel) finding no evidence for any broadening from extended emission superposed upon the PSF.
 
\begin{figure}
\resizebox{\hsize}{!}{\includegraphics[width=\textwidth]{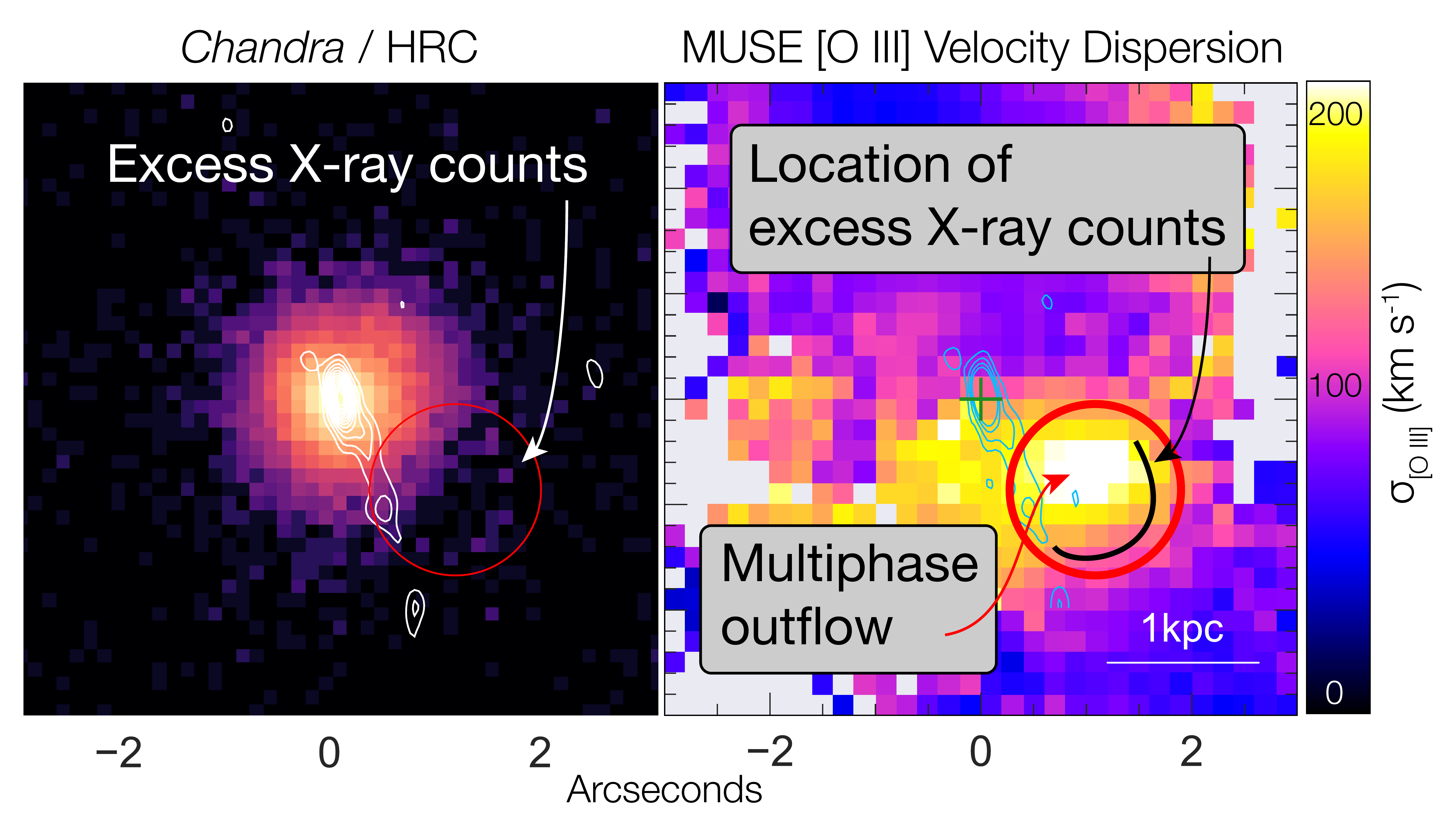}}
\caption{Comparison of the \textit{Chandra} X-ray image with the location of the outflow. This figure highlights a $\sim$10 X-ray count excess over the local azimuthal (background plus PSF wing) average, arranged in a semicircular arc $\sim$2\arcsec SW of the nucleus. The feature is labeled with a white arrow in the left panel, which shows the exposure-corrected \textit{Chandra}/HRC image at the native pixel scale (0\farcs13 per pixel). While the significance of $<$2$\sigma$ of this feature is too low for confident detection,  it is co-spatial with the rim of the multiphase outflow discussed in this paper. This is highlighted comparing the red (identical, WCS-matched) circles in both the \textit{Chandra} and MUSE gas velocity dispersion panels (right and left, respectively).}
\label{fig:chandra_excess}
\end{figure}

There is, however, a semicircular arc, $\sim2\arcsec$ SW of the nucleus, along which we find a $\sim$10 count excess over the local semi-azimuthal average (see Fig.~\ref{fig:chandra_excess}). Depending on the comparison aperture used, this feature represents only a $\sim$1.3--2$\sigma$ excess over the background-plus-PSF-wing level, and therefore cannot be ruled out as a background fluctuation. Nevertheless, the arc-like structure imposed by the excess counts is exactly co-spatial with the rim of the multiphase outflow discussed in this paper (compare, for example, the WCS-matched red circles in both panels of Fig.~\ref{fig:chandra_excess}). If there is indeed an expanding hot cocoon or bubble draped about the rim of the cooler phases of the outflow, the faint semicircular arc arrangement of X-ray counts detected by \textit{Chandra}/HRC are exactly where one would expect them to be.  This feature is not consistent with any known HRC PSF artifact, and we have run a series of deep \texttt{ChaRT} ray tracing simulations to ensure that it might not be a subtle effect of, e.g., misaligned mirror pairs. The location and arrangement of these excess counts is therefore too compelling a coincidence to ignore, and certainly motivation for a deeper future observation with \textit{Chandra}. While we in no way claim that this $<$2$\sigma$ feature is real, we suggest it is \textit{possible} that we have detected signal from a hot bubble associated with the outflow. A significantly deeper observation with \textit{Chandra} is certainly required to confirm this possibility.

\section{Archival multi-band data}\label{sect:archival_data}
\subsection{Mid-IR imaging with WISE}\label{sect:WISE_appendix}
The Wide-field Infrared Survey Explorer \citep[{\it WISE}, ][]{Wright:2010} satellite performed an all-sky survey in three mid-IR pass bands at 3.4, 4.6, 12 and 22 $\mu$m at a spatial resolution of 6.1\arcsec, 6.4\arcsec, 6.5\arcsec, and 12.0\arcsec, respectively. We retrieved the {\it WISE} cut-out images centered on HE~1353$-$1917 from the NASA/IPAC Infrared Science Archive. The images in the first three bands are shown in Fig.~\ref{fig:overview} in which the disc contribution below the bright AGN are well recognized even at the relatively low spatial resolution.

Additionally, we retrieve the single-frame PSF images that we use as template for modelling the images. Estimation of physical fluxes from the {\it WISE} images is performed following the {\it WISE} Photometric calibration following \citep{Wright:2010} and explained in section IV.h.i of the explanatory supplement\footnote{\url{http://wise2.ipac.caltech.edu/docs/release/allsky/expsup/}} in that paper.

\subsection{Far-IR imaging with Herschel}\label{sect:Herschel_appendix}
Far-IR imaging of HE~1353$-$1017 was obtained with {\it Herschel} \citep{Pilbratt:2010} with the instruments PACS \citep{Poglitsch:2010} at 70 and 160\,$\mu$m and SPIRE \citep{Griffin:2010} at 250, 350 and 500\,$\mu$m. The observations were taken in January 2012 as part of the programme ``Determining the bolometric luminosities of AGN'' (PI: R. Mushotzky)  with the Observation IDs 1342237734, 1342237735, and 1342236189. We retrieve the public data through the {\it Herschel} science archive\footnote{\url{http://www.cosmos.esa.int/web/herschel/science-archive}}. The PSF increased with wavelength from $\sim$9\,\arcsec\ to $\sim$36\,\arcsec over the entire wavelength range. We therefore only show the PACS images in Fig.~\ref{fig:overview} for which the spatial resolution is still sufficient to resolve the galaxy.

We performed simple aperture photometry of this isolated source to obtain integrated FIR fluxes for the galaxy. The galaxy is well detected in all band with $>5\sigma$ significance. We infer integrated 70 and 160\,$\mu$m PACS fluxes of $204\pm18$\,mJy and $659\pm82$\,mJy, respectively. In the three SPIRE bands we measure integrated 250, 350 and 500\,$\mu$m fluxes of $444\pm53$\,mJy, $215\pm32$\,mJy and $85\pm17$\,mJy, respectively.

\subsection{Optical imaging from Pan-STARRS}\label{sect:PNASTARRS_appendix}
The Panoramic Survey Telescope and Rapid Response System \citep[Pan-STARRS,][]{Chambers:2016} also covered HE~1353$-$1917 in the $g,r,i,z,y$ filters. From the Pan-STARRS data release 1\footnote{available at \url{http://panstarrs.stsci.edu}}, we retrieved the fully processed and stacked cut-out image   \citep{Waters:2016}. The sampling of the images is 0\farcs25 with integration times of 792\,s, 1372\,s, 1639\,s, 1080\,s, and 1160\,s for the $g$, $r$, $i$, $z$, $y$ filters, respectively. Adopting the same elliptical aperture as for the SOAR images, we obtained brightnesses of  $m_g=15.55$\,mag(AB), $m_r=14.67$\,mag(AB), $m_i=14.39$\,mag(AB), $m_z=14.10$\,mag(AB) and $m_y=13.90$mag(AB) with a photometric error of roughly $0.02$\,mag \citep{Magnier:2016}.

\subsection{UV photometry with GALEX}\label{sect:GALEX_appendix}
The Galaxy Evolution Explorer \citep[GALEX,][]{Martin:2005} has performed an all-sky imaging survey (AIS) in two UV bands, 1350--1750\AA\ (FUV channel) and 1750--2750\AA\ (NUV channel). The AIS data is shallow with a sensitivity down to $m_\mathrm{AB}\lesssim20.5$\,mag. The resolution implied by the point-spread function in the two bands is $4\farcs9$ and $4\farcs2$ in the NUV and FUV channel, which means that the images of HE~1353$-$1917 are dominated by the point-like emission of the AGN. From the GALEX Release 6/7 photometric catalogs \citep{Bianchi:2014} we retrieved the UV photometry for HE~1353$-$1917 in the two bands. With a total integration time per band of 225\,s in the NUV and 114\,s in the FUV, the reported fluxes are $0.180\pm0.006$\,mJy in the NUV and $0.054\pm0.009$\,mJy in the FUV channel.

\end{appendix}

\end{document}

%% file: table1.tex
\begin{tabular}{lcccccccccc}\hline\hline
  & \multicolumn{3}{c}{narrow core} & & \multicolumn{3}{c}{broad wing} & &\\\cline{2-4}\cline{6-8}
Line & $f_\mathrm{line}$\tablefootmark{a} & $v_\mathrm{line}$ & $\sigma_\mathrm{line}$ & &$f_\mathrm{line}$\tablefootmark{a} & $v_\mathrm{line}$ & $\sigma_\mathrm{line}$ & $v_{10}$ & $W_{80}$ & $v_\mathrm{max}$ \\
 & & [$\mathrm{km}\,\mathrm{s}^{-1}$] & [$\mathrm{km}\,\mathrm{s}^{-1}$] & & & [$\mathrm{km}\,\mathrm{s}^{-1}$] & [$\mathrm{km}\,\mathrm{s}^{-1}$] & [$\mathrm{km}\,\mathrm{s}^{-1}$] & [$\mathrm{km}\,\mathrm{s}^{-1}$] & [$\mathrm{km}\,\mathrm{s}^{-1}$]\\\hline
\multicolumn{11}{c}{south west region}\\\hline
$\mathrm{H}\beta$ & $6.2\pm0.2$ & $-60\pm10$ & $143\pm3$ & & $4.0\pm0.2$ & $-310\pm20$ & $360\pm30$ & $-550\pm30$ & $680\pm30$ & $1040\pm50$ \\
$\mathrm{[OIII]\ }\lambda$5007\AA & $56.7\pm0.9$ & $-60\pm10$ & $143\pm3$ & & $29.2\pm0.6$ & $-310\pm20$ & $360\pm30$ & $-510\pm20$ & $640\pm20$ & $1040\pm60$ \\
$\mathrm{H}\alpha$ & $37.3\pm0.4$ & $-80\pm10$ & $143\pm3$ & & $20.3\pm0.5$ & $-350\pm20$ & $360\pm30$ & $-560\pm20$ & $660\pm20$ & $1080\pm60$ \\
$\mathrm{[NII]\ }\lambda$6548\AA & $11.7\pm0.2$ & $-80\pm10$ & $143\pm3$ & & $8.1\pm0.1$ & $-350\pm20$ & $360\pm30$ & $-600\pm20$ & $710\pm30$ & $1080\pm60$ \\
$\mathrm{[SII]\ }\lambda$6717\AA & $12.3\pm0.2$ & $-90\pm10$ & $143\pm3$ & & $5.2\pm0.8$ & $-440\pm60$ & $360\pm30$ & $-590\pm70$ & $680\pm60$ & $1170\pm80$ \\
$\mathrm{[SII]\ }\lambda$6730\AA & $10.1\pm0.3$ & $-90\pm10$ & $143\pm3$ & & $7.1\pm0.6$ & $-440\pm60$ & $360\pm30$ & $-690\pm70$ & $770\pm60$ & $1170\pm80$ \\
Pa$\beta$ & $5.0\pm0.1$ & $70\pm10$ & $97\pm3$ & & ... & ... & ... & ... & ... & ...\\
$\mathrm{H}_2$ [1-0 S(1)]  & $2.8\pm0.1$ & $-20\pm10$ & $144\pm7$ & & ... & ... & ... & ... & ... & ... \\
CO(1-0) & $0.66\pm0.04$ & $-40\pm10$ & $25\pm1$ & & $0.30\pm0.09$ & $-160\pm20$ & $70\pm30$ & $-190\pm30$  & $180\pm30$ & $290\pm50$ \\
\noalign{\smallskip}\hline
\multicolumn{11}{c}{north east region}\\\hline
$\mathrm{H}\beta$ & $4.2\pm0.3$ & $60\pm10$ & $66\pm3$ & & $<2.0$ & $160\pm10$ & $180\pm20$ & $163\pm4$ & $190\pm20$ & $530\pm30$ \\
$\mathrm{[OIII]\ }\lambda$5007\AA & $37.8\pm0.8$ & $60\pm10$ & $66\pm3$ & & $11.5\pm0.8$ & $160\pm10$ & $180\pm20$ & $213\pm5$ & $250\pm10$ & $530\pm30$ \\
$\mathrm{H}\alpha$ & $17.2\pm0.7$ & $70\pm10$ & $66\pm3$ & & $17.6\pm0.7$ & $120\pm10$ & $180\pm20$ & $280\pm10$ & $340\pm20$ & $490\pm30$ \\
$\mathrm{[NII]\ }\lambda$6548\AA & $4.2\pm0.2$ & $70\pm10$ & $66\pm3$ & & $6.2\pm0.2$ & $120\pm10$ & $180\pm20$ & $300\pm10$ & $370\pm30$ & $490\pm30$ \\
$\mathrm{[SII]\ }\lambda$6717\AA & $4.1\pm0.3$ & $70\pm10$ & $66\pm3$ & & $5.6\pm0.5$ & $120\pm20$ & $180\pm20$ & $300\pm20$ & $360\pm30$ & $490\pm40$ \\
$\mathrm{[SII]\ }\lambda$6730\AA & $3.4\pm0.3$ & $70\pm10$ & $66\pm3$ & & $4.7\pm0.4$ & $120\pm20$ & $180\pm20$ & $300\pm20$ & $360\pm30$ & $490\pm40$ \\
Pa$\beta$ & $3.7\pm0.1$ & $180\pm10$ & $77\pm30$ & & ... & ... & ... & ... & ... & ...\\
$\mathrm{H}_2$ [1-0 S(1)]  & $2.1\pm0.1$ & $170\pm10$ & $97\pm4$ & & ... & ... & ... & ... & ... & ... \\
CO(1-0) & $0.32\pm0.07$ & $70\pm10$ & $25\pm3$ & & $0.37\pm0.10$ & $130\pm40$ & $100\pm30$ & $220\pm30$  & $190\pm50$ & $330\pm70$ \\
\noalign{\smallskip}\hline
\end{tabular}
\tablefoot{
\tablefoottext{a}{Emission line fluxes are in units $10^{-16}\mathrm{erg}\,\mathrm{s}^{-1}\mathrm{cm}^{-2}$ except for the CO(1-0) for which the units are $\mathrm{Jy\,km}^{-1}$.}}

%% file: table2.tex
\begin{tabular}{lccccccccccccccc}\hline\hline
 & \multicolumn{7}{c}{cone geometry} & &\multicolumn{7}{c}{shell geometry}\\\cline{2-8}\cline{10-16}
 & \multicolumn{3}{c}{ionized} & & \multicolumn{3}{c}{molecular} & &\multicolumn{3}{c}{ionized} & &\multicolumn{3}{c}{molecular} \\\cline{2-4}\cline{6-8}\cline{10-12}\cline{14-16}
 & $\log(\dot{M})$ & $\log(\dot{p})$& $\log(\dot{E})$ & & $\log(\dot{M})$ & $\log(\dot{p})$ & $\log(\dot{E})$& & $\log(\dot{M})$& $\log(\dot{p})$ & $\log(\dot{E})$ & & $\log(\dot{M})$ & $\log(\dot{p})$ & $\log(\dot{E})$\\\hline
$v_\mathrm{br}$ & $-0.24$ & $32.80$ & $40.12$ & & $1.89$ & $34.85$ & $41.85$ & & $-0.02$ & $33.02$ & $40.34$ & &  $2.11$ & $35.07$ & $42.07$\\
$v_\mathrm{max}$ & $0.28$ & $33.84$ & $41.32$ & & $2.23$ & $35.53$ & $42.72$ & & $0.50$ & $34.06$ & $41.54$ & & $2.26$ & $35.43$ & $42.56$\\
$v_\mathrm{10}$ & $0.41$ & $33.83$ & $41.19$ & & $2.42$ & $35.50$ & $42.49$ & & $0.64$ & $34.05$ & $41.41$ & & $2.64$ & $35.73$ & $42.72$\\
$W_\mathrm{80}$ & $0.50$ & $34.02$ & $41.49$ & & $2.40$ & $35.47$ & $42.44$ & & $0.72$ &  $34.24$ & $41.71$ & & $2.62$ & $35.69$ & $42.66$\\
\noalign{\smallskip}\hline
\end{tabular}
\tablefoot{
Units for mass outflow rates ($\dot M$) are $M_{\sun}\,\mathrm{yr}^{-1}$, for the momentum injection rates ($\dot p$) are dyne, and for energy injection rates ($\dot E$) are $\mathrm{erg\,s}^{-1}$. We do not provide the formal measurement errors here since they are significantly smaller than the spread implied by the different asummptions.}

%% file: CARS_HE1353_1917_Husemann.bbl
\begin{thebibliography}{254}
\expandafter\ifx\csname natexlab\endcsname\relax\def\natexlab#1{#1}\fi

\bibitem[{{Alloin} {et~al.}(1979){Alloin}, {Collin-Souffrin}, {Joly}, \&
  {Vigroux}}]{Alloin:1979}
{Alloin}, D., {Collin-Souffrin}, S., {Joly}, M., \& {Vigroux}, L. 1979, \aap,
  78, 200

\bibitem[{{Alonso-Herrero} {et~al.}(2018){Alonso-Herrero}, {Pereira-Santaella},
  {Garc{\'{\i}}a-Burillo}, {Davies}, {Combes}, {Asmus}, {Bunker},
  {D{\'{\i}}az-Santos}, {Gandhi}, {Gonz{\'a}lez-Mart{\'{\i}}n},
  {Hern{\'a}n-Caballero}, {Hicks}, {H{\"o}nig}, {Labiano}, {Levenson},
  {Packham}, {Ramos Almeida}, {Ricci}, {Rigopoulou}, {Rosario}, {Sani}, \&
  {Ward}}]{Alonso-Herrero:2018}
{Alonso-Herrero}, A., {Pereira-Santaella}, M., {Garc{\'{\i}}a-Burillo}, S.,
  {et~al.} 2018, \apj, 859, 144

\bibitem[{{Antonucci}(1993)}]{Antonucci:1993}
{Antonucci}, R. 1993, \araa, 31, 473

\bibitem[{{Arnaud}(1996)}]{Arnaud:1996}
{Arnaud}, K.~A. 1996, in Astronomical Data Analysis Software and Systems V,
  Vol. 101, 17

\bibitem[{{Bacon} {et~al.}(2010){Bacon}, {Accardo}, {Adjali}, {Anwand},
  {Bauer}, {Biswas}, {Blaizot}, {Boudon}, {Brau-Nogue}, {Brinchmann},
  {Caillier}, {Capoani}, {Carollo}, {Contini}, {Couderc}, {Daguis{\'e}},
  {Deiries}, {Delabre}, {Dreizler}, {Dubois}, {Dupieux}, {Dupuy}, {Emsellem},
  {Fechner}, {Fleischmann}, {Fran{\c c}ois}, {Gallou}, {Gharsa}, {Glindemann},
  {Gojak}, {Guiderdoni}, {Hansali}, {Hahn}, {Jarno}, {Kelz}, {Koehler},
  {Kosmalski}, {Laurent}, {Le Floch}, {Lilly}, {Lizon}, {Loupias}, {Manescau},
  {Monstein}, {Nicklas}, {Olaya}, {Pares}, {Pasquini}, {P{\'e}contal-Rousset},
  {Pell{\'o}}, {Petit}, {Popow}, {Reiss}, {Remillieux}, {Renault}, {Roth},
  {Rupprecht}, {Serre}, {Schaye}, {Soucail}, {Steinmetz}, {Streicher}, {Stuik},
  {Valentin}, {Vernet}, {Weilbacher}, {Wisotzki}, \& {Yerle}}]{Bacon:2010}
{Bacon}, R., {Accardo}, M., {Adjali}, L., {et~al.} 2010, SPIE Conf. Ser., 7735,
  8

\bibitem[{{Bacon} {et~al.}(2017){Bacon}, {Conseil}, {Mary}, {Brinchmann},
  {Shepherd}, {Akhlaghi}, {Weilbacher}, {Piqueras}, {Wisotzki}, {Lagattuta},
  {Epinat}, {Guerou}, {Inami}, {Cantalupo}, {Courbot}, {Contini}, {Richard},
  {Maseda}, {Bouwens}, {Bouch{\'e}}, {Kollatschny}, {Schaye}, {Marino},
  {Pello}, {Herenz}, {Guiderdoni}, \& {Carollo}}]{Bacon:2017}
{Bacon}, R., {Conseil}, S., {Mary}, D., {et~al.} 2017, \aap, 608, A1

\bibitem[{{Bacon} {et~al.}(2014){Bacon}, {Vernet}, {Borisova}, {Bouch{\'e}},
  {Brinchmann}, {Carollo}, {Carton}, {Caruana}, {Cerda}, {Contini}, {Franx},
  {Girard}, {Guerou}, {Haddad}, {Hau}, {Herenz}, {Herrera}, \&
  {Husemann}}]{Bacon:2014a}
{Bacon}, R., {Vernet}, J., {Borisova}, E., {et~al.} 2014, The Messenger, 157,
  13

\bibitem[{{Bae} \& {Woo}(2016)}]{Bae:2016}
{Bae}, H.-J. \& {Woo}, J.-H. 2016, \apj, 828, 97

\bibitem[{{Baldwin} {et~al.}(1981){Baldwin}, {Phillips}, \&
  {Terlevich}}]{Baldwin:1981}
{Baldwin}, J.~A., {Phillips}, M.~M., \& {Terlevich}, R. 1981, \pasp, 93, 5

\bibitem[{{Balmaverde} {et~al.}(2016){Balmaverde}, {Marconi}, {Brusa},
  {Carniani}, {Cresci}, {Lusso}, {Maiolino}, {Mannucci}, \&
  {Nagao}}]{Balmaverde:2016}
{Balmaverde}, B., {Marconi}, A., {Brusa}, M., {et~al.} 2016, \aap, 585, A148

\bibitem[{{Baron} \& {Netzer}(2019)}]{Baron:2019}
{Baron}, D. \& {Netzer}, H. 2019, \mnras, 482, 3915

\bibitem[{{Baumeister} {et~al.}(2008){Baumeister}, {Alter}, {C{\'a}rdenas
  V{\'a}zquez}, {Fernandez}, {Fried}, {Helmling}, {Huber}, {Ib{\'a}{\~n}ez
  Mengual}, {Rodr{\'{\i}}guez G{\'o}mez}, {Laun}, {Lenzen}, {Mall}, {Naranjo},
  {Ramos}, {Rohloff}, {Garc{\'{\i}}a Segura}, {Storz}, {Ubierna}, \&
  {Wagner}}]{Baumeister:2008}
{Baumeister}, H., {Alter}, M., {C{\'a}rdenas V{\'a}zquez}, M.~C., {et~al.}
  2008, in \procspie, Vol. 7014, Ground-based and Airborne Instrumentation for
  Astronomy II, 70142R

\bibitem[{{Bell} {et~al.}(2003){Bell}, {McIntosh}, {Katz}, \&
  {Weinberg}}]{Bell:2003}
{Bell}, E.~F., {McIntosh}, D.~H., {Katz}, N., \& {Weinberg}, M.~D. 2003, \apjs,
  149, 289

\bibitem[{{Bennert} {et~al.}(2002){Bennert}, {Falcke}, {Schulz}, {Wilson}, \&
  {Wills}}]{Bennert:2002}
{Bennert}, N., {Falcke}, H., {Schulz}, H., {Wilson}, A.~S., \& {Wills}, B.~J.
  2002, \apjl, 574, L105

\bibitem[{{Benson} {et~al.}(2003){Benson}, {Bower}, {Frenk}, {Lacey}, {Baugh},
  \& {Cole}}]{Benson:2003}
{Benson}, A.~J., {Bower}, R.~G., {Frenk}, C.~S., {et~al.} 2003, \apj, 599, 38

\bibitem[{{Bertram} {et~al.}(2007){Bertram}, {Eckart}, {Fischer}, {Zuther},
  {Straubmeier}, {Wisotzki}, \& {Krips}}]{Bertram:2007}
{Bertram}, T., {Eckart}, A., {Fischer}, S., {et~al.} 2007, \aap, 470, 571

\bibitem[{{Bianchi} {et~al.}(2014){Bianchi}, {Conti}, \&
  {Shiao}}]{Bianchi:2014}
{Bianchi}, L., {Conti}, A., \& {Shiao}, B. 2014, Advances in Space Research,
  53, 900

\bibitem[{{Bieri} {et~al.}(2017){Bieri}, {Dubois}, {Rosdahl}, {Wagner}, {Silk},
  \& {Mamon}}]{Bieri:2017}
{Bieri}, R., {Dubois}, Y., {Rosdahl}, J., {et~al.} 2017, \mnras, 464, 1854

\bibitem[{{Bigiel} {et~al.}(2011){Bigiel}, {Leroy}, {Walter}, {Brinks}, {de
  Blok}, {Kramer}, {Rix}, {Schruba}, {Schuster}, {Usero}, \&
  {Wiesemeyer}}]{Bigiel:2011}
{Bigiel}, F., {Leroy}, A.~K., {Walter}, F., {et~al.} 2011, \apjl, 730, L13

\bibitem[{{Bischetti} {et~al.}(2017){Bischetti}, {Piconcelli}, {Vietri},
  {Bongiorno}, {Fiore}, {Sani}, {Marconi}, {Duras}, {Zappacosta}, {Brusa},
  {Comastri}, {Cresci}, {Feruglio}, {Giallongo}, {La Franca}, {Mainieri},
  {Mannucci}, {Martocchia}, {Ricci}, {Schneider}, {Testa}, \&
  {Vignali}}]{Bischetti:2017}
{Bischetti}, M., {Piconcelli}, E., {Vietri}, G., {et~al.} 2017, \aap, 598, A122

\bibitem[{{Blundell} \& {Beasley}(1998)}]{Blundell:1998}
{Blundell}, K.~M. \& {Beasley}, A.~J. 1998, \mnras, 299, 165

\bibitem[{{Bower} {et~al.}(2006){Bower}, {Benson}, {Malbon}, {Helly}, {Frenk},
  {Baugh}, {Cole}, \& {Lacey}}]{Bower:2006}
{Bower}, R.~G., {Benson}, A.~J., {Malbon}, R., {et~al.} 2006, \mnras, 370, 645

\bibitem[{{Brinchmann} {et~al.}(2004){Brinchmann}, {Charlot}, {White},
  {Tremonti}, {Kauffmann}, {Heckman}, \& {Brinkmann}}]{Brinchmann:2004}
{Brinchmann}, J., {Charlot}, S., {White}, S.~D.~M., {et~al.} 2004, \mnras, 351,
  1151

\bibitem[{{Buta} \& {Combes}(1996)}]{Buta:1996}
{Buta}, R. \& {Combes}, F. 1996, \fcp, 17, 95

\bibitem[{{Calistro Rivera} {et~al.}(2016){Calistro Rivera}, {Lusso},
  {Hennawi}, \& {Hogg}}]{CalistroRivera:2016}
{Calistro Rivera}, G., {Lusso}, E., {Hennawi}, J.~F., \& {Hogg}, D.~W. 2016,
  \apj, 833, 98

\bibitem[{{Cappellari}(2002)}]{Cappellari:2002}
{Cappellari}, M. 2002, \mnras, 333, 400

\bibitem[{{Cappellari} \& {Copin}(2003)}]{Cappellari:2003}
{Cappellari}, M. \& {Copin}, Y. 2003, \mnras, 342, 345

\bibitem[{{Cardelli} {et~al.}(1989){Cardelli}, {Clayton}, \&
  {Mathis}}]{Cardelli:1989}
{Cardelli}, J.~A., {Clayton}, G.~C., \& {Mathis}, J.~S. 1989, \apj, 345, 245

\bibitem[{{Carniani} {et~al.}(2015){Carniani}, {Marconi}, {Maiolino},
  {Balmaverde}, {Brusa}, {Cano-D{\'{\i}}az}, {Cicone}, {Comastri}, {Cresci},
  {Fiore}, {Feruglio}, {La Franca}, {Mainieri}, {Mannucci}, {Nagao}, {Netzer},
  {Piconcelli}, {Risaliti}, {Schneider}, \& {Shemmer}}]{Carniani:2015}
{Carniani}, S., {Marconi}, A., {Maiolino}, R., {et~al.} 2015, \aap, 580, A102

\bibitem[{{Carniani} {et~al.}(2016){Carniani}, {Marconi}, {Maiolino},
  {Balmaverde}, {Brusa}, {Cano-D{\'{\i}}az}, {Cicone}, {Comastri}, {Cresci},
  {Fiore}, {Feruglio}, {La Franca}, {Mainieri}, {Mannucci}, {Nagao}, {Netzer},
  {Piconcelli}, {Risaliti}, {Schneider}, \& {Shemmer}}]{Carniani:2016}
{Carniani}, S., {Marconi}, A., {Maiolino}, R., {et~al.} 2016, \aap, 591, A28

\bibitem[{{Cattaneo} {et~al.}(2006){Cattaneo}, {Dekel}, {Devriendt},
  {Guiderdoni}, \& {Blaizot}}]{Cattaneo:2006}
{Cattaneo}, A., {Dekel}, A., {Devriendt}, J., {Guiderdoni}, B., \& {Blaizot},
  J. 2006, \mnras, 370, 1651

\bibitem[{{Cavagnolo} {et~al.}(2010){Cavagnolo}, {McNamara}, {Nulsen},
  {Carilli}, {Jones}, \& {B{\^i}rzan}}]{Cavagnolo:2010}
{Cavagnolo}, K.~W., {McNamara}, B.~R., {Nulsen}, P.~E.~J., {et~al.} 2010, \apj,
  720, 1066

\bibitem[{{Cecil} {et~al.}(2001){Cecil}, {Bland-Hawthorn}, {Veilleux}, \&
  {Filippenko}}]{Cecil:2001}
{Cecil}, G., {Bland-Hawthorn}, J., {Veilleux}, S., \& {Filippenko}, A.~V. 2001,
  \apj, 555, 338

\bibitem[{{Chambers} {et~al.}(2016){Chambers}, {Magnier}, {Metcalfe},
  {Flewelling}, {Huber}, {Waters}, {Denneau}, {Draper}, {Farrow}, {Finkbeiner},
  {Holmberg}, {Koppenhoefer}, {Price}, {Saglia}, {Schlafly}, {Smartt},
  {Sweeney}, {Wainscoat}, {Burgett}, {Grav}, {Heasley}, {Hodapp}, {Jedicke},
  {Kaiser}, {Kudritzki}, {Luppino}, {Lupton}, {Monet}, {Morgan}, {Onaka},
  {Stubbs}, {Tonry}, {Banados}, {Bell}, {Bender}, {Bernard}, {Botticella},
  {Casertano}, {Chastel}, {Chen}, {Chen}, {Cole}, {Deacon}, {Frenk},
  {Fitzsimmons}, {Gezari}, {Goessl}, {Goggia}, {Goldman}, {Grebel}, {Hambly},
  {Hasinger}, {Heavens}, {Heckman}, {Henderson}, {Henning}, {Holman}, {Hopp},
  {Ip}, {Isani}, {Keyes}, {Koekemoer}, {Kotak}, {Long}, {Lucey}, {Liu},
  {Martin}, {McLean}, {Morganson}, {Murphy}, {Nieto-Santisteban}, {Norberg},
  {Peacock}, {Pier}, {Postman}, {Primak}, {Rae}, {Rest}, {Riess}, {Riffeser},
  {Rix}, {Roser}, {Schilbach}, {Schultz}, {Scolnic}, {Szalay}, {Seitz},
  {Shiao}, {Small}, {Smith}, {Soderblom}, {Taylor}, {Thakar}, {Thiel},
  {Thilker}, {Urata}, {Valenti}, {Walter}, {Watters}, {Werner}, {White},
  {Wood-Vasey}, \& {Wyse}}]{Chambers:2016}
{Chambers}, K.~C., {Magnier}, E.~A., {Metcalfe}, N., {et~al.} 2016, ArXiv
  e-prints [\eprint[arXiv]{1612.05560}]

\bibitem[{{Cicone} {et~al.}(2018{\natexlab{a}}){Cicone}, {Brusa}, {Ramos
  Almeida}, {Cresci}, {Husemann}, \& {Mainieri}}]{Cicone:2018}
{Cicone}, C., {Brusa}, M., {Ramos Almeida}, C., {et~al.} 2018{\natexlab{a}},
  Nature Astronomy, 2, 176

\bibitem[{{Cicone} {et~al.}(2014){Cicone}, {Maiolino}, {Sturm},
  {Graci{\'a}-Carpio}, {Feruglio}, {Neri}, {Aalto}, {Davies}, {Fiore},
  {Fischer}, {Garc{\'{\i}}a-Burillo}, {Gonz{\'a}lez-Alfonso},
  {Hailey-Dunsheath}, {Piconcelli}, \& {Veilleux}}]{Cicone:2014}
{Cicone}, C., {Maiolino}, R., {Sturm}, E., {et~al.} 2014, \aap, 562, A21

\bibitem[{{Cicone} {et~al.}(2018{\natexlab{b}}){Cicone}, {Severgnini},
  {Papadopoulos}, {Maiolino}, {Feruglio}, {Treister}, {Privon}, {Zhang}, {Della
  Ceca}, {Fiore}, {Schawinski}, \& {Wagg}}]{Cicone:2018b}
{Cicone}, C., {Severgnini}, P., {Papadopoulos}, P.~P., {et~al.}
  2018{\natexlab{b}}, \apj, 863, 143

\bibitem[{{Cielo} {et~al.}(2018){Cielo}, {Bieri}, {Volonteri}, {Wagner}, \&
  {Dubois}}]{Cielo:2018}
{Cielo}, S., {Bieri}, R., {Volonteri}, M., {Wagner}, A.~Y., \& {Dubois}, Y.
  2018, \mnras, 477, 1336

\bibitem[{{Cohen} {et~al.}(2007){Cohen}, {Lister}, {Homan}, {Kadler},
  {Kellermann}, {Kovalev}, \& {Vermeulen}}]{Cohen:2007}
{Cohen}, M.~H., {Lister}, M.~L., {Homan}, D.~C., {et~al.} 2007, \apj, 658, 232

\bibitem[{{Combes} {et~al.}(2013){Combes}, {Garc{\'{\i}}a-Burillo}, {Casasola},
  {Hunt}, {Krips}, {Baker}, {Boone}, {Eckart}, {Marquez}, {Neri}, {Schinnerer},
  \& {Tacconi}}]{Combes:2013}
{Combes}, F., {Garc{\'{\i}}a-Burillo}, S., {Casasola}, V., {et~al.} 2013, \aap,
  558, A124

\bibitem[{{Costa} {et~al.}(2014){Costa}, {Sijacki}, \& {Haehnelt}}]{Costa:2014}
{Costa}, T., {Sijacki}, D., \& {Haehnelt}, M.~G. 2014, \mnras, 444, 2355

\bibitem[{{Crain} {et~al.}(2015){Crain}, {Schaye}, {Bower}, {Furlong},
  {Schaller}, {Theuns}, {Dalla Vecchia}, {Frenk}, {McCarthy}, {Helly},
  {Jenkins}, {Rosas-Guevara}, {White}, \& {Trayford}}]{Crain:2015}
{Crain}, R.~A., {Schaye}, J., {Bower}, R.~G., {et~al.} 2015, \mnras, 450, 1937

\bibitem[{{Crenshaw} \& {Kraemer}(2000)}]{Crenshaw:2000}
{Crenshaw}, D.~M. \& {Kraemer}, S.~B. 2000, \apjl, 532, L101

\bibitem[{{Cresci} {et~al.}(2015{\natexlab{a}}){Cresci}, {Mainieri}, {Brusa},
  {Marconi}, {Perna}, {Mannucci}, {Piconcelli}, {Maiolino}, {Feruglio},
  {Fiore}, {Bongiorno}, {Lanzuisi}, {Merloni}, {Schramm}, {Silverman}, \&
  {Civano}}]{Cresci:2015}
{Cresci}, G., {Mainieri}, V., {Brusa}, M., {et~al.} 2015{\natexlab{a}}, \apj,
  799, 82

\bibitem[{{Cresci} {et~al.}(2015{\natexlab{b}}){Cresci}, {Marconi}, {Zibetti},
  {Risaliti}, {Carniani}, {Mannucci}, {Gallazzi}, {Maiolino}, {Balmaverde},
  {Brusa}, {Capetti}, {Cicone}, {Feruglio}, {Bland-Hawthorn}, {Nagao}, {Oliva},
  {Salvato}, {Sani}, {Tozzi}, {Urrutia}, \& {Venturi}}]{Cresci:2015b}
{Cresci}, G., {Marconi}, A., {Zibetti}, S., {et~al.} 2015{\natexlab{b}}, \aap,
  582, A63

\bibitem[{{Croton} {et~al.}(2006){Croton}, {Springel}, {White}, {De Lucia},
  {Frenk}, {Gao}, {Jenkins}, {Kauffmann}, {Navarro}, \&
  {Yoshida}}]{Croton:2006}
{Croton}, D.~J., {Springel}, V., {White}, S.~D.~M., {et~al.} 2006, \mnras, 365,
  11

\bibitem[{{da Silva} {et~al.}(2011){da Silva}, {Prochaska}, {Rosario},
  {Tumlinson}, \& {Tripp}}]{daSilva:2011}
{da Silva}, R.~L., {Prochaska}, J.~X., {Rosario}, D., {Tumlinson}, J., \&
  {Tripp}, T.~M. 2011, \apj, 735, 54

\bibitem[{{Dame}(2011)}]{Dame:2011}
{Dame}, T.~M. 2011, ArXiv e-prints [\eprint[arXiv]{1101.1499}]

\bibitem[{{Das} {et~al.}(2006){Das}, {Crenshaw}, {Kraemer}, \&
  {Deo}}]{Das:2006}
{Das}, V., {Crenshaw}, D.~M., {Kraemer}, S.~B., \& {Deo}, R.~P. 2006, \aj, 132,
  620

\bibitem[{{Dasyra} {et~al.}(2015){Dasyra}, {Bostrom}, {Combes}, \&
  {Vlahakis}}]{Dasyra:2015}
{Dasyra}, K.~M., {Bostrom}, A.~C., {Combes}, F., \& {Vlahakis}, N. 2015, \apj,
  815, 34

\bibitem[{{Dasyra} {et~al.}(2016){Dasyra}, {Combes}, {Oosterloo}, {Oonk},
  {Morganti}, {Salom{\'e}}, \& {Vlahakis}}]{Dasyra:2016}
{Dasyra}, K.~M., {Combes}, F., {Oosterloo}, T., {et~al.} 2016, \aap, 595, L7

\bibitem[{{Dav{\'e}} {et~al.}(2011){Dav{\'e}}, {Oppenheimer}, \&
  {Finlator}}]{Dave:2011}
{Dav{\'e}}, R., {Oppenheimer}, B.~D., \& {Finlator}, K. 2011, \mnras, 415, 11

\bibitem[{{Davies} {et~al.}(2015){Davies}, {Robotham}, {Driver}, {Alpaslan},
  {Baldry}, {Bland-Hawthorn}, {Brough}, {Brown}, {Cluver}, {Drinkwater},
  {Foster}, {Grootes}, {Konstantopoulos}, {Lara-L{\'o}pez},
  {L{\'o}pez-S{\'a}nchez}, {Loveday}, {Meyer}, {Moffett}, {Norberg}, {Owers},
  {Popescu}, {De Propris}, {Sharp}, {Tuffs}, {Wang}, {Wilkins}, {Dunne},
  {Bourne}, \& {Smith}}]{Davies:2015}
{Davies}, L.~J.~M., {Robotham}, A.~S.~G., {Driver}, S.~P., {et~al.} 2015,
  \mnras, 452, 616

\bibitem[{{Davies} {et~al.}(2016){Davies}, {Groves}, {Kewley}, {Dopita},
  {Hampton}, {Shastri}, {Scharw{\"a}chter}, {Sutherland}, {Kharb}, {Bhatt},
  {Jin}, {Banfield}, {Zaw}, {James}, {Juneau}, \& {Srivastava}}]{Davies:2016}
{Davies}, R.~L., {Groves}, B., {Kewley}, L.~J., {et~al.} 2016, \mnras, 462,
  1616

\bibitem[{{Davis} {et~al.}(2013{\natexlab{a}}){Davis}, {Alatalo}, {Bureau},
  {Cappellari}, {Scott}, {Young}, {Blitz}, {Crocker}, {Bayet}, {Bois},
  {Bournaud}, {Davies}, {de Zeeuw}, {Duc}, {Emsellem}, {Khochfar},
  {Krajnovi{\'c}}, {Kuntschner}, {Lablanche}, {McDermid}, {Morganti}, {Naab},
  {Oosterloo}, {Sarzi}, {Serra}, \& {Weijmans}}]{Davis:2013}
{Davis}, T.~A., {Alatalo}, K., {Bureau}, M., {et~al.} 2013{\natexlab{a}},
  \mnras, 429, 534

\bibitem[{{Davis} {et~al.}(2013{\natexlab{b}}){Davis}, {Bureau}, {Cappellari},
  {Sarzi}, \& {Blitz}}]{Davis:2013b}
{Davis}, T.~A., {Bureau}, M., {Cappellari}, M., {Sarzi}, M., \& {Blitz}, L.
  2013{\natexlab{b}}, \nat, 494, 328

\bibitem[{{Davis} {et~al.}(2015){Davis}, {Rowlands}, {Allison}, {Shabala},
  {Ting}, {Lagos}, {Kaviraj}, {Bourne}, {Dunne}, {Eales}, {Ivison}, {Maddox},
  {Smith}, {Smith}, \& {Temi}}]{Davis:2015}
{Davis}, T.~A., {Rowlands}, K., {Allison}, J.~R., {et~al.} 2015, \mnras, 449,
  3503

\bibitem[{{Denney} {et~al.}(2009){Denney}, {Peterson}, {Dietrich},
  {Vestergaard}, \& {Bentz}}]{Denney:2009}
{Denney}, K.~D., {Peterson}, B.~M., {Dietrich}, M., {Vestergaard}, M., \&
  {Bentz}, M.~C. 2009, \apj, 692, 246

\bibitem[{{DiPompeo} {et~al.}(2018){DiPompeo}, {Hickox}, {Carroll}, {Runnoe},
  {Mullaney}, \& {Fischer}}]{DiPompeo:2018}
{DiPompeo}, M.~A., {Hickox}, R.~C., {Carroll}, C.~M., {et~al.} 2018, \apj, 856,
  76

\bibitem[{{Doi} {et~al.}(2013){Doi}, {Asada}, {Fujisawa}, {Nagai}, {Hagiwara},
  {Wajima}, \& {Inoue}}]{Doi:2013}
{Doi}, A., {Asada}, K., {Fujisawa}, K., {et~al.} 2013, \apj, 765, 69

\bibitem[{{Duric} \& {Seaquist}(1988)}]{Duric:1988}
{Duric}, N. \& {Seaquist}, E.~R. 1988, \apj, 326, 574

\bibitem[{{Emonts} {et~al.}(2005){Emonts}, {Morganti}, {Tadhunter},
  {Oosterloo}, {Holt}, \& {van der Hulst}}]{Emonts:2005}
{Emonts}, B.~H.~C., {Morganti}, R., {Tadhunter}, C.~N., {et~al.} 2005, \mnras,
  362, 931

\bibitem[{{Emsellem} {et~al.}(1994){Emsellem}, {Monnet}, \&
  {Bacon}}]{Emsellem:1994}
{Emsellem}, E., {Monnet}, G., \& {Bacon}, R. 1994, \aap, 285, 723

\bibitem[{{Fabian}(2012)}]{Fabian:2012}
{Fabian}, A.~C. 2012, \araa, 50, 455

\bibitem[{{Farrah} {et~al.}(2012){Farrah}, {Urrutia}, {Lacy}, {Efstathiou},
  {Afonso}, {Coppin}, {Hall}, {Lonsdale}, {Jarrett}, {Bridge}, {Borys}, \&
  {Petty}}]{Farrah:2012}
{Farrah}, D., {Urrutia}, T., {Lacy}, M., {et~al.} 2012, \apj, 745, 178

\bibitem[{{Faucher-Gigu{\`e}re} \& {Quataert}(2012)}]{Faucher-Giguere:2012}
{Faucher-Gigu{\`e}re}, C.-A. \& {Quataert}, E. 2012, \mnras, 425, 605

\bibitem[{{Ferruit} {et~al.}(1999){Ferruit}, {Wilson}, {Falcke}, {Simpson},
  {P{\'e}contal}, \& {Durret}}]{Ferruit:1999}
{Ferruit}, P., {Wilson}, A.~S., {Falcke}, H., {et~al.} 1999, \mnras, 309, 1

\bibitem[{{Feruglio} {et~al.}(2015){Feruglio}, {Fiore}, {Carniani},
  {Piconcelli}, {Zappacosta}, {Bongiorno}, {Cicone}, {Maiolino}, {Marconi},
  {Menci}, {Puccetti}, \& {Veilleux}}]{Feruglio:2015}
{Feruglio}, C., {Fiore}, F., {Carniani}, S., {et~al.} 2015, \aap, 583, A99

\bibitem[{{Feruglio} {et~al.}(2010){Feruglio}, {Maiolino}, {Piconcelli},
  {Menci}, {Aussel}, {Lamastra}, \& {Fiore}}]{Feruglio:2010}
{Feruglio}, C., {Maiolino}, R., {Piconcelli}, E., {et~al.} 2010, \aap, 518,
  L155

\bibitem[{{Fiore} {et~al.}(2017){Fiore}, {Feruglio}, {Shankar}, {Bischetti},
  {Bongiorno}, {Brusa}, {Carniani}, {Cicone}, {Duras}, {Lamastra}, {Mainieri},
  {Marconi}, {Menci}, {Maiolino}, {Piconcelli}, {Vietri}, \&
  {Zappacosta}}]{Fiore:2017}
{Fiore}, F., {Feruglio}, C., {Shankar}, F., {et~al.} 2017, \aap, 601, A143

\bibitem[{{Fischer} {et~al.}(2013){Fischer}, {Crenshaw}, {Kraemer}, \&
  {Schmitt}}]{Fischer:2013}
{Fischer}, T.~C., {Crenshaw}, D.~M., {Kraemer}, S.~B., \& {Schmitt}, H.~R.
  2013, \apjs, 209, 1

\bibitem[{{Fluetsch} {et~al.}(2019){Fluetsch}, {Maiolino}, {Carniani},
  {Marconi}, {Cicone}, {Bourne}, {Costa}, {Fabian}, {Ishibashi}, \&
  {Venturi}}]{Fluetsch:2019}
{Fluetsch}, A., {Maiolino}, R., {Carniani}, S., {et~al.} 2019, \mnras, 483,
  4586

\bibitem[{{Fu} \& {Stockton}(2009)}]{Fu:2009}
{Fu}, H. \& {Stockton}, A. 2009, \apj, 690, 953

\bibitem[{{Gabor} \& {Bournaud}(2014)}]{Gabor:2014}
{Gabor}, J.~M. \& {Bournaud}, F. 2014, \mnras, 441, 1615

\bibitem[{{Gallimore} {et~al.}(2016){Gallimore}, {Elitzur}, {Maiolino},
  {Marconi}, {O'Dea}, {Lutz}, {Baum}, {Nikutta}, {Impellizzeri}, {Davies},
  {Kimball}, \& {Sani}}]{Gallimore:2016}
{Gallimore}, J.~F., {Elitzur}, M., {Maiolino}, R., {et~al.} 2016, \apjl, 829,
  L7

\bibitem[{{Garc{\'i}a-Burillo} {et~al.}(2014){Garc{\'i}a-Burillo}, {Combes},
  {Usero}, {Aalto}, {Krips}, {Viti}, {Alonso-Herrero}, {Hunt}, {Schinnerer},
  {Baker}, {Boone}, {Casasola}, {Colina}, {Costagliola}, {Eckart}, {Fuente},
  {Henkel}, {Labiano}, {Mart{\'{\i}}n}, {M{\'a}rquez}, {Muller}, {Planesas},
  {Ramos Almeida}, {Spaans}, {Tacconi}, \& {van der
  Werf}}]{Garcia-Burillo:2014}
{Garc{\'i}a-Burillo}, S., {Combes}, F., {Usero}, A., {et~al.} 2014, \aap, 567,
  A125

\bibitem[{{Gaspari} {et~al.}(2013){Gaspari}, {Brighenti}, \&
  {Ruszkowski}}]{Gaspari:2013}
{Gaspari}, M., {Brighenti}, F., \& {Ruszkowski}, M. 2013, Astronomische
  Nachrichten, 334, 394

\bibitem[{{Gaspari} {et~al.}(2018){Gaspari}, {McDonald}, {Hamer}, {Brighenti},
  {Temi}, {Gendron-Marsolais}, {Hlavacek-Larrondo}, {Edge}, {Werner}, {Tozzi},
  {Sun}, {Stone}, {Tremblay}, {Hogan}, {Eckert}, {Ettori}, {Yu}, {Biffi}, \&
  {Planelles}}]{Gaspari:2018}
{Gaspari}, M., {McDonald}, M., {Hamer}, S.~L., {et~al.} 2018, \apj, 854, 167

\bibitem[{{Gaspari} \& {S{\k{a}}dowski}(2017)}]{Gaspari:2017b}
{Gaspari}, M. \& {S{\k{a}}dowski}, A. 2017, \apj, 837, 149

\bibitem[{{Gaspari} {et~al.}(2017){Gaspari}, {Temi}, \&
  {Brighenti}}]{Gaspari:2017}
{Gaspari}, M., {Temi}, P., \& {Brighenti}, F. 2017, \mnras, 466, 677

\bibitem[{{Genel} {et~al.}(2014){Genel}, {Vogelsberger}, {Springel}, {Sijacki},
  {Nelson}, {Snyder}, {Rodriguez-Gomez}, {Torrey}, \& {Hernquist}}]{Genel:2014}
{Genel}, S., {Vogelsberger}, M., {Springel}, V., {et~al.} 2014, \mnras, 445,
  175

\bibitem[{{Gibson} {et~al.}(2009){Gibson}, {Jiang}, {Brandt}, {Hall}, {Shen},
  {Wu}, {Anderson}, {Schneider}, {Vanden Berk}, {Gallagher}, {Fan}, \&
  {York}}]{Gibson:2009}
{Gibson}, R.~R., {Jiang}, L., {Brandt}, W.~N., {et~al.} 2009, \apj, 692, 758

\bibitem[{{Greene} \& {Ho}(2005)}]{Greene:2005}
{Greene}, J.~E. \& {Ho}, L.~C. 2005, \apj, 630, 122

\bibitem[{{Greene} {et~al.}(2011){Greene}, {Zakamska}, {Ho}, \&
  {Barth}}]{Greene:2011}
{Greene}, J.~E., {Zakamska}, N.~L., {Ho}, L.~C., \& {Barth}, A.~J. 2011, \apj,
  732, 9

\bibitem[{{Greene} {et~al.}(2012){Greene}, {Zakamska}, \&
  {Smith}}]{Greene:2012}
{Greene}, J.~E., {Zakamska}, N.~L., \& {Smith}, P.~S. 2012, \apj, 746, 86

\bibitem[{{Griffin} {et~al.}(2010){Griffin}, {Abergel}, {Abreu}, {Ade},
  {Andr{\'e}}, {Augueres}, {Babbedge}, {Bae}, {Baillie}, {Baluteau}, {Barlow},
  {Bendo}, {Benielli}, {Bock}, {Bonhomme}, {Brisbin}, {Brockley-Blatt},
  {Caldwell}, {Cara}, {Castro-Rodriguez}, {Cerulli}, {Chanial}, {Chen},
  {Clark}, {Clements}, {Clerc}, {Coker}, {Communal}, {Conversi}, {Cox},
  {Crumb}, {Cunningham}, {Daly}, {Davis}, {de Antoni}, {Delderfield}, {Devin},
  {di Giorgio}, {Didschuns}, {Dohlen}, {Donati}, {Dowell}, {Dowell}, {Duband},
  {Dumaye}, {Emery}, {Ferlet}, {Ferrand}, {Fontignie}, {Fox}, {Franceschini},
  {Frerking}, {Fulton}, {Garcia}, {Gastaud}, {Gear}, {Glenn}, {Goizel},
  {Griffin}, {Grundy}, {Guest}, {Guillemet}, {Hargrave}, {Harwit}, {Hastings},
  {Hatziminaoglou}, {Herman}, {Hinde}, {Hristov}, {Huang}, {Imhof}, {Isaak},
  {Israelsson}, {Ivison}, {Jennings}, {Kiernan}, {King}, {Lange}, {Latter},
  {Laurent}, {Laurent}, {Leeks}, {Lellouch}, {Levenson}, {Li}, {Li},
  {Lilienthal}, {Lim}, {Liu}, {Lu}, {Madden}, {Mainetti}, {Marliani}, {McKay},
  {Mercier}, {Molinari}, {Morris}, {Moseley}, {Mulder}, {Mur}, {Naylor},
  {Nguyen}, {O'Halloran}, {Oliver}, {Olofsson}, {Olofsson}, {Orfei}, {Page},
  {Pain}, {Panuzzo}, {Papageorgiou}, {Parks}, {Parr-Burman}, {Pearce},
  {Pearson}, {P{\'e}rez-Fournon}, {Pinsard}, {Pisano}, {Podosek}, {Pohlen},
  {Polehampton}, {Pouliquen}, {Rigopoulou}, {Rizzo}, {Roseboom}, {Roussel},
  {Rowan-Robinson}, {Rownd}, {Saraceno}, {Sauvage}, {Savage}, {Savini},
  {Sawyer}, {Scharmberg}, {Schmitt}, {Schneider}, {Schulz}, {Schwartz},
  {Shafer}, {Shupe}, {Sibthorpe}, {Sidher}, {Smith}, {Smith}, {Smith},
  {Spencer}, {Stobie}, {Sudiwala}, {Sukhatme}, {Surace}, {Stevens}, {Swinyard},
  {Trichas}, {Tourette}, {Triou}, {Tseng}, {Tucker}, {Turner}, {Vaccari},
  {Valtchanov}, {Vigroux}, {Virique}, {Voellmer}, {Walker}, {Ward}, {Waskett},
  {Weilert}, {Wesson}, {White}, {Whitehouse}, {Wilson}, {Winter}, {Woodcraft},
  {Wright}, {Xu}, {Zavagno}, {Zemcov}, {Zhang}, \& {Zonca}}]{Griffin:2010}
{Griffin}, M.~J., {Abergel}, A., {Abreu}, A., {et~al.} 2010, \aap, 518, L3

\bibitem[{{Groves} {et~al.}(2006){Groves}, {Heckman}, \&
  {Kauffmann}}]{Groves:2006}
{Groves}, B.~A., {Heckman}, T.~M., \& {Kauffmann}, G. 2006, \mnras, 371, 1559

\bibitem[{{Guillard} {et~al.}(2012){Guillard}, {Ogle}, {Emonts}, {Appleton},
  {Morganti}, {Tadhunter}, {Oosterloo}, {Evans}, \& {Evans}}]{Guillard:2012}
{Guillard}, P., {Ogle}, P.~M., {Emonts}, B.~H.~C., {et~al.} 2012, \apj, 747, 95

\bibitem[{{Hainline} {et~al.}(2013){Hainline}, {Hickox}, {Greene}, {Myers}, \&
  {Zakamska}}]{Hainline:2013}
{Hainline}, K.~N., {Hickox}, R., {Greene}, J.~E., {Myers}, A.~D., \&
  {Zakamska}, N.~L. 2013, \apj, 774, 145

\bibitem[{{Harrison} {et~al.}(2012){Harrison}, {Alexander}, {Mullaney},
  {Altieri}, {Coia}, {Charmandaris}, {Daddi}, {Dannerbauer}, {Dasyra}, {Del
  Moro}, {Dickinson}, {Hickox}, {Ivison}, {Kartaltepe}, {Le Floc'h}, {Leiton},
  {Magnelli}, {Popesso}, {Rovilos}, {Rosario}, \& {Swinbank}}]{Harrison:2012b}
{Harrison}, C.~M., {Alexander}, D.~M., {Mullaney}, J.~R., {et~al.} 2012, \apjl,
  760, L15

\bibitem[{{Harrison} {et~al.}(2016){Harrison}, {Alexander}, {Mullaney},
  {Stott}, {Swinbank}, {Arumugam}, {Bauer}, {Bower}, {Bunker}, \&
  {Sharples}}]{Harrison:2016b}
{Harrison}, C.~M., {Alexander}, D.~M., {Mullaney}, J.~R., {et~al.} 2016,
  \mnras, 456, 1195

\bibitem[{{Harrison} {et~al.}(2014){Harrison}, {Alexander}, {Mullaney}, \&
  {Swinbank}}]{Harrison:2014}
{Harrison}, C.~M., {Alexander}, D.~M., {Mullaney}, J.~R., \& {Swinbank}, A.~M.
  2014, \mnras, 441, 3306

\bibitem[{{Harrison} {et~al.}(2018){Harrison}, {Costa}, {Tadhunter},
  {Fl{\"u}tsch}, {Kakkad}, {Perna}, \& {Vietri}}]{Harrison:2018}
{Harrison}, C.~M., {Costa}, T., {Tadhunter}, C.~N., {et~al.} 2018, Nature
  Astronomy, 2, 198

\bibitem[{{Harrison} {et~al.}(2015){Harrison}, {Thomson}, {Alexander}, {Bauer},
  {Edge}, {Hogan}, {Mullaney}, \& {Swinbank}}]{Harrison:2015}
{Harrison}, C.~M., {Thomson}, A.~P., {Alexander}, D.~M., {et~al.} 2015, \apj,
  800, 45

\bibitem[{{Hayward} {et~al.}(2014){Hayward}, {Lanz}, {Ashby}, {Fazio},
  {Hernquist}, {Mart{\'{\i}}nez-Galarza}, {Noeske}, {Smith}, {Wuyts}, \&
  {Zezas}}]{Hayward:2014}
{Hayward}, C.~C., {Lanz}, L., {Ashby}, M.~L.~N., {et~al.} 2014, \mnras, 445,
  1598

\bibitem[{{Heckman} {et~al.}(1981){Heckman}, {Miley}, {van Breugel}, \&
  {Butcher}}]{Heckman:1981}
{Heckman}, T.~M., {Miley}, G.~K., {van Breugel}, W.~J.~M., \& {Butcher}, H.~R.
  1981, \apj, 247, 403

\bibitem[{{Ho} {et~al.}(2014){Ho}, {Kewley}, {Dopita}, {Medling}, {Allen},
  {Bland-Hawthorn}, {Bloom}, {Bryant}, {Croom}, {Fogarty}, {Goodwin}, {Green},
  {Konstantopoulos}, {Lawrence}, {L{\'o}pez-S{\'a}nchez}, {Owers}, {Richards},
  \& {Sharp}}]{Ho:2014}
{Ho}, I.-T., {Kewley}, L.~J., {Dopita}, M.~A., {et~al.} 2014, \mnras, 444, 3894

\bibitem[{{Ho} {et~al.}(2016){Ho}, {Medling}, {Bland-Hawthorn}, {Groves},
  {Kewley}, {Kobayashi}, {Dopita}, {Leslie}, {Sharp}, {Allen}, {Bourne},
  {Bryant}, {Cortese}, {Croom}, {Dunne}, {Fogarty}, {Goodwin}, {Green},
  {Konstantopoulos}, {Lawrence}, {Lorente}, {Owers}, {Richards}, {Sweet},
  {Tescari}, \& {Valiante}}]{Ho:2016}
{Ho}, I.-T., {Medling}, A.~M., {Bland-Hawthorn}, J., {et~al.} 2016, \mnras,
  457, 1257

\bibitem[{{Holt} {et~al.}(2008){Holt}, {Tadhunter}, \& {Morganti}}]{Holt:2008}
{Holt}, J., {Tadhunter}, C.~N., \& {Morganti}, R. 2008, \mnras, 387, 639

\bibitem[{{Hopkins} \& {Elvis}(2010)}]{Hopkins:2010b}
{Hopkins}, P.~F. \& {Elvis}, M. 2010, \mnras, 401, 7

\bibitem[{{Humphrey} {et~al.}(2010){Humphrey}, {Villar-Mart{\'{\i}}n},
  {S{\'a}nchez}, {Mart{\'{\i}}nez-Sansigre}, {Delgado}, {P{\'e}rez},
  {Tadhunter}, \& {P{\'e}rez-Torres}}]{Humphrey:2010}
{Humphrey}, A., {Villar-Mart{\'{\i}}n}, M., {S{\'a}nchez}, S.~F., {et~al.}
  2010, \mnras, 408, L1

\bibitem[{{Husemann} {et~al.}(2016{\natexlab{a}}){Husemann}, {Bennert},
  {Scharw{\"a}chter}, {Woo}, \& {Choudhury}}]{Husemann:2016a}
{Husemann}, B., {Bennert}, V.~N., {Scharw{\"a}chter}, J., {Woo}, J.-H., \&
  {Choudhury}, O.~S. 2016{\natexlab{a}}, \mnras, 455, 1905

\bibitem[{{Husemann} {et~al.}(2014){Husemann}, {Jahnke}, {S{\'a}nchez},
  {Wisotzki}, {Nugroho}, {Kupko}, \& {Schramm}}]{Husemann:2014}
{Husemann}, B., {Jahnke}, K., {S{\'a}nchez}, S.~F., {et~al.} 2014, \mnras, 443,
  755

\bibitem[{{Husemann} {et~al.}(2016{\natexlab{b}}){Husemann},
  {Scharw{\"a}chter}, {Bennert}, {Mainieri}, {Woo}, \&
  {Kakkad}}]{Husemann:2016c}
{Husemann}, B., {Scharw{\"a}chter}, J., {Bennert}, V.~N., {et~al.}
  2016{\natexlab{b}}, \aap, 594, A44

\bibitem[{{Husemann} {et~al.}(2017){Husemann}, {Tremblay}, {Davis}, {Busch},
  {McElroy}, {Neumann}, {Urrutia}, {Krumpe}, {Scharw{\"a}chter}, {Powell},
  {Perez-Torres}, \& {The CARS Team}}]{Husemann:2017b}
{Husemann}, B., {Tremblay}, G., {Davis}, T., {et~al.} 2017, The Messenger, 169,
  42

\bibitem[{{Husemann} {et~al.}(2011){Husemann}, {Wisotzki}, {Jahnke}, \&
  {S{\'a}nchez}}]{Husemann:2011}
{Husemann}, B., {Wisotzki}, L., {Jahnke}, K., \& {S{\'a}nchez}, S.~F. 2011,
  \aap, 535, A72

\bibitem[{{Husemann} {et~al.}(2013){Husemann}, {Wisotzki}, {S{\'a}nchez}, \&
  {Jahnke}}]{Husemann:2013a}
{Husemann}, B., {Wisotzki}, L., {S{\'a}nchez}, S.~F., \& {Jahnke}, K. 2013,
  \aap, 549, A43

\bibitem[{{Hwang} {et~al.}(2018){Hwang}, {Zakamska}, {Alexandroff}, {Hamann},
  {Greene}, {Perrotta}, \& {Richards}}]{Hwang:2018}
{Hwang}, H.-C., {Zakamska}, N.~L., {Alexandroff}, R.~M., {et~al.} 2018, \mnras,
  477, 830

\bibitem[{{Ib{\'a}{\~n}ez} {et~al.}(2012){Ib{\'a}{\~n}ez}, {Garc{\'{\i}}a
  Segura}, {Storz}, {Fried}, {Fern{\'a}ndez}, {Rodr{\'{\i}}guez G{\'o}mez},
  {Terr{\'o}n}, \& {C{\'a}rdenas}}]{Ibanez:2012}
{Ib{\'a}{\~n}ez}, J.-M., {Garc{\'{\i}}a Segura}, A.~J., {Storz}, C., {et~al.}
  2012, in \procspie, Vol. 8451, Software and Cyberinfrastructure for Astronomy
  II, 84511E

\bibitem[{{Irwin} \& {Sofue}(1992)}]{Irwin:1992}
{Irwin}, J.~A. \& {Sofue}, Y. 1992, \apjl, 396, L75

\bibitem[{{Jahnke} {et~al.}(2004){Jahnke}, {Kuhlbrodt}, \&
  {Wisotzki}}]{Jahnke:2004b}
{Jahnke}, K., {Kuhlbrodt}, B., \& {Wisotzki}, L. 2004, \mnras, 352, 399

\bibitem[{{Johnson} {et~al.}(2018){Johnson}, {Chen}, {Straka}, {Schaye},
  {Cantalupo}, {Wendt}, {Muzahid}, {Bouch{\'e}}, {Herenz}, {Kollatschny},
  {Mulchaey}, {Marino}, {Maseda}, \& {Wisotzki}}]{Johnson:2018}
{Johnson}, S.~D., {Chen}, H.-W., {Straka}, L.~A., {et~al.} 2018, \apj, 869, L1

\bibitem[{{Kakkad} {et~al.}(2018){Kakkad}, {Groves}, {Dopita}, {Thomas},
  {Davies}, {Mainieri}, {Kharb}, {Scharw{\"a}chter}, {Hampton}, \&
  {Ho}}]{Kakkad:2018}
{Kakkad}, D., {Groves}, B., {Dopita}, M., {et~al.} 2018, \aap, 618, A6

\bibitem[{{Kakkad} {et~al.}(2016){Kakkad}, {Mainieri}, {Padovani}, {Cresci},
  {Husemann}, {Carniani}, {Brusa}, {Lamastra}, {Lanzuisi}, {Piconcelli}, \&
  {Schramm}}]{Kakkad:2016}
{Kakkad}, D., {Mainieri}, V., {Padovani}, P., {et~al.} 2016, \aap, 592, A148

\bibitem[{{Kang} \& {Woo}(2018)}]{Kang:2018}
{Kang}, D. \& {Woo}, J.-H. 2018, \apj, 864, 124

\bibitem[{{Karouzos} {et~al.}(2016){Karouzos}, {Woo}, \& {Bae}}]{Karouzos:2016}
{Karouzos}, M., {Woo}, J.-H., \& {Bae}, H.-J. 2016, \apj, 819, 148

\bibitem[{{Kauffmann} {et~al.}(2003){Kauffmann}, {Heckman}, {Tremonti},
  {Brinchmann}, {Charlot}, {White}, {Ridgway}, {Brinkmann}, {Fukugita}, {Hall},
  {Ivezi{\'c}}, {Richards}, \& {Schneider}}]{Kauffmann:2003}
{Kauffmann}, G., {Heckman}, T.~M., {Tremonti}, C., {et~al.} 2003, \mnras, 346,
  1055

\bibitem[{{Kewley} \& {Dopita}(2002)}]{Kewley:2002}
{Kewley}, L.~J. \& {Dopita}, M.~A. 2002, \apjs, 142, 35

\bibitem[{{Kewley} \& {Ellison}(2008)}]{Kewley:2008}
{Kewley}, L.~J. \& {Ellison}, S.~L. 2008, \apj, 681, 1183

\bibitem[{{Kewley} {et~al.}(2006){Kewley}, {Groves}, {Kauffmann}, \&
  {Heckman}}]{Kewley:2006}
{Kewley}, L.~J., {Groves}, B., {Kauffmann}, G., \& {Heckman}, T. 2006, \mnras,
  372, 961

\bibitem[{{King}(2003)}]{King:2003}
{King}, A. 2003, \apjl, 596, L27

\bibitem[{{King} {et~al.}(2011){King}, {Zubovas}, \& {Power}}]{King:2011}
{King}, A.~R., {Zubovas}, K., \& {Power}, C. 2011, \mnras, 415, L6

\bibitem[{{K{\"o}nig} {et~al.}(2009){K{\"o}nig}, {Eckart},
  {Garc{\'{\i}}a-Mar{\'{\i}}n}, \& {Huchtmeier}}]{Koenig:2009}
{K{\"o}nig}, S., {Eckart}, A., {Garc{\'{\i}}a-Mar{\'{\i}}n}, M., \&
  {Huchtmeier}, W.~K. 2009, \aap, 507, 757

\bibitem[{{Kova{\v c}evi{\'c}} {et~al.}(2010){Kova{\v c}evi{\'c}},
  {Popovi{\'c}}, \& {Dimitrijevi{\'c}}}]{Kovacevic:2010}
{Kova{\v c}evi{\'c}}, J., {Popovi{\'c}}, L.~{\v C}., \& {Dimitrijevi{\'c}},
  M.~S. 2010, \apjs, 189, 15

\bibitem[{{Krause} \& {Alexander}(2007)}]{Krause:2007}
{Krause}, M. \& {Alexander}, P. 2007, \mnras, 376, 465

\bibitem[{{Kukula} {et~al.}(1998){Kukula}, {Dunlop}, {Hughes}, \&
  {Rawlings}}]{Kukula:1998}
{Kukula}, M.~J., {Dunlop}, J.~S., {Hughes}, D.~H., \& {Rawlings}, S. 1998,
  \mnras, 297, 366

\bibitem[{{Kulkarni} {et~al.}(1998){Kulkarni}, {Calzetti}, {Bergeron}, {Rieke},
  {Axon}, {Skinner}, {Colina}, {Sparks}, {Daou}, {Gilmore}, {Holfeltz},
  {MacKenty}, {Noll}, {Ritchie}, {Schneider}, {Schultz}, {Storrs}, {Suchkov},
  \& {Thompson}}]{Kulkarni:1998}
{Kulkarni}, V.~P., {Calzetti}, D., {Bergeron}, L., {et~al.} 1998, \apjl, 492,
  L121

\bibitem[{{Laing}(1988)}]{Laing:1988}
{Laing}, R.~A. 1988, \nat, 331, 149

\bibitem[{{Leipski} \& {Bennert}(2006)}]{Leipski:2006a}
{Leipski}, C. \& {Bennert}, N. 2006, \aap, 448, 165

\bibitem[{{Leipski} {et~al.}(2006){Leipski}, {Falcke}, {Bennert}, \&
  {H{\"u}ttemeister}}]{Leipski:2006b}
{Leipski}, C., {Falcke}, H., {Bennert}, N., \& {H{\"u}ttemeister}, S. 2006,
  \aap, 455, 161

\bibitem[{{Lequeux}(1983)}]{Lequeux:1983}
{Lequeux}, J. 1983, \aap, 125, 394

\bibitem[{{Leroy} {et~al.}(2008){Leroy}, {Walter}, {Brinks}, {Bigiel}, {de
  Blok}, {Madore}, \& {Thornley}}]{Leroy:2008}
{Leroy}, A.~K., {Walter}, F., {Brinks}, E., {et~al.} 2008, \aj, 136, 2782

\bibitem[{{Leung} {et~al.}(2017){Leung}, {Coil}, {Azadi}, {Aird}, {Shapley},
  {Kriek}, {Mobasher}, {Reddy}, {Siana}, {Freeman}, {Price}, {Sanders}, \&
  {Shivaei}}]{Leung:2017}
{Leung}, G.~C.~K., {Coil}, A.~L., {Azadi}, M., {et~al.} 2017, \apj, 849, 48

\bibitem[{{Liu} {et~al.}(2014){Liu}, {Zakamska}, \& {Greene}}]{Liu:2014}
{Liu}, G., {Zakamska}, N.~L., \& {Greene}, J.~E. 2014, \mnras, 442, 1303

\bibitem[{{Liu} {et~al.}(2013{\natexlab{a}}){Liu}, {Zakamska}, {Greene},
  {Nesvadba}, \& {Liu}}]{Liu:2013}
{Liu}, G., {Zakamska}, N.~L., {Greene}, J.~E., {Nesvadba}, N.~P.~H., \& {Liu},
  X. 2013{\natexlab{a}}, \mnras, 430, 2327

\bibitem[{{Liu} {et~al.}(2013{\natexlab{b}}){Liu}, {Zakamska}, {Greene},
  {Nesvadba}, \& {Liu}}]{Liu:2013b}
{Liu}, G., {Zakamska}, N.~L., {Greene}, J.~E., {Nesvadba}, N.~P.~H., \& {Liu},
  X. 2013{\natexlab{b}}, \mnras, 436, 2576

\bibitem[{{Magnier} {et~al.}(2016){Magnier}, {Schlafly}, {Finkbeiner}, {Tonry},
  {Goldman}, {R{\"o}ser}, {Schilbach}, {Chambers}, {Flewelling}, {Huber},
  {Price}, {Sweeney}, {Waters}, {Denneau}, {Draper}, {Hodapp}, {Jedicke},
  {Kudritzki}, {Metcalfe}, {Stubbs}, \& {Wainscoast}}]{Magnier:2016}
{Magnier}, E.~A., {Schlafly}, E.~F., {Finkbeiner}, D.~P., {et~al.} 2016, ArXiv
  e-prints [\eprint[arXiv]{1612.05242}]

\bibitem[{{Mahony} {et~al.}(2013){Mahony}, {Morganti}, {Emonts}, {Oosterloo},
  \& {Tadhunter}}]{Mahony:2013}
{Mahony}, E.~K., {Morganti}, R., {Emonts}, B.~H.~C., {Oosterloo}, T.~A., \&
  {Tadhunter}, C. 2013, \mnras, 435, L58

\bibitem[{{Mahony} {et~al.}(2016){Mahony}, {Oonk}, {Morganti}, {Tadhunter},
  {Bessiere}, {Short}, {Emonts}, \& {Oosterloo}}]{Mahony:2016}
{Mahony}, E.~K., {Oonk}, J.~B.~R., {Morganti}, R., {et~al.} 2016, \mnras, 455,
  2453

\bibitem[{{Marconi} {et~al.}(2004){Marconi}, {Risaliti}, {Gilli}, {Hunt},
  {Maiolino}, \& {Salvati}}]{Marconi:2004}
{Marconi}, A., {Risaliti}, G., {Gilli}, R., {et~al.} 2004, \mnras, 351, 169

\bibitem[{{Marino} {et~al.}(2013){Marino}, {Rosales-Ortega}, {S{\'a}nchez},
  {Gil de Paz}, {V{\'{\i}}lchez}, {Miralles-Caballero}, {Kehrig},
  {P{\'e}rez-Montero}, {Stanishev}, {Iglesias-P{\'a}ramo}, {D{\'{\i}}az},
  {Castillo-Morales}, {Kennicutt}, {L{\'o}pez-S{\'a}nchez}, {Galbany},
  {Garc{\'{\i}}a-Benito}, {Mast}, {Mendez-Abreu}, {Monreal-Ibero}, {Husemann},
  {Walcher}, {Garc{\'{\i}}a-Lorenzo}, {Masegosa}, {Del Olmo Orozco},
  {Mour{\~a}o}, {Ziegler}, {Moll{\'a}}, {Papaderos},
  {S{\'a}nchez-Bl{\'a}zquez}, {Gonz{\'a}lez Delgado}, {Falc{\'o}n-Barroso},
  {Roth}, {van de Ven}, \& {Califa Team}}]{Marino:2013}
{Marino}, R.~A., {Rosales-Ortega}, F.~F., {S{\'a}nchez}, S.~F., {et~al.} 2013,
  \aap, 559, A114

\bibitem[{{Martin}(2005)}]{Martin:2005}
{Martin}, C.~L. 2005, \apj, 621, 227

\bibitem[{{Mart{{\'i}}n} {et~al.}(2015){Mart{{\'i}}n}, {Kohno}, {Izumi},
  {Krips}, {Meier}, {Aladro}, {Matsushita}, {Takano}, {Turner}, {Espada},
  {Nakajima}, {Terashima}, {Fathi}, {Hsieh}, {Imanishi}, {Lundgren}, {Nakai},
  {Schinnerer}, {Sheth}, \& {Wiklind}}]{Martin:2015}
{Mart{{\'i}}n}, S., {Kohno}, K., {Izumi}, T., {et~al.} 2015, \aap, 573, A116

\bibitem[{{Martinsson} {et~al.}(2013){Martinsson}, {Verheijen}, {Westfall},
  {Bershady}, {Andersen}, \& {Swaters}}]{Martinsson:2013}
{Martinsson}, T.~P.~K., {Verheijen}, M.~A.~W., {Westfall}, K.~B., {et~al.}
  2013, \aap, 557, A131

\bibitem[{{McElroy} {et~al.}(2015){McElroy}, {Croom}, {Pracy}, {Sharp}, {Ho},
  \& {Medling}}]{McElroy:2015}
{McElroy}, R., {Croom}, S.~M., {Pracy}, M., {et~al.} 2015, \mnras, 446, 2186

\bibitem[{{McGregor} {et~al.}(2003){McGregor}, {Hart}, {Conroy}, {Pfitzner},
  {Bloxham}, {Jones}, {Downing}, {Dawson}, {Young}, {Jarnyk}, \& {Van
  Harmelen}}]{McGregor:2003}
{McGregor}, P.~J., {Hart}, J., {Conroy}, P.~G., {et~al.} 2003, in \procspie,
  Vol. 4841, Instrument Design and Performance for Optical/Infrared
  Ground-based Telescopes, ed. M.~{Iye} \& A.~F.~M. {Moorwood}, 1581--1591

\bibitem[{{McMullin} {et~al.}(2007){McMullin}, {Waters}, {Schiebel}, {Young},
  \& {Golap}}]{McMullin:2007}
{McMullin}, J.~P., {Waters}, B., {Schiebel}, D., {Young}, W., \& {Golap}, K.
  2007, in Astronomical Society of the Pacific Conference Series, Vol. 376,
  Astronomical Data Analysis Software and Systems XVI, ed. R.~A. {Shaw},
  F.~{Hill}, \& D.~J. {Bell}, 127

\bibitem[{{Mezcua} {et~al.}(2015){Mezcua}, {Prieto}, {Fern{\'a}ndez-Ontiveros},
  {Tristram}, {Neumayer}, \& {Kotilainen}}]{Mezcua:2015}
{Mezcua}, M., {Prieto}, M.~A., {Fern{\'a}ndez-Ontiveros}, J.~A., {et~al.} 2015,
  \mnras, 452, 4128

\bibitem[{{Middelberg} {et~al.}(2007){Middelberg}, {Agudo}, {Roy}, \&
  {Krichbaum}}]{Middelberg:2007}
{Middelberg}, E., {Agudo}, I., {Roy}, A.~L., \& {Krichbaum}, T.~P. 2007,
  \mnras, 377, 731

\bibitem[{{Mingozzi} {et~al.}(2019){Mingozzi}, {Cresci}, {Venturi}, {Marconi},
  {Mannucci}, {Perna}, {Belfiore}, {Carniani}, {Balmaverde}, {Brusa}, {Cicone},
  {Feruglio}, {Gallazzi}, {Mainieri}, {Maiolino}, {Nagao}, {Nardini}, {Sani},
  {Tozzi}, \& {Zibetti}}]{Mingozzi:2019}
{Mingozzi}, M., {Cresci}, G., {Venturi}, G., {et~al.} 2019, \aap, 622, A146

\bibitem[{{Mommert}(2017)}]{Mommert:2017}
{Mommert}, M. 2017, Astronomy and Computing, 18, 47

\bibitem[{{Morganti} {et~al.}(2013){Morganti}, {Frieswijk}, {Oonk},
  {Oosterloo}, \& {Tadhunter}}]{Morganti:2013}
{Morganti}, R., {Frieswijk}, W., {Oonk}, R.~J.~B., {Oosterloo}, T., \&
  {Tadhunter}, C. 2013, \aap, 552, L4

\bibitem[{{Morganti} {et~al.}(2015){Morganti}, {Oosterloo}, {Oonk},
  {Frieswijk}, \& {Tadhunter}}]{Morganti:2015}
{Morganti}, R., {Oosterloo}, T., {Oonk}, J.~B.~R., {Frieswijk}, W., \&
  {Tadhunter}, C. 2015, \aap, 580, A1

\bibitem[{{Morganti} {et~al.}(2005){Morganti}, {Tadhunter}, \&
  {Oosterloo}}]{Morganti:2005}
{Morganti}, R., {Tadhunter}, C.~N., \& {Oosterloo}, T.~A. 2005, \aap, 444, L9

\bibitem[{{Mukherjee} {et~al.}(2018){Mukherjee}, {Bicknell}, {Wagner},
  {Sutherland}, \& {Silk}}]{Mukherjee:2018}
{Mukherjee}, D., {Bicknell}, G.~V., {Wagner}, A.~Y., {Sutherland}, R.~S., \&
  {Silk}, J. 2018, \mnras, 479, 5544

\bibitem[{{Mullaney} {et~al.}(2015){Mullaney}, {Alexander}, {Aird}, {Bernhard},
  {Daddi}, {Del Moro}, {Dickinson}, {Elbaz}, {Harrison}, {Juneau}, {Liu},
  {Pannella}, {Rosario}, {Santini}, {Sargent}, {Schreiber}, {Simpson}, \&
  {Stanley}}]{Mullaney:2015}
{Mullaney}, J.~R., {Alexander}, D.~M., {Aird}, J., {et~al.} 2015, \mnras, 453,
  L83

\bibitem[{{Mullaney} {et~al.}(2013){Mullaney}, {Alexander}, {Fine}, {Goulding},
  {Harrison}, \& {Hickox}}]{Mullaney:2013}
{Mullaney}, J.~R., {Alexander}, D.~M., {Fine}, S., {et~al.} 2013, \mnras, 433,
  622

\bibitem[{{Neininger} {et~al.}(1998){Neininger}, {Guelin}, {Klein},
  {Garcia-Burillo}, \& {Wielebinski}}]{Neininger:1998}
{Neininger}, N., {Guelin}, M., {Klein}, U., {Garcia-Burillo}, S., \&
  {Wielebinski}, R. 1998, \aap, 339, 737

\bibitem[{{Nesvadba} {et~al.}(2006){Nesvadba}, {Lehnert}, {Eisenhauer},
  {Gilbert}, {Tecza}, \& {Abuter}}]{Nesvadba:2006}
{Nesvadba}, N.~P.~H., {Lehnert}, M.~D., {Eisenhauer}, F., {et~al.} 2006, \apj,
  650, 693

\bibitem[{{Netzer} {et~al.}(2004){Netzer}, {Shemmer}, {Maiolino}, {Oliva},
  {Croom}, {Corbett}, \& {di Fabrizio}}]{Netzer:2004}
{Netzer}, H., {Shemmer}, O., {Maiolino}, R., {et~al.} 2004, \apj, 614, 558

\bibitem[{{Noll} {et~al.}(2014){Noll}, {Kausch}, {Kimeswenger}, {Barden},
  {Jones}, {Modigliani}, {Szyszka}, \& {Taylor}}]{Noll:2014}
{Noll}, S., {Kausch}, W., {Kimeswenger}, S., {et~al.} 2014, \aap, 567, A25

\bibitem[{{O'Dea} {et~al.}(2002){O'Dea}, {de Vries}, {Koekemoer}, {Baum},
  {Morganti}, {Fanti}, {Capetti}, {Tadhunter}, {Barthel}, {Axon}, \&
  {Gelderman}}]{ODea:2002}
{O'Dea}, C.~P., {de Vries}, W.~H., {Koekemoer}, A.~M., {et~al.} 2002, \aj, 123,
  2333

\bibitem[{{Oh} {et~al.}(2018){Oh}, {Koss}, {Markwardt}, {Schawinski},
  {Baumgartner}, {Barthelmy}, {Cenko}, {Gehrels}, {Mushotzky}, {Petulante},
  {Ricci}, {Lien}, \& {Trakhtenbrot}}]{Oh:2018}
{Oh}, K., {Koss}, M., {Markwardt}, C.~B., {et~al.} 2018, \apjs, 235, 4

\bibitem[{{Oosterloo} {et~al.}(2017){Oosterloo}, {Raymond Oonk}, {Morganti},
  {Combes}, {Dasyra}, {Salom{\'e}}, {Vlahakis}, \&
  {Tadhunter}}]{Oosterloo:2017}
{Oosterloo}, T., {Raymond Oonk}, J.~B., {Morganti}, R., {et~al.} 2017, \aap,
  608, A38

\bibitem[{{Osterbrock} \& {Ferland}(2006)}]{Osterbrock:2006}
{Osterbrock}, D.~E. \& {Ferland}, G.~J. 2006, {Astrophysics of gaseous nebulae
  and active galactic nuclei} (2nd.~ed.~by D.E.~Osterbrock and
  G.J.~Ferland.~Sausalito, CA: University Science Books, 2006)

\bibitem[{{Ostriker} {et~al.}(2010){Ostriker}, {Choi}, {Ciotti}, {Novak}, \&
  {Proga}}]{Ostriker:2010}
{Ostriker}, J.~P., {Choi}, E., {Ciotti}, L., {Novak}, G.~S., \& {Proga}, D.
  2010, \apj, 722, 642

\bibitem[{{P{\'e}rez-Montero} \& {Contini}(2009)}]{Perez-Montero:2009}
{P{\'e}rez-Montero}, E. \& {Contini}, T. 2009, \mnras, 398, 949

\bibitem[{{Perna} {et~al.}(2015){Perna}, {Brusa}, {Cresci}, {Comastri},
  {Lanzuisi}, {Lusso}, {Marconi}, {Salvato}, {Zamorani}, {Bongiorno},
  {Mainieri}, {Maiolino}, \& {Mignoli}}]{Perna:2015}
{Perna}, M., {Brusa}, M., {Cresci}, G., {et~al.} 2015, \aap, 574, A82

\bibitem[{{Perna} {et~al.}(2017){Perna}, {Lanzuisi}, {Brusa}, {Mignoli}, \&
  {Cresci}}]{Perna:2017}
{Perna}, M., {Lanzuisi}, G., {Brusa}, M., {Mignoli}, M., \& {Cresci}, G. 2017,
  \aap, 603, A99

\bibitem[{{Pettini} \& {Pagel}(2004)}]{Pettini:2004}
{Pettini}, M. \& {Pagel}, B.~E.~J. 2004, \mnras, 348, L59

\bibitem[{{Pilbratt} {et~al.}(2010){Pilbratt}, {Riedinger}, {Passvogel},
  {Crone}, {Doyle}, {Gageur}, {Heras}, {Jewell}, {Metcalfe}, {Ott}, \&
  {Schmidt}}]{Pilbratt:2010}
{Pilbratt}, G.~L., {Riedinger}, J.~R., {Passvogel}, T., {et~al.} 2010, \aap,
  518, L1

\bibitem[{{Pogge}(1988)}]{Pogge:1988a}
{Pogge}, R.~W. 1988, \apj, 328, 519

\bibitem[{{Poglitsch} {et~al.}(2010){Poglitsch}, {Waelkens}, {Geis},
  {Feuchtgruber}, {Vandenbussche}, {Rodriguez}, {Krause}, {Renotte}, {van
  Hoof}, {Saraceno}, {Cepa}, {Kerschbaum}, {Agn{\`e}se}, {Ali}, {Altieri},
  {Andreani}, {Augueres}, {Balog}, {Barl}, {Bauer}, {Belbachir}, {Benedettini},
  {Billot}, {Boulade}, {Bischof}, {Blommaert}, {Callut}, {Cara}, {Cerulli},
  {Cesarsky}, {Contursi}, {Creten}, {De Meester}, {Doublier}, {Doumayrou},
  {Duband}, {Exter}, {Genzel}, {Gillis}, {Gr{\"o}zinger}, {Henning},
  {Herreros}, {Huygen}, {Inguscio}, {Jakob}, {Jamar}, {Jean}, {de Jong},
  {Katterloher}, {Kiss}, {Klaas}, {Lemke}, {Lutz}, {Madden}, {Marquet},
  {Martignac}, {Mazy}, {Merken}, {Montfort}, {Morbidelli}, {M{\"u}ller},
  {Nielbock}, {Okumura}, {Orfei}, {Ottensamer}, {Pezzuto}, {Popesso},
  {Putzeys}, {Regibo}, {Reveret}, {Royer}, {Sauvage}, {Schreiber}, {Stegmaier},
  {Schmitt}, {Schubert}, {Sturm}, {Thiel}, {Tofani}, {Vavrek}, {Wetzstein},
  {Wieprecht}, \& {Wiezorrek}}]{Poglitsch:2010}
{Poglitsch}, A., {Waelkens}, C., {Geis}, N., {et~al.} 2010, \aap, 518, L2

\bibitem[{{Pounds} {et~al.}(2003){Pounds}, {Reeves}, {King}, {Page}, {O'Brien},
  \& {Turner}}]{Pounds:2003}
{Pounds}, K.~A., {Reeves}, J.~N., {King}, A.~R., {et~al.} 2003, \mnras, 345,
  705

\bibitem[{{Revalski} {et~al.}(2018){Revalski}, {Dashtamirova}, {Crenshaw},
  {Kraemer}, {Fischer}, {Schmitt}, {Gnilka}, {Schmidt}, {Elvis}, {Fabbiano},
  {Storchi-Bergmann}, {Maksym}, \& {Gandhi}}]{Revalski:2018}
{Revalski}, M., {Dashtamirova}, D., {Crenshaw}, D.~M., {et~al.} 2018, \apj,
  867, 88

\bibitem[{{Richards} {et~al.}(2006){Richards}, {Lacy}, {Storrie-Lombardi},
  {Hall}, {Gallagher}, {Hines}, {Fan}, {Papovich}, {Vanden Berk}, {Trammell},
  {Schneider}, {Vestergaard}, {York}, {Jester}, {Anderson}, {Budav{\'a}ri}, \&
  {Szalay}}]{Richards:2006}
{Richards}, G.~T., {Lacy}, M., {Storrie-Lombardi}, L.~J., {et~al.} 2006, \apjs,
  166, 470

\bibitem[{{Riffel} {et~al.}(2013){Riffel}, {Storchi-Bergmann}, \&
  {Winge}}]{Riffel:2013}
{Riffel}, R.~A., {Storchi-Bergmann}, T., \& {Winge}, C. 2013, \mnras, 430, 2249

\bibitem[{{Roche} {et~al.}(2016){Roche}, {Humphrey}, {Lagos}, {Papaderos},
  {Silva}, {Cardoso}, \& {Gomes}}]{Roche:2016}
{Roche}, N., {Humphrey}, A., {Lagos}, P., {et~al.} 2016, \mnras, 459, 4259

\bibitem[{{Rosario} {et~al.}(2018){Rosario}, {Burtscher}, {Davies}, {Koss},
  {Ricci}, {Lutz}, {Riffel}, {Alexander}, {Genzel}, {Hicks}, {Lin},
  {Maciejewski}, {M{\"u}ller-S{\'a}nchez}, {Orban de Xivry}, {Riffel},
  {Schartmann}, {Schawinski}, {Schnorr-M{\"u}ller}, {Saintonge}, {Shimizu},
  {Sternberg}, {Storchi-Bergmann}, {Sturm}, {Tacconi}, {Treister}, \&
  {Veilleux}}]{Rosario:2018}
{Rosario}, D.~J., {Burtscher}, L., {Davies}, R.~I., {et~al.} 2018, \mnras, 473,
  5658

\bibitem[{{Rose} {et~al.}(2018){Rose}, {Tadhunter}, {Ramos Almeida},
  {Rodr{\'{\i}}guez Zaur{\'{\i}}n}, {Santoro}, \& {Spence}}]{Rose:2018}
{Rose}, M., {Tadhunter}, C., {Ramos Almeida}, C., {et~al.} 2018, \mnras, 474,
  128

\bibitem[{{Rupke} {et~al.}(2017){Rupke}, {G{\"u}ltekin}, \&
  {Veilleux}}]{Rupke:2017}
{Rupke}, D.~S.~N., {G{\"u}ltekin}, K., \& {Veilleux}, S. 2017, \apj, 850, 40

\bibitem[{{Rupke} \& {Veilleux}(2013)}]{Rupke:2013}
{Rupke}, D.~S.~N. \& {Veilleux}, S. 2013, \apjl, 775, L15

\bibitem[{{Sabbadin} {et~al.}(1977){Sabbadin}, {Minello}, \&
  {Bianchini}}]{Sabbadin:1977}
{Sabbadin}, F., {Minello}, S., \& {Bianchini}, A. 1977, \aap, 60, 147

\bibitem[{{Saintonge} {et~al.}(2011){Saintonge}, {Kauffmann}, {Kramer},
  {Tacconi}, {Buchbender}, {Catinella}, {Fabello}, {Graci{\'a}-Carpio}, {Wang},
  {Cortese}, {Fu}, {Genzel}, {Giovanelli}, {Guo}, {Haynes}, {Heckman},
  {Krumholz}, {Lemonias}, {Li}, {Moran}, {Rodriguez-Fernandez}, {Schiminovich},
  {Schuster}, \& {Sievers}}]{Saintonge:2011}
{Saintonge}, A., {Kauffmann}, G., {Kramer}, C., {et~al.} 2011, \mnras, 415, 32

\bibitem[{{Salucci} {et~al.}(1991){Salucci}, {Ashman}, \&
  {Persic}}]{Salucci:1991}
{Salucci}, P., {Ashman}, K.~M., \& {Persic}, M. 1991, \apj, 379, 89

\bibitem[{{S{\'a}nchez} {et~al.}(2014){S{\'a}nchez}, {Rosales-Ortega},
  {Iglesias-P{\'a}ramo}, {Moll{\'a}}, {Barrera-Ballesteros}, {Marino},
  {P{\'e}rez}, {S{\'a}nchez-Blazquez}, {Gonz{\'a}lez Delgado}, {Cid Fernandes},
  {de Lorenzo-C{\'a}ceres}, {Mendez-Abreu}, {Galbany}, {Falcon-Barroso},
  {Miralles-Caballero}, {Husemann}, {Garc{\'{\i}}a-Benito}, {Mast}, {Walcher},
  {Gil de Paz}, {Garc{\'{\i}}a-Lorenzo}, {Jungwiert}, {V{\'{\i}}lchez},
  {J{\'{\i}}lkov{\'a}}, {Lyubenova}, {Cortijo-Ferrero}, {D{\'{\i}}az},
  {Wisotzki}, {M{\'a}rquez}, {Bland-Hawthorn}, {Ellis}, {van de Ven}, {Jahnke},
  {Papaderos}, {Gomes}, {Mendoza}, \& {L{\'o}pez-S{\'a}nchez}}]{Sanchez:2014}
{S{\'a}nchez}, S.~F., {Rosales-Ortega}, F.~F., {Iglesias-P{\'a}ramo}, J.,
  {et~al.} 2014, \aap, 563, A49

\bibitem[{{Santoro} {et~al.}(2018){Santoro}, {Rose}, {Morganti}, {Tadhunter},
  {Oosterloo}, \& {Holt}}]{Santoro:2018}
{Santoro}, F., {Rose}, M., {Morganti}, R., {et~al.} 2018, \aap, 617, A139

\bibitem[{{Schirmer} {et~al.}(2013){Schirmer}, {Diaz}, {Holhjem}, {Levenson},
  \& {Winge}}]{Schirmer:2013}
{Schirmer}, M., {Diaz}, R., {Holhjem}, K., {Levenson}, N.~A., \& {Winge}, C.
  2013, \apj, 763, 60

\bibitem[{{Schmitt} {et~al.}(2003){Schmitt}, {Donley}, {Antonucci},
  {Hutchings}, {Kinney}, \& {Pringle}}]{Schmitt:2003b}
{Schmitt}, H.~R., {Donley}, J.~L., {Antonucci}, R.~R.~J., {et~al.} 2003, \apj,
  597, 768

\bibitem[{{Shangguan} {et~al.}(2018){Shangguan}, {Ho}, \&
  {Xie}}]{Shangguan:2018}
{Shangguan}, J., {Ho}, L.~C., \& {Xie}, Y. 2018, \apj, 854, 158

\bibitem[{{Shimizu} {et~al.}(2015){Shimizu}, {Mushotzky}, {Mel{\'e}ndez},
  {Koss}, \& {Rosario}}]{Shimizu:2015}
{Shimizu}, T.~T., {Mushotzky}, R.~F., {Mel{\'e}ndez}, M., {Koss}, M., \&
  {Rosario}, D.~J. 2015, \mnras, 452, 1841

\bibitem[{{Silk} \& {Rees}(1998)}]{Silk:1998}
{Silk}, J. \& {Rees}, M.~J. 1998, \aap, 331, L1

\bibitem[{{Skrutskie} {et~al.}(2006){Skrutskie}, {Cutri}, {Stiening},
  {Weinberg}, {Schneider}, {Carpenter}, {Beichman}, {Capps}, {Chester},
  {Elias}, {Huchra}, {Liebert}, {Lonsdale}, {Monet}, {Price}, {Seitzer},
  {Jarrett}, {Kirkpatrick}, {Gizis}, {Howard}, {Evans}, {Fowler}, {Fullmer},
  {Hurt}, {Light}, {Kopan}, {Marsh}, {McCallon}, {Tam}, {Van Dyk}, \&
  {Wheelock}}]{Skrutskie:2006}
{Skrutskie}, M.~F., {Cutri}, R.~M., {Stiening}, R., {et~al.} 2006, \aj, 131,
  1163

\bibitem[{{Somerville} {et~al.}(2008){Somerville}, {Hopkins}, {Cox},
  {Robertson}, \& {Hernquist}}]{Somerville:2008}
{Somerville}, R.~S., {Hopkins}, P.~F., {Cox}, T.~J., {Robertson}, B.~E., \&
  {Hernquist}, L. 2008, \mnras, 391, 481

\bibitem[{{Soto} {et~al.}(2016){Soto}, {Lilly}, {Bacon}, {Richard}, \&
  {Conseil}}]{Soto:2016}
{Soto}, K.~T., {Lilly}, S.~J., {Bacon}, R., {Richard}, J., \& {Conseil}, S.
  2016, \mnras, 458, 3210

\bibitem[{{Spoon} {et~al.}(2013){Spoon}, {Farrah}, {Lebouteiller},
  {Gonz{\'a}lez-Alfonso}, {Bernard-Salas}, {Urrutia}, {Rigopoulou},
  {Westmoquette}, {Smith}, {Afonso}, {Pearson}, {Cormier}, {Efstathiou},
  {Borys}, {Verma}, {Etxaluze}, \& {Clements}}]{Spoon:2013}
{Spoon}, H.~W.~W., {Farrah}, D., {Lebouteiller}, V., {et~al.} 2013, \apj, 775,
  127

\bibitem[{{Stasi{\'n}ska} {et~al.}(2008){Stasi{\'n}ska}, {Vale Asari}, {Cid
  Fernandes}, {Gomes}, {Schlickmann}, {Mateus}, {Schoenell}, \&
  {Sodr{\'e}}}]{Stasinska:2008}
{Stasi{\'n}ska}, G., {Vale Asari}, N., {Cid Fernandes}, R., {et~al.} 2008,
  \mnras, 391, L29

\bibitem[{{Stern} \& {Laor}(2013)}]{Stern:2013}
{Stern}, J. \& {Laor}, A. 2013, \mnras, 431, 836

\bibitem[{{Storchi-Bergmann} {et~al.}(2018){Storchi-Bergmann}, {Dall'Agnol de
  Oliveira}, {Longo Micchi}, {Schmitt}, {Fischer}, {Kraemer}, {Crenshaw},
  {Maksym}, {Elvis}, {Fabbiano}, \& {Colina}}]{Storchi-Bergmann:2018}
{Storchi-Bergmann}, T., {Dall'Agnol de Oliveira}, B., {Longo Micchi}, L.~F.,
  {et~al.} 2018, \apj, 868, 14

\bibitem[{{Storchi-Bergmann} {et~al.}(1998){Storchi-Bergmann}, {Schmitt},
  {Calzetti}, \& {Kinney}}]{Storchi-Bergmann:1998}
{Storchi-Bergmann}, T., {Schmitt}, H.~R., {Calzetti}, D., \& {Kinney}, A.~L.
  1998, \aj, 115, 909

\bibitem[{{Storchi-Bergmann} {et~al.}(1992){Storchi-Bergmann}, {Wilson}, \&
  {Baldwin}}]{Storchi-Bergmann:1992}
{Storchi-Bergmann}, T., {Wilson}, A.~S., \& {Baldwin}, J.~A. 1992, \apj, 396,
  45

\bibitem[{{Storey} \& {Zeippen}(2000)}]{Storey:2000}
{Storey}, P.~J. \& {Zeippen}, C.~J. 2000, \mnras, 312, 813

\bibitem[{{Sturm} {et~al.}(2011){Sturm}, {Gonz{\'a}lez-Alfonso}, {Veilleux},
  {Fischer}, {Graci{\'a}-Carpio}, {Hailey-Dunsheath}, {Contursi}, {Poglitsch},
  {Sternberg}, {Davies}, {Genzel}, {Lutz}, {Tacconi}, {Verma}, {Maiolino}, \&
  {de Jong}}]{Sturm:2011}
{Sturm}, E., {Gonz{\'a}lez-Alfonso}, E., {Veilleux}, S., {et~al.} 2011, \apjl,
  733, L16

\bibitem[{{Suh} {et~al.}(2017){Suh}, {Civano}, {Hasinger}, {Lusso}, {Lanzuisi},
  {Marchesi}, {Trakhtenbrot}, {Allevato}, {Cappelluti}, {Capak}, {Elvis},
  {Griffiths}, {Laigle}, {Lira}, {Riguccini}, {Rosario}, {Salvato},
  {Schawinski}, \& {Vignali}}]{Suh:2017}
{Suh}, H., {Civano}, F., {Hasinger}, G., {et~al.} 2017, \apj, 841, 102

\bibitem[{{Sun} {et~al.}(2017){Sun}, {Greene}, \& {Zakamska}}]{Sun:2017}
{Sun}, A.-L., {Greene}, J.~E., \& {Zakamska}, N.~L. 2017, \apj, 835, 222

\bibitem[{{Sutherland} \& {Bicknell}(2007)}]{Sutherland:2007}
{Sutherland}, R.~S. \& {Bicknell}, G.~V. 2007, \apjs, 173, 37

\bibitem[{{Tadhunter} {et~al.}(2014){Tadhunter}, {Morganti}, {Rose}, {Oonk}, \&
  {Oosterloo}}]{Tadhunter:2014}
{Tadhunter}, C., {Morganti}, R., {Rose}, M., {Oonk}, J.~B.~R., \& {Oosterloo},
  T. 2014, \nat, 511, 440

\bibitem[{{Tadhunter} {et~al.}(2018){Tadhunter}, {Rodr{\'{\i}}guez
  Zaur{\'{\i}}n}, {Rose}, {Spence}, {Batcheldor}, {Berg}, {Ramos Almeida},
  {Spoon}, {Sparks}, \& {Chiaberge}}]{Tadhunter:2018}
{Tadhunter}, C., {Rodr{\'{\i}}guez Zaur{\'{\i}}n}, J., {Rose}, M., {et~al.}
  2018, \mnras, 478, 1558

\bibitem[{{Tombesi} {et~al.}(2010{\natexlab{a}}){Tombesi}, {Cappi}, {Reeves},
  {Palumbo}, {Yaqoob}, {Braito}, \& {Dadina}}]{Tombesi:2010}
{Tombesi}, F., {Cappi}, M., {Reeves}, J.~N., {et~al.} 2010{\natexlab{a}}, \aap,
  521, A57

\bibitem[{{Tombesi} {et~al.}(2015){Tombesi}, {Mel{\'e}ndez}, {Veilleux},
  {Reeves}, {Gonz{\'a}lez-Alfonso}, \& {Reynolds}}]{Tombesi:2015}
{Tombesi}, F., {Mel{\'e}ndez}, M., {Veilleux}, S., {et~al.} 2015, \nat, 519,
  436

\bibitem[{{Tombesi} {et~al.}(2010{\natexlab{b}}){Tombesi}, {Sambruna},
  {Reeves}, {Braito}, {Ballo}, {Gofford}, {Cappi}, \&
  {Mushotzky}}]{Tombesi:2010b}
{Tombesi}, F., {Sambruna}, R.~M., {Reeves}, J.~N., {et~al.} 2010{\natexlab{b}},
  \apj, 719, 700

\bibitem[{{Tremblay} {et~al.}(2018){Tremblay}, {Combes}, {Oonk}, {Russell},
  {McDonald}, {Gaspari}, {Husemann}, {Nulsen}, {McNamara}, {Hamer}, {O'Dea},
  {Baum}, {Davis}, {Donahue}, {Voit}, {Edge}, {Blanton}, {Bremer}, {Bulbul},
  {Clarke}, {David}, {Edwards}, {Eggerman}, {Fabian}, {Forman}, {Jones},
  {Kerman}, {Kraft}, {Li}, {Powell}, {Randall}, {Salom{\'e}}, {Simionescu},
  {Su}, {Sun}, {Urry}, {Vantyghem}, {Wilkes}, \& {ZuHone}}]{Tremblay:2018}
{Tremblay}, G.~R., {Combes}, F., {Oonk}, J.~B.~R., {et~al.} 2018, \apj, 865, 13

\bibitem[{{Tremonti} {et~al.}(2004){Tremonti}, {Heckman}, {Kauffmann},
  {Brinchmann}, {Charlot}, {White}, {Seibert}, {Peng}, {Schlegel}, {Uomoto},
  {Fukugita}, \& {Brinkmann}}]{Tremonti:2004}
{Tremonti}, C.~A., {Heckman}, T.~M., {Kauffmann}, G., {et~al.} 2004, \apj, 613,
  898

\bibitem[{{Ulvestad} {et~al.}(2005){Ulvestad}, {Antonucci}, \&
  {Barvainis}}]{Ulvestad:2005}
{Ulvestad}, J.~S., {Antonucci}, R.~R.~J., \& {Barvainis}, R. 2005, \apj, 621,
  123

\bibitem[{{Urry} \& {Padovani}(1995)}]{Urry:1995}
{Urry}, C.~M. \& {Padovani}, P. 1995, \pasp, 107, 803

\bibitem[{{Vaddi} {et~al.}(2016){Vaddi}, {O'Dea}, {Baum}, {Whitmore}, {Ahmed},
  {Pierce}, \& {Leary}}]{Vaddi:2016}
{Vaddi}, S., {O'Dea}, C.~P., {Baum}, S.~A., {et~al.} 2016, \apj, 818, 182

\bibitem[{{Veilleux} {et~al.}(2017){Veilleux}, {Bolatto}, {Tombesi},
  {Mel{\'e}ndez}, {Sturm}, {Gonz{\'a}lez-Alfonso}, {Fischer}, \&
  {Rupke}}]{Veilleux:2017}
{Veilleux}, S., {Bolatto}, A., {Tombesi}, F., {et~al.} 2017, \apj, 843, 18

\bibitem[{{Veilleux} {et~al.}(1994){Veilleux}, {Cecil}, {Bland-Hawthorn},
  {Tully}, {Filippenko}, \& {Sargent}}]{Veilleux:1994}
{Veilleux}, S., {Cecil}, G., {Bland-Hawthorn}, J., {et~al.} 1994, \apj, 433, 48

\bibitem[{{Veilleux} {et~al.}(2013){Veilleux}, {Mel{\'e}ndez}, {Sturm},
  {Gracia-Carpio}, {Fischer}, {Gonz{\'a}lez-Alfonso}, {Contursi}, {Lutz},
  {Poglitsch}, {Davies}, {Genzel}, {Tacconi}, {de Jong}, {Sternberg}, {Netzer},
  {Hailey-Dunsheath}, {Verma}, {Rupke}, {Maiolino}, {Teng}, \&
  {Polisensky}}]{Veilleux:2013}
{Veilleux}, S., {Mel{\'e}ndez}, M., {Sturm}, E., {et~al.} 2013, \apj, 776, 27

\bibitem[{{Veilleux} \& {Osterbrock}(1987)}]{Veilleux:1987}
{Veilleux}, S. \& {Osterbrock}, D.~E. 1987, \apjs, 63, 295

\bibitem[{{Verner} {et~al.}(1996){Verner}, {Ferland}, {Korista}, \&
  {Yakovlev}}]{Verner:1996}
{Verner}, D.~A., {Ferland}, G.~J., {Korista}, K.~T., \& {Yakovlev}, D.~G. 1996,
  \apj, 465, 487

\bibitem[{{Vestergaard} \& {Peterson}(2006)}]{Vestergaard:2006}
{Vestergaard}, M. \& {Peterson}, B.~M. 2006, \apj, 641, 689

\bibitem[{{Viironen} {et~al.}(2007){Viironen}, {Delgado-Inglada}, {Mampaso},
  {Magrini}, \& {Corradi}}]{Viironen:2007}
{Viironen}, K., {Delgado-Inglada}, G., {Mampaso}, A., {Magrini}, L., \&
  {Corradi}, R.~L.~M. 2007, \mnras, 381, 1719

\bibitem[{{Villar-Mart{\'{\i}}n} {et~al.}(2016){Villar-Mart{\'{\i}}n},
  {Arribas}, {Emonts}, {Humphrey}, {Tadhunter}, {Bessiere}, {Cabrera Lavers},
  \& {Ramos Almeida}}]{Villar-Martin:2016}
{Villar-Mart{\'{\i}}n}, M., {Arribas}, S., {Emonts}, B., {et~al.} 2016, \mnras,
  460, 130

\bibitem[{{Villar-Mart{\'{\i}}n} {et~al.}(1999){Villar-Mart{\'{\i}}n},
  {Binette}, \& {Fosbury}}]{Villar-Martin:1999b}
{Villar-Mart{\'{\i}}n}, M., {Binette}, L., \& {Fosbury}, R.~A.~E. 1999, \aap,
  346, 7

\bibitem[{{Villar-Mart{\'{\i}}n} {et~al.}(2018){Villar-Mart{\'{\i}}n},
  {Cabrera-Lavers}, {Humphrey}, {Silva}, {Ramos Almeida}, {Piqueras-L{\'o}pez},
  \& {Emonts}}]{Villar-Martin:2018}
{Villar-Mart{\'{\i}}n}, M., {Cabrera-Lavers}, A., {Humphrey}, A., {et~al.}
  2018, \mnras, 474, 2302

\bibitem[{{Villar-Mart{\'{\i}}n} {et~al.}(2017){Villar-Mart{\'{\i}}n},
  {Emonts}, {Cabrera Lavers}, {Tadhunter}, {Mukherjee}, {Humphrey},
  {Rodr{\'{\i}}guez Zaur{\'{\i}}n}, {Ramos Almeida}, {P{\'e}rez Torres}, \&
  {Bessiere}}]{Villar-Martin:2017}
{Villar-Mart{\'{\i}}n}, M., {Emonts}, B., {Cabrera Lavers}, A., {et~al.} 2017,
  \mnras, 472, 4659

\bibitem[{{Villar Mart{\'{\i}}n} {et~al.}(2014){Villar Mart{\'{\i}}n},
  {Emonts}, {Humphrey}, {Cabrera Lavers}, \& {Binette}}]{Villar-Martin:2014}
{Villar Mart{\'{\i}}n}, M., {Emonts}, B., {Humphrey}, A., {Cabrera Lavers}, A.,
  \& {Binette}, L. 2014, \mnras, 440, 3202

\bibitem[{{Villar-Mart{\'{\i}}n} {et~al.}(2010){Villar-Mart{\'{\i}}n},
  {Tadhunter}, {P{\'e}rez}, {Humphrey}, {Mart{\'{\i}}nez-Sansigre}, {Delgado},
  \& {P{\'e}rez-Torres}}]{Villar-Martin:2010}
{Villar-Mart{\'{\i}}n}, M., {Tadhunter}, C., {P{\'e}rez}, E., {et~al.} 2010,
  \mnras, 407, L6

\bibitem[{{Wagner} \& {Bicknell}(2011)}]{Wagner:2011}
{Wagner}, A.~Y. \& {Bicknell}, G.~V. 2011, \apj, 728, 29

\bibitem[{{Wagner} {et~al.}(2012){Wagner}, {Bicknell}, \&
  {Umemura}}]{Wagner:2012}
{Wagner}, A.~Y., {Bicknell}, G.~V., \& {Umemura}, M. 2012, \apj, 757, 136

\bibitem[{{Walcher} {et~al.}(2015){Walcher}, {Coelho}, {Gallazzi}, {Bruzual},
  {Charlot}, \& {Chiappini}}]{Walcher:2015}
{Walcher}, C.~J., {Coelho}, P.~R.~T., {Gallazzi}, A., {et~al.} 2015, \aap, 582,
  A46

\bibitem[{Walker {et~al.}(2003)Walker, Boccas, Bonati, Galvez, Martinez,
  Schurter, Schmidt, Ashe, Delgado, \& Tighe}]{Walker:2003}
Walker, A.~R., Boccas, M., Bonati, M., {et~al.} 2003, The SOAR Optical Imager

\bibitem[{{Walter} {et~al.}(2017){Walter}, {Bolatto}, {Leroy}, {Veilleux},
  {Warren}, {Hodge}, {Levy}, {Meier}, {Ostriker}, {Ott}, {Rosolowsky},
  {Scoville}, {Weiss}, {Zschaechner}, \& {Zwaan}}]{Walter:2017}
{Walter}, F., {Bolatto}, A.~D., {Leroy}, A.~K., {et~al.} 2017, \apj, 835, 265

\bibitem[{{Waters} {et~al.}(2016){Waters}, {Magnier}, {Price}, {Chambers},
  {Burgett}, {Draper}, {Flewelling}, {Hodapp}, {Huber}, {Jedicke}, {Kaiser},
  {Kudritzki}, {Lupton}, {Metcalfe}, {Rest}, {Sweeney}, {Tonry}, {Wainscoat},
  {Wood-Vasey}, \& {Builders}}]{Waters:2016}
{Waters}, C.~Z., {Magnier}, E.~A., {Price}, P.~A., {et~al.} 2016, ArXiv
  e-prints [\eprint[arXiv]{1612.05245}]

\bibitem[{{Weaver} {et~al.}(2018){Weaver}, {Husemann}, {Kuntschner},
  {Mart{\'{\i}}n-Navarro}, {Bournaud}, {Duc}, {Emsellem}, {Krajnovi{\'c}},
  {Lyubenova}, \& {McDermid}}]{Weaver:2018}
{Weaver}, J., {Husemann}, B., {Kuntschner}, H., {et~al.} 2018, \aap, 614, A32

\bibitem[{{Weilbacher} {et~al.}(2012){Weilbacher}, {Streicher}, {Urrutia},
  {Jarno}, {P{\'e}contal-Rousset}, {Bacon}, \& {B{\"o}hm}}]{Weilbacher:2012}
{Weilbacher}, P.~M., {Streicher}, O., {Urrutia}, T., {et~al.} 2012, SPIE Conf.
  Ser., 8451

\bibitem[{{Weilbacher} {et~al.}(2014){Weilbacher}, {Streicher}, {Urrutia},
  {P{\'e}contal-Rousset}, {Jarno}, \& {Bacon}}]{Weilbacher:2014}
{Weilbacher}, P.~M., {Streicher}, O., {Urrutia}, T., {et~al.} 2014, in
  Astronomical Society of the Pacific Conference Series, Vol. 485, Astronomical
  Data Analysis Software and Systems XXIII, ed. N.~{Manset} \& P.~{Forshay},
  451

\bibitem[{{Whittle}(1985)}]{Whittle:1985a}
{Whittle}, M. 1985, \mnras, 213, 1

\bibitem[{{Wilms} {et~al.}(2000){Wilms}, {Allen}, \& {McCray}}]{Wilms:2000}
{Wilms}, J., {Allen}, A., \& {McCray}, R. 2000, \apj, 542, 914

\bibitem[{{Wisotzki} {et~al.}(2000){Wisotzki}, {Christlieb}, {Bade},
  {Beckmann}, {K{\"o}hler}, {Vanelle}, \& {Reimers}}]{Wisotzki:2000}
{Wisotzki}, L., {Christlieb}, N., {Bade}, N., {et~al.} 2000, \aap, 358, 77

\bibitem[{{Woo} {et~al.}(2016){Woo}, {Bae}, {Son}, \& {Karouzos}}]{Woo:2016}
{Woo}, J.-H., {Bae}, H.-J., {Son}, D., \& {Karouzos}, M. 2016, \apj, 817, 108

\bibitem[{{Woo} {et~al.}(2017){Woo}, {Son}, \& {Bae}}]{Woo:2017}
{Woo}, J.-H., {Son}, D., \& {Bae}, H.-J. 2017, \apj, 839, 120

\bibitem[{{Woo} {et~al.}(2015){Woo}, {Yoon}, {Park}, {Park}, \&
  {Kim}}]{Woo:2015}
{Woo}, J.-H., {Yoon}, Y., {Park}, S., {Park}, D., \& {Kim}, S.~C. 2015, \apj,
  801, 38

\bibitem[{{Wright} {et~al.}(2010){Wright}, {Eisenhardt}, {Mainzer}, {Ressler},
  {Cutri}, {Jarrett}, {Kirkpatrick}, {Padgett}, {McMillan}, {Skrutskie},
  {Stanford}, {Cohen}, {Walker}, {Mather}, {Leisawitz}, {Gautier}, {McLean},
  {Benford}, {Lonsdale}, {Blain}, {Mendez}, {Irace}, {Duval}, {Liu}, {Royer},
  {Heinrichsen}, {Howard}, {Shannon}, {Kendall}, {Walsh}, {Larsen}, {Cardon},
  {Schick}, {Schwalm}, {Abid}, {Fabinsky}, {Naes}, \& {Tsai}}]{Wright:2010}
{Wright}, E.~L., {Eisenhardt}, P.~R.~M., {Mainzer}, A.~K., {et~al.} 2010, \aj,
  140, 1868

\bibitem[{{Wylezalek} \& {Morganti}(2018)}]{Wylezalek:2018b}
{Wylezalek}, D. \& {Morganti}, R. 2018, Nature Astronomy, 2, 0181

\bibitem[{{Wylezalek} \& {Zakamska}(2016)}]{Wylezalek:2016}
{Wylezalek}, D. \& {Zakamska}, N.~L. 2016, \mnras, 461, 3724

\bibitem[{{Xu} {et~al.}(2015){Xu}, {Rieke}, {Egami}, {Haines}, {Pereira}, \&
  {Smith}}]{Xu:2015}
{Xu}, L., {Rieke}, G.~H., {Egami}, E., {et~al.} 2015, \apj, 808, 159

\bibitem[{{Yim} {et~al.}(2014){Yim}, {Wong}, {Xue}, {Rand}, {Rosolowsky}, {van
  der Hulst}, {Benjamin}, \& {Murphy}}]{Yim:2014}
{Yim}, K., {Wong}, T., {Xue}, R., {et~al.} 2014, \aj, 148, 127

\bibitem[{{Yusef-Zadeh} {et~al.}(2009){Yusef-Zadeh}, {Hewitt}, {Arendt},
  {Whitney}, {Rieke}, {Wardle}, {Hinz}, {Stolovy}, {Lang}, {Burton}, \&
  {Ramirez}}]{Yusef-Zadeh:2009}
{Yusef-Zadeh}, F., {Hewitt}, J.~W., {Arendt}, R.~G., {et~al.} 2009, \apj, 702,
  178

\bibitem[{{Zakamska} \& {Greene}(2014)}]{Zakamska:2014}
{Zakamska}, N.~L. \& {Greene}, J.~E. 2014, \mnras, 442, 784

\bibitem[{{Zakamska} {et~al.}(2016){Zakamska}, {Hamann}, {P{\^a}ris}, {Brandt},
  {Greene}, {Strauss}, {Villforth}, {Wylezalek}, {Alexandroff}, \&
  {Ross}}]{Zakamska:2016}
{Zakamska}, N.~L., {Hamann}, F., {P{\^a}ris}, I., {et~al.} 2016, \mnras, 459,
  3144

\bibitem[{{Zhang} {et~al.}(2016){Zhang}, {Shi}, {Rieke}, {Xia}, {Wang}, {Sun},
  \& {Wan}}]{Zhang:2016}
{Zhang}, Z., {Shi}, Y., {Rieke}, G.~H., {et~al.} 2016, \apjl, 819, L27

\bibitem[{{Zschaechner} {et~al.}(2018){Zschaechner}, {Bolatto}, {Walter},
  {Leroy}, {Herrera}, {Krieger}, {Kruijssen}, {Meier}, {Mills}, {Ott},
  {Veilleux}, \& {Weiss}}]{Zschaechner:2018}
{Zschaechner}, L.~K., {Bolatto}, A.~D., {Walter}, F., {et~al.} 2018, \apj, 867,
  111

\end{thebibliography}
